%% file: main.tex
\documentclass[preprint]{elsarticle}

\usepackage{booktabs} 
\usepackage[utf8]{inputenc}
\usepackage{amsmath}
\usepackage{amsthm}
\usepackage{amssymb}
\usepackage{mathrsfs} 
\usepackage{lscape}
\usepackage{multirow} 
\usepackage{comment}  
\usepackage{balance}
\usepackage{url}

\usepackage{kantlipsum}
\usepackage{dblfloatfix}

\usepackage{siunitx}

\usepackage{relsize,amsmath}

\usepackage{cases}

\usepackage{bbm}
\usepackage{subfig}
\usepackage{color}				
\usepackage{graphicx}	
\usepackage{epsfig}	
\usepackage{algorithm}
\usepackage[normalem]{ulem}
\usepackage[noend]{algpseudocode}
\newtheorem{theorem}{Theorem}
\newtheorem{proposition}[theorem]{Proposition}

\newtheorem{definition}[theorem]{Definition}

\newcommand{\pubj}{_{ij}} 

\begin{document}
\title{Fairness in Online Social Network Timelines: \\ Measurements, Models and Mechanism Design}

\author[add1]{Eduardo Hargreaves\corref{cor1}}
\ead{eduardo@hargreaves.tech}
\author[add2]{Claudio Agosti}
\ead{claudio.agosti@tracking.exposed}
\author[add1]{Daniel Menasché}
\ead{sadoc@dcc.ufrj.br}
\author[add3]{Giovanni Neglia}
\ead{giovanni.neglia@inria.fr}
\author[add4]{Alexandre Reiffers-Masson}
\ead{reiffers.alexandre@gmail.com}
\author[add3]{Eitan Altman}
\ead{eitan.altman@inria.fr}

\address[add1]{Federal University of Rio de Janeiro,Rio de Janeiro, Brazil} 

\address[add2]{University of Amsterdam (Amsterdam,Netherlands)}

\address[add3]{INRIA (Sophia Antipolis, France)}

\address[add4]{Indian Institute of Science (Bangalore, India)}
\cortext[cor1]{Corresponding author}

\input{abstract}

\maketitle

\input{intro}

\input{methodology}
\input{empirical}

\input{model}

\input{validation}

\input{mechanism}

\input{fairness}

\input{related}
\input{conclusion}

\clearpage
\pagebreak

\bibliographystyle{ACM-Reference-Format}

\bibliography{newsfeed-bibliography}

\input{appendix}

\end{document}

%% file: abstract.tex
\begin{abstract}
 Facebook News Feed personalization  algorithm  has a significant impact, on a daily basis, on
 the lifestyle, mood and opinion of millions of Internet users. Nonetheless, the behavior of such  algorithm  lacks transparency,   motivating measurements, modeling and analysis in order to understand  and improve its properties.  In this paper, we propose a reproducible  methodology encompassing measurements,  an analytical model and a fairness-based News Feed design.  The  model leverages the versatility and analytical tractability of time-to-live (TTL)  counters to capture the visibility and occupancy of publishers over a News Feed.   Measurements  are used to parameterize and to validate the expressive power of the proposed model.  Then, we conduct a what-if analysis to assess the visibility and occupancy  bias incurred by users against  a baseline derived from the model.  Our results indicate that a significant bias exists and it is more prominent at the top position of the News Feed.  In addition, we find that the bias is non-negligible even for  users that are  deliberately set as neutral with respect to their political views, motivating the proposal of  a novel and  more transparent  fairness-based News Feed design.  

\end{abstract}


%% file: intro.tex
\section{Introduction}

\emph{Background.} Online social networks (OSNs) have an increasingly important influence in the life of millions of Internet users, shaping their mood, tastes and political views~\cite{bond201261, jones2017social}.  In essence, the goal of OSNs is to allow users to connect and to efficiently  share meaningful information.  To this aim, one of the key building blocks of OSNs is their filtering algorithm, which personalizes content made available to each user of the system.   Facebook, for instance, developed the \emph{News Feed} algorithm for that purpose~\cite{Facebook_2bi, Facebook_top_stories}.


\emph{Challenges.}
 The News Feed algorithm is a  recommendation system that shows posts to users based on  inferred  users' preferences, trading among   possibly conflicting factors while prioritizing posts~\cite{facebook1, facebook2,Krishnasamy2016}.  Hence,  the News Feed algorithm shares common features  with traditional recommender systems, such as those used by Netflix and Spotify, to recommend movies and music.   
 For instance, in all such systems  users typically do not provide explicit feedback about  recommendations.
Nonetheless, due to the nature of OSNs, the News Feed algorithm also poses its own set of challenges, related to the measurement, modeling and control of publishers' visibilities.  
 
Facebook users may be unaware of the influence of the filtering they are subject to~\cite{Eslami2015a}. Such lack of awareness   favors the creation of  a \emph{filter bubble} that reinforces  users' opinions by selecting the users information diets~\cite{pariser2011filter,juhiInfoDiets,Rossi2018}. While researchers are willing to understand the influence of Facebook through users' News Feed~\cite{eslami2016first},  the News Feed algorithm uses sensitive data about preferences, which precludes the sharing of datasets.   Public datasets, in turn, are needed in order to parameterize models to reason about  how the News Feed is populated.

Models are instrumental to perform what-if analysis, e.g., to understand how the News Feed would behave in the presence of different filtering algorithms.  In addition, analytical models can also serve as building blocks towards novel mechanisms to design News Feed algorithms.     Such  foundational development of  principled mechanisms to populate the News Feed is key to build transparency into the system.


\emph{Prior art.} The literature on the News Feed algorithm  includes  measurements~\cite{Cheng2018,BESSI2016319,bessi2016users,Bucher2012}, models~\cite{altman2013competition,dhounchak2017viral} and  user awareness surveys~\cite{Eslami2015}.  
%
Nonetheless, most of the prior work that quantifies the effect of OSNs on information diffusion with large datasets~\cite{Bakshy2012,Bakshy2015,Guillory2014,Sun2009,bond201261} relies on measurements obtained through restrictive non-disclosure agreements that are  not made publicly  available to other researchers and practitioners.   As the data analyzed in such studies is very sensitive, and their sources are not audited, there are multiple potential factors and confounding variables that are unreachable to the general public.   

 \emph{Goals.} Our goal is to provide insights on the filtering that occurs in OSNs through  a  reproducible methodology and a dataset  made  publicly available.\footnote{https://github.com/EduardoHargreaves/Effect-of-the-OSN-on-the-elections} 
Given such measurements, we pose the following questions:
\begin{enumerate}
\item what would be the occupancies of the various sources under alternative scenarios wherein different filtering algorithms are in place? 
\item how to design mechanisms to populate  timelines in a principled fashion, accounting for users preferences and providing content diversity, e.g., under a fairness-based framework? 
\end{enumerate}

To address the first question above, we propose an analytical model for the News Feed.  The model   allows us to derive the occupancy and visibility of each publisher at users' timelines, as a function of the considered  filtering process.  Using the model, we conduct what-if analysis, e.g., to assess  publishers' visibilities in a scenario without filters.

Then, we use the proposed model to build fairness-driven mechanisms to populate  timelines.  Utilities are used to capture the preferences of users with respect to the exposure to posts from different publishers.  The  mechanism leverages  results on utility-based cache design~\cite{Dehghan2016a}, and accounts for fairness among publishers through  utility functions.

\emph{Contributions.}  In this paper we take important steps towards measuring, modeling, auditing and designing timelines. Our key contributions are summarized as follows.

\textbf{A measurement methodology } 
 is introduced to publicly and transparently audit the OSN ecosystem, focusing on the Facebook News Feed algorithm.  The methodology encompasses an Internet browser extension to autonomously and independently collect information on  the posts presented to users by the News Feed algorithm (Section \ref{sec:methodology}). Such  information  is not available through the   Facebook API. 

\textbf{Empirical findings} are reported using data collected from a measurement campaign conducted during the 2018 Italian elections.
%
We observed that $a)$ the 
filtering algorithm 
is impacted by the profile of pages that  a user ``likes'', $b)$ this effect is more prominent at the topmost News Feed position and $c)$ neutral users are also exposed to non-uniform filtering (Section~\ref{sec:empiricalfindings}). 

\textbf{An analytical model } is proposed to quantify the visibility and occupancy of publishers in the users' News Feeds. The model allows us to conduct a what-if   analysis, to assess the metrics of interest under different filtering mechanisms and is validated using data from the Italian election experiment (Sections~\ref{sec:model} and~\ref{sec:validation}).

\textbf{A fairness-driven mechanism  design } is proposed, leveraging the proposed model and measurements (Section~\ref{sec:mechanism}). Given a user profile, that ``likes'' a certain subset of publishers,  the measurements are used to parameterize a simple instance of the model. Then, a family of $\alpha$-fair utility functions are used to allocate resources to publishers subject to users preferences under a utility maximization framework.  

\textbf{A model-based bias assessment } is conducted where the News Feed occupancies are contrasted against an unfiltered resource allocation baseline to quantify the \emph{bias}, i.e., how publishers' occupancies are affected by the News Feed algorithm (Section~\ref{sec:utilitaly}).

%% file: methodology.tex
\section{Measurement methodology}

\label{sec:methodology}

The goal of our experiments is to assess the bias experienced by OSN users through a reproducible method. 
To this aim, we  created controlled virtual users that have no social ties and that follow  the same set of publishers.   By considering minimalistic user profiles, we can assess how preferences towards publishers affect  posts presented to users removing, for instance, the effect of social connections.    

\subsection{Terminology}

Next, we introduce some basic terminology. 
\emph{Publishers} produce \emph{posts} that are fed into users' News Feeds. Each user consumes posts from his/her News Feed. A \emph{News Feed} is an ordered list of posts, also referred to as a timeline, presented to a given user.  News Feed may refer to the algorithm used by Facebook to fill the timeline, or to the timeline itself.

Users \emph{follow} publishers that they are interested in.  The News Feed of a user is filled up with posts from publishers that they follow.   A user may follow a publisher  to have posts from that publisher in the user's News Feed. 
A user who \emph{likes}  a publisher automatically follows that  publisher. A user likes a publisher to show general support for  its posts. In our work,  users orientations are established by letting    them \emph{like}  a subset of the preselected publishers.

\subsection{Data collection methodology}

Next, we present our measurement methodology. 
Although this methodology is general, for concreteness our description is based on the 2018 Italian Parliament election, which constitutes  the key case study considered in this paper.  The Italian  election was held on March 4,  2018, and our experiment was conducted between January~10, 2018 and March 6, 2018, encompassing the preparation for  the election campaign and the reactions to its outcome.

 We asked some Italian voters to select a set of thirty representative  public  Facebook pages, six for each of the following five political orientations: center-left, far-right, left, five-star movement (M5S) and right. Appendix~\ref{appendix:list_publishers} contains the selection of representative  pages and their respective political orientations mapping.
 The classification of publishers into political categories is  debatable, 
 but our focus in this paper is on the methodology rather than on specific political conclusions. Moreover,  most of our results are detailed on a per-publisher basis, as a measurement-based political orientation classification is out of scope of this paper(see, e.g~\cite{mediabiasmonitor}). 
Then, we created six  virtual Facebook 
users, henceforth also referred to as \emph{bots}. Each bot followed \emph{all} the thirty selected \emph{pages}. We gave to five bots a specific political orientation, by making each of them ``like'' pages from the corresponding publishers. The sixth bot does not ``like'' any page, i.e., it has no political orientation. We call it  \emph{undecided}.

Each  bot   kept  open an Internet   browser window  (Firefox or Chrome)  accessing the Facebook page.  The bots were instrumented to collect data on the posts to which they were exposed.  
To that aim, a browser extension named Facebook Tracking Exposed
~\cite{fbtrex} was developed.
The extension auto-scrolls the Facebook window at pre-established instants of the day.  Every auto-scroll produces a  set of posts which are stored at a local database.  Each set of posts is referred to as a \emph{snapshot}.  Each bot was scheduled to collect thirteen snapshots per day. Snapshots were collected once every hour, 
from 7 am to 7 pm (Italian local time). 
Each post appearing    in a snapshot counts as  a post  \emph{impression}.   
At each bot, 
the developed browser extension collects all impressions and records their corresponding publisher,  publication  time, impression time, content, number of ``likes'' and number of shares.
We also have a second dataset which contains the set of \emph{all} posts published by the thirty pages during the interval of interest, as provided by the Facebook API.  This dataset is  used to study what  users would experience in the absence of  filters, or in the presence of alternative filters.
%
%
%
%
%
%

Information about impressions used to be available in 2015, in a deprecated version of the Facebook API.   In any case, that information was not necessarily reliable as recognized by  Facebook itself~\cite{Facebook_API_home}. 
For such reasons, we believe that the developed browser extension  and the methodology described in this section constitute  important building blocks to promote transparency  in the Facebook ecosystem. 

\subsection{Measurement challenges}

\label{sec:gapmeasures}
\emph{Gaps in measurements:}      During our 
 measurement campaign, we experienced measurement  gaps due to two reasons: 1) the computer running a bot went down, due to unforeseen glitches such as lack of power and 2) at random points in time, either Facebook or related applications, such as the browser itself,    solicit human interaction (e.g., by showing a pop-up  requiring  users to answer simple questions before proceeding). 
%
We denote by $S_i$ the number of snapshots collected by the $i$-th bot. 
In our experiments,  the bots are indexed from 1 to 6, denoting  center-left, far-right, left, M5S, right and undecided orientations. The values of $S_i$ equal 
577, 504, 623,
674, 655, 576, for $i=1, \ldots, 6$. To account for the different number of snapshots, all the reported results rely on  values averaged across snapshots rather than quantities that depend on the absolute number of snapshots.

\emph{Small number of  bots:} we use six bots to capture different perspectives on the Facebook dynamics. 
Each bot provides a \emph{personal perspective} on the system, which is well aligned with our goals.  
Although we considered a small population size,  we believe that  the limited points of view provided by the six bots already shed  important insights on the biases introduced by Facebook.  In particular, the consistent biases observed in our dataset, reported in the sections that follow,  indicates that  the collected sample is  representative.    

%

\subsection{Metrics of interest}
\label{sec:metricsofinterest}

We define our key metrics of interest that will be obtained from the dataset generated by the experiment. We consider the top $K$ positions of the News Feed of each user.  
\begin{definition}[visibility]
Let $\pi_{ij}$ be the fraction   
of snapshots from user $i$ that contain at least one post from publisher $j$.   
\end{definition}

\begin{definition}[occupancy]
Let $N_{ij}$ be the average number of posts of publisher $j$ in the News Feed of user $i$.  
\end{definition}
We refer to $\pi_{ij}$ and $N_{ij}$ as the \emph{visibility} and the \emph{occupancy}  of publisher $j$ at News Feed $i$, respectively.  
The \emph{normalized occupancy}
 is given by $N_{ij}/K$.  
 The visibility and the (normalized) occupancy are two metrics of exposure of publishers to users~\cite{Bakshy2015}.
 
 \begin{definition}[hit probability]
Let $h_{ij}$ be the probability that user $i$ sees (or clicks) on a post of publisher $j$. 
\end{definition}

We  refer to $h_{ij}$ as the hit probability of publisher $j$ at user $i$. Then, $h_{ij}=N_{ij}/K$  if user $i$ goes  through all the top posts in the News Feed, and $h_{ij}=\pi_{ij}$ if he/she picks uniformly at random a single post in the News Feed. 


%% file: empirical.tex
\section{Empirical findings}
\label{sec:empiricalfindings}

\begin{figure*}[t!]
\begin{tabular}{cc}
  \includegraphics[width=1\textwidth]{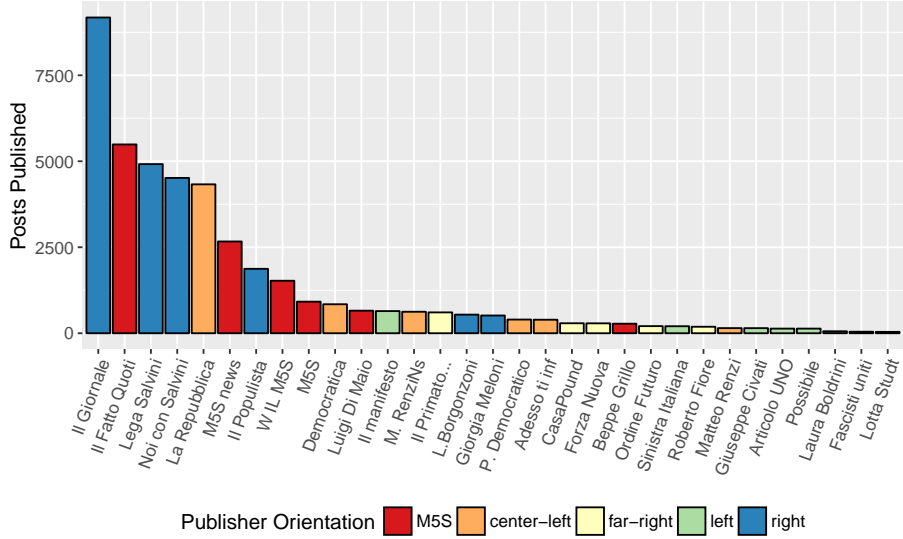} & \\
  (a)\\
  \includegraphics[width=1\textwidth]{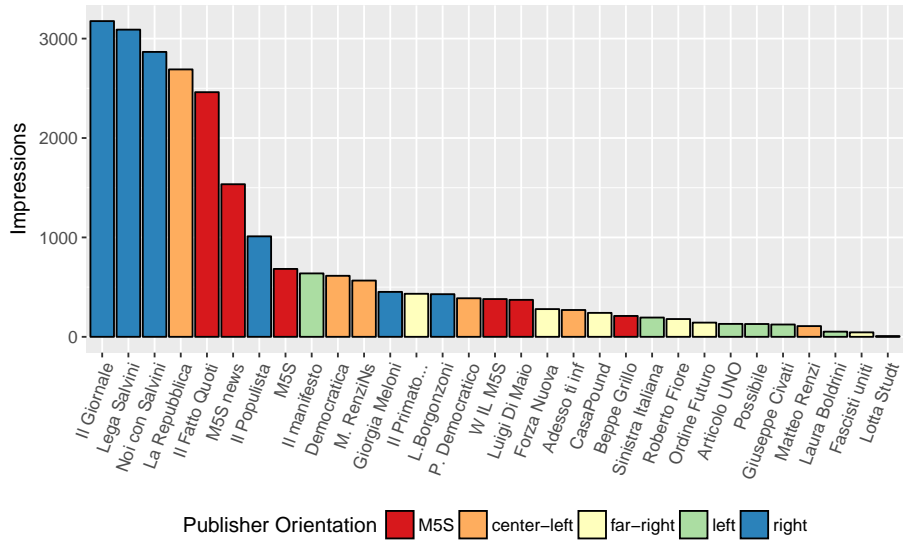} \\ 
     (b) \\ 
\end{tabular}
\caption{(a) Posts per publisher, (b) posts per publisher that were seen by at least one of our bots.}
\label{fig:pubfontes}
\end{figure*}

In the following two sections, we report our empirical findings from the perspective of publishers and users.

\subsection{The effects of filtering on publishers}

Next, we report findings on the behavior of the publishers  and their general effect on users' News Feeds.   Figure~\ref{fig:pubfontes}(a)  shows the number of unique posts per publisher.  We denote by $C_j$ the number of posts of publisher $j$.  This information was collected directly from the Facebook API. A few publishers generated thousands of posts during the considered time frame, whereas the majority generated tens of posts. 

Figure~\ref{fig:pubfontes}(b) shows the number of impressions per publisher. 
This information was collected from our Facebook extension. 
Publishers are ordered based on the number of posts generated and seen in Figures~\ref{fig:pubfontes}(a) and \ref{fig:pubfontes}(b), respectively.   It is worth noting that the distinct order at which publishers appear  in those figures is fruit of the filtering  experienced by the users.  In what follows, such filtering is further analyzed through measurements (Section~\ref{sec:effec}), models (Section~\ref{sec:model}) and a combination of the two (Sections~\ref{sec:validation} and~\ref{sec:mechanism}).


\subsection{The effects of filtering on users}
 \label{sec:effec}


\paragraph{The effect of filtering is stronger at the topmost News Feed position} \label{sec:topstrong}

\begin{figure*}[t!]
\begin{center}
\begin{tabular}{cc}    
\includegraphics[width=1\textwidth]{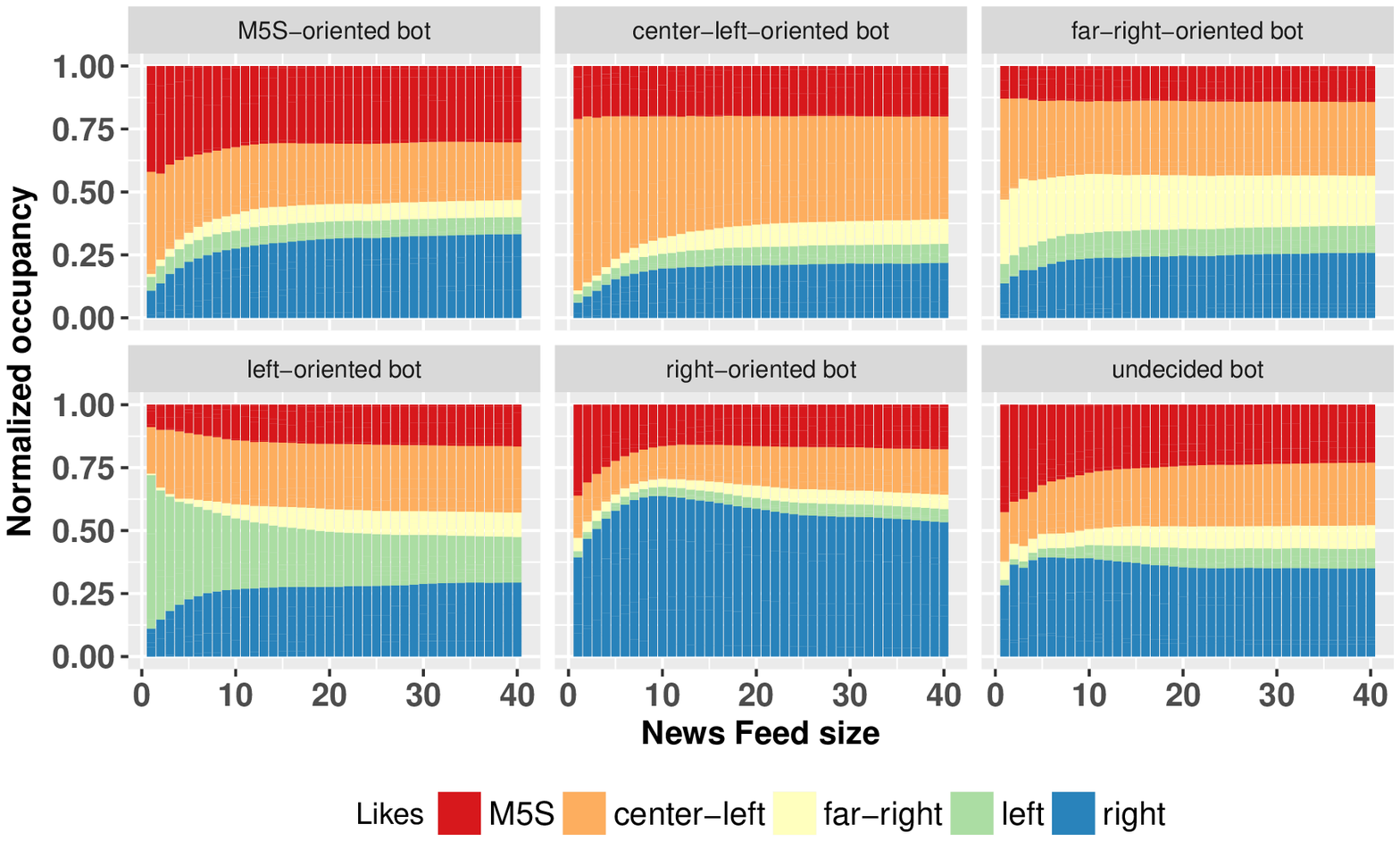} & \\
(a)\\
 \includegraphics[width=1\textwidth]{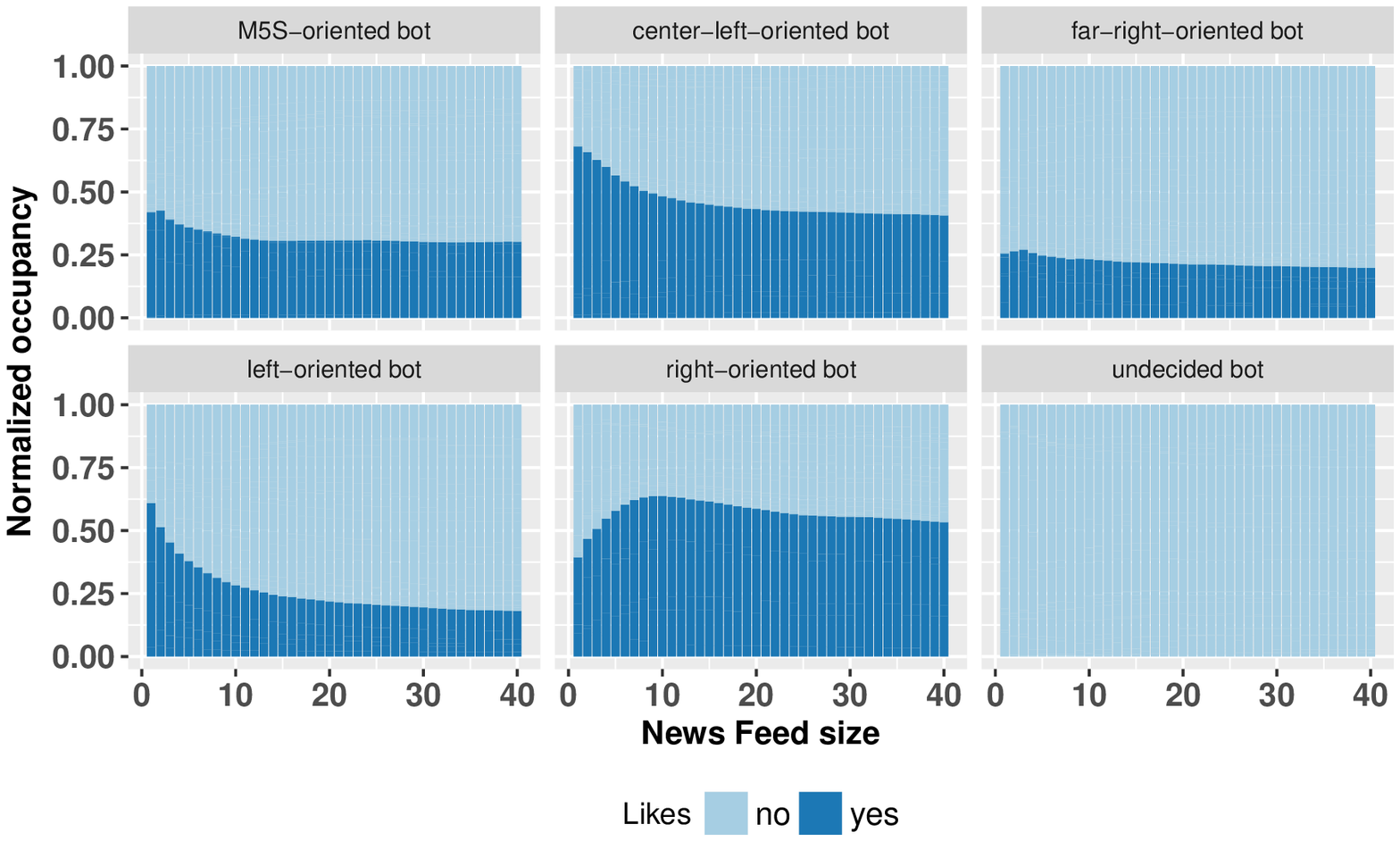}  \\
         (b) 
\end{tabular}
\end{center}
\caption{Normalized occupancy  as a function  of   $K$, classified by $(a)$ publisher orientations and $(b)$ user preferences. }   
\label{fig:visib_occ_poli_over_k}
\end{figure*}

Figures~\ref{fig:visib_occ_poli_over_k}(a) and~\ref{fig:visib_occ_poli_over_k}(b) 
 show the normalized publisher occupancy,   as a function of the News Feed size (the corresponding   visibilities are reported in Appendix~\ref{sec:visibilities}). The publishers are colored according to their political orientation (Fig.~\ref{fig:visib_occ_poli_over_k}(a)) and user preferences (Fig.~\ref{fig:visib_occ_poli_over_k}(b)). Figure~\ref{fig:visib_occ_poli_over_k}(a)  shows that the occupancy distribution over the five orientations changes with the considered News Feed size.

In Figure~\ref{fig:visib_occ_poli_over_k}(b), a  publisher is colored in blue (resp., red) at a given bot if the bot  ``likes'' (resp., does not ``like'') the corresponding publisher.   Note  that, except for the right and far-right bots, the normalized occupancy of publishers that users ``like'' is maximum at the topmost position, 
achieving more than $70\%$ at the center-left oriented bot. The right-oriented bot achieved a similar normalized occupancy when $K=10$. The noteworthy  bias on the topmost  position must be placed under scrutiny, as  there is a strong correlation between post positions and click rates~\cite{Bakshy2015,message}. These figures also reveal that the amount of exposure to cross-cutting contents depends on the size of the News Feed.  
 

\emph{Occupancy is impacted by orientation.}  Figure~\ref{fig:visib_occ_poli_over_k}(a)
also shows that the occupancies  are impacted by the orientation of the bots. For instance, the News Feed of the bot with a center-left orientation was occupied mostly by center-left (red) publishers.  As a notable exception, center-left posts were prevalent in the News Feed of the bot with a far-right orientation, where far-right posts are responsible for roughly 25\% of the normalized occupancy. Nonetheless,  the occupancy of  far-right posts in that bot was still the highest among all bots.

\begin{figure*}[h!]
        \includegraphics[width=1.0\textwidth]     {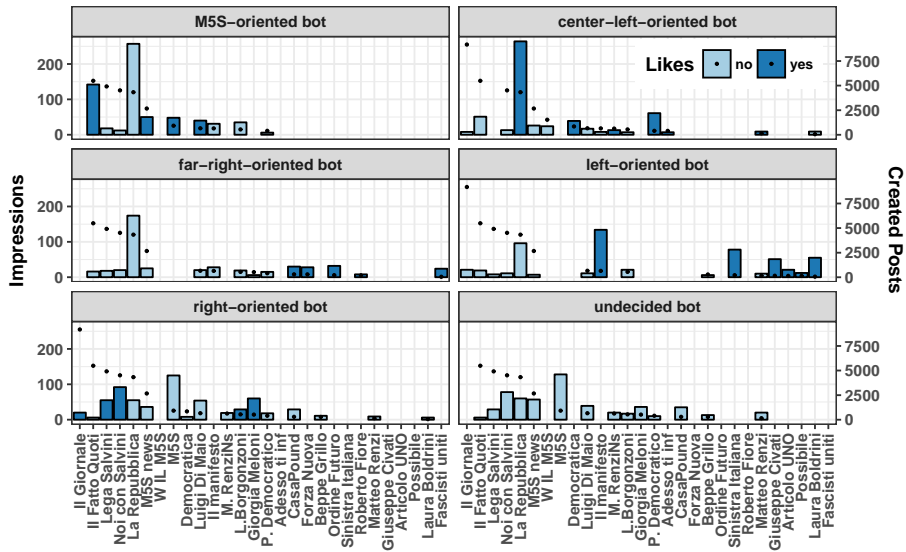} 
\caption{Publishers impressions at the six bots  (bars colored by preferences) and number of created posts (black dots). The number of created posts is  represented when at least one impression from the corresponding publisher was observed. }
\label{fig:pubfontes1}
\end{figure*}

\emph{Noticeable publishers selection.} The bars in Figure~\ref{fig:pubfontes1} show the total number of impressions per publisher in the topmost position of the News Feed of each bot (the color of the bars indicates orientation). For the sake of readability, only publishers that achieved a normalized occupancy larger than $5\%$ are represented in this figure.  The black dots correspond to  the number of posts created by each publisher  
(the publishers are  ordered by the number of posts generated).  Figure~\ref{fig:pubfontes1} shows that only a small subset of publishers are represented in  topmost positions. For example, the center-left bot sees 
primarily posts from two of the publishers that it ``likes''.  Moreover, the number of impressions per publisher is not proportional to the number of posts the publisher generated, a further indication of a filtering effect from News Feed algorithm.  


\emph{Neutral users are also exposed to non-uniform filtering.} 
It is worth noting that filtering affects also the “undecided” bot, with some publishers over-represented in the News feed.

%% file: model.tex
\section{News Feed model}

\label{sec:model}

Next, we present the proposed News Feed model to derive occupancy and visibility metrics.   We start by presenting the basic rationale behind the model in Section~\ref{sec:insights}. Then, the model is introduced in Section~\ref{sec:ttlnews}.

\subsection{Insights on  News Feed modeling}
\label{sec:insights}
Next, we introduce some of the key ideas that inspire the analytical model introduced in the following section.

\subsubsection{Queues, caches and the News Feed} \label{sec:queues}
%
%
%
In the simplest  setting, posts are organized at each News Feed in a \emph{first in, first out} (FIFO) fashion. Then, the personalization algorithm at the News Feed of user $i$  
filters posts from each of the publishers.   Given the  rate  $\Lambda_j$ at which publisher $j$ creates posts, we denote by $\lambda_{ij}$ the corresponding  \emph{effective arrival rate}   at which posts from publisher $j$ are published  at the News Feed of user $i$.



We assume that a News Feed has $K$ slots.  
Under the FIFO approximation described above, new posts are inserted at the top of the News Feed and each new arrival shifts older posts one position lower.
A post is evicted from the News Feed when it is shifted from position $K$.  
Although this is a preliminary step to capture the real News Feed operation, it cannot capture  the stronger filtering at the topmost News Feed positions that we observed in Section~\ref{sec:empiricalfindings}.

In the remainder of this paper, we consider a generalization of the FIFO model, which  accommodates  different residence times for different posts using time-to-live (TTL)  counters~\cite{Dehghan2016a,Kelly1997,netecon}.    Under the TTL model, every time a post is inserted into a News Feed, it is associated to a timer (TTL), and the content remains in the News Feed until its timer expires.      In what follows, we further detail the similarities and differences between TTL caches and the News Feed.



\subsubsection{The News Feed is a publisher-driven cache}

Next, we leverage a recently established connection between timelines in OSNs and caches~\cite{netecon}, as summarized in Table~\ref{tab:diferences}.  %

\emph{Triggering events}  In the News Feed, the insertion of new posts is triggered by their creation. In caches, in contrast, user requests typically lead to content insertion and eviction.    When proactive caching and prefetching of content is considered, the proactive caching is still usually fruit of  content requests \cite{tadrous2016optimal, bastug2014living,golrezaei2013femtocaching}.

\emph{Nature of contents and requests}
When News Feed users  search for  user-generated content, they  are typically  interested in a class of items related to a given category. 
The demand for News Feeds posts is elastic. Consider, for instance, a user interested in the latest headlines from his favorite newspaper, or a college student willing to learn about the latest developments of his football team.    There may be multiple posts that satisfy the demands of such users.
The literature of caches, in contrast, presumes that each of the cached items is  uniquely identified and non-substitutable, and that demand is inelastic.   



\emph{Classes of items}  The occupancy of caches is given by the items that it stores, where each item is uniquely identified. Caches do not store repeated items.  For the purposes of this work, in contrast, posts stored in a News Feed are distinguished  solely by their publisher, and  the News Feed may store multiple posts from the same publisher.  Whereas the News Feed  behaves as a cache that can store multiple copies of  items of the same class,   in traditional caches each item corresponds to its own class.

\emph{Capacity} For all practical purposes, the News Feed can be assumed to be infinite in size,  i.e., the News Feed can admit all the  published posts. Nonetheless, it is well known that  the topmost positions of the News Feed receive higher visibility.  This, in turn, motivates our study of the (average) topmost $K$ positions of the News Feed.

  
\begin{table}[t]
\centering
\footnotesize
\caption{Comparison between News Feed and caches }
\label{tab:diferences}
\begin{tabular}{l||l|l}
\hline 
 
 & News Feed & Cache \\
\hline
\hline 
Trigger event  & post publishing & content request\\
\hline 
Insertions and  & after a post creation & after a miss \\
evictions & or user engagement  & \\
 \hline
 \multicolumn{3}{c}{Nature of contents and requests} \\
\hline 
 Cache stores & multiple items of content & at most one copy \\
 &  from given class  & of a specific item \\
 \hline
 Requests  for & general content classes  &  specific content items\\
 \hline
  \multicolumn{3}{c}{Classes of items} \\
\hline 
Control & of items by &  of specific content\\ 
occupancy & given publisher & items \\
\hline
 \multicolumn{3}{c}{Capacity} \\
\hline 
Capacity & infinite (average $K$ topmost & finite \\
&   positions more relevant) & \\
 \hline
\end{tabular} \label{table:compare}
\vspace{-0.2in}
\end{table}

\subsection{TTL News Feed model} \label{sec:ttlnews}

Next, we introduce  the proposed News Feed model.  To each content posted in a News Feed we associate a time-to-live (TTL) timer.  The timer is set to $T$ when the content is inserted, and is decremented at fixed time intervals. Once  the timer reaches zero, it expires and the associated content is removed from the News Feed.
 
 \subsubsection{Why TTL model?}

 


It is well known that Facebook uses recency as a parameter to show posts to users~\cite{How_Facebook_NewsFeed_Works,insidenewsfeed}. TTL counters are a natural way to capture the perishable  nature of  posts. Furthermore, TTL counters are a flexible way to extend FIFO schemes.  
 In general, TTL-based models are well-suited to represent objects with expiration times \cite{Jung2003a}.
 

While proposing the TTL model of a News Feed, our aim is not to argue whether  Facebook  deploys TTL counters, which is out of the scope of this paper.  Instead, our goal is to show that a simple model  can already capture the dynamics of  Facebook News Feed.  Then, we leverage the flexibility of TTL counters to propose novel News Feed algorithms.

\subsubsection{Model description}



Let $\mathcal{I}$ be 
the set of $I$ users, and let  $\mathcal{J}$ 
be the set of $J$  publishers: $\mathcal{J}_i$ denotes the set of publishers 
followed by user $i \in \mathcal{I}$.  
Publisher $j \in \mathcal{J}$  publishes posts according to a Poisson process with rate $\Lambda_j$. The total publishing rate is $\Lambda=\sum_{j=1}^J \Lambda_j$. 
%
%
Whenever a content is generated, it is immediately sent to the News Feed of all users.  In what follows, we provide further details about  how user $i$ reacts to the content arrival.

\emph{Content and publisher classes}  We consider $C$ content classes.  Each content class corresponds to a set of posts published at a given user News Feed. In the most general case, each user-publisher pair is associated to a given content class. In that case, the  class associated to the $i$-th user and $j$-publisher is denoted by the ordered pair $(i,j)$.  

Alternatively, we associate  each user-publisher pair to one of two classes.  Class $l_i$ (resp., $\overline{l}_i$) is the class of contents generated by publishers that the $i$-th user  ``likes'' (resp., does not ``like'').    We denote by $L(i,j)$ the indicator variable which characterizes the set of publishers that a user likes,
\begin{equation}
L(i,j) = \left\{
\begin{array}{ll}
1, & \textrm{if user } i \textrm{ ``likes''  publisher } j \\
0, & \textrm{otherwise}
\end{array}  \right. \label{eq:likeij}
\end{equation}


\emph{Order of posts} The simplest instance of the proposed model corresponds to a FIFO queue, wherein contents are ordered in the News Feed based on the instant at which they are posted, and  new arrivals shift older posts (Section~\ref{sec:queues}).  The  TTL model, in contrast, does not presume any pre-established ordering of posts in the News Feed.  In particular, it is flexible to account for  eventual  rearrangements  of posts. Note that  the TTL model can be parameterized to capture the behavior illustrated in Figure~\ref{fig:visib_occ_poli_over_k}, wherein the bias present in the topmost positions is different from that seen in the remainder of the News Feed.   This occurs, for instance,  if posts with larger TTL  are placed on the top of the News Feed, producing different biases at different News Feed positions.


\emph{Timer classes} We consider per-class content dynamics. For concreteness, except otherwise noted we let $L(i,j)$ be the class of contents generated by publisher $j$ at the News Feed of the  $i$-th user.

Whenever a content from class $L(i,j)$ is generated, it is inserted in the News Feed of the $i$-th user and a timer with value  $T_{L(i,j)}$    is associated to that content.  Even though we assume, for simplicity, that the initial timer values are set to a fixed constant, our analysis also holds if the initial values of the timers are  sampled from a probability distribution with mean $T_{L(i,j)}$.

We expect that $T_1 \ge T_0$, i.e., contents generated by publishers that the user ``likes'' remain longer in the News Feed, when compared against those that the user does not ``like''.  Nonetheless, we do not explicitly   assume any relationship between $T_1$ and $T_0$.  Instead, we perform simple    consistency  checks using the collected measurement data (Section~\ref{sec:validation}).  

\subsubsection{Metrics of interest} Next, we derive the metrics of interest corresponding to an infinite capacity News Feed. 
Recall that we assume that  users scroll up to an average of $K$ News Feed positions, i.e., we study the visibility and occupancy of an average of $K$  topmost positions. 
%
%
To simplify notation  we drop the explicit dependence  of metrics and corresponding variables on the value of $K$, e.g.,  denoting $N_{ij}(K)$ and $T_{L(i,j)}(K)$ simply as $N_{ij}$ and $T_{L(i,j)}$.

The occupancy of the $j$-th publisher at the $i$-th News Feed, $N_{ij}$, follows from Little's Law and  is given by
\begin{equation}
N_{ij}= \Lambda_{j}  T_{L(i,j)} \label{eq:little1}
\end{equation}
The expected number of slots occupied in a News Feed is given by the sum of $ N_{ij}$, for all $j$,
\begin{align}
\sum_{j\in\mathcal{J}} N_{ij} &= K.
\label{eq:K_decomposition}
\end{align}

The visibility of publisher $j$ at the News Feed of user $i$, $\pi\pubj$, is given by 
\begin{equation}
\pi_{ij}=1-e^{- N_{ij}}.
\label{theo:visibility_general}
\end{equation}
The equation above follows from the observation that the dynamics of posts by the $j$-th publisher at the  News Feed of user $i$ are given by  an  M/G$/\infty$ queue.  Arrivals of posts occur with rate $\Lambda_j$, and each arrival remains in the News Feed for an average of $T_{L(i,j)}$ time units.  The probability that there is at least one customer in the  M/G$/\infty$ equals the probability that there is at least one post from publisher $j$ at the News Feed of user $i$, and is given by~\eqref{theo:visibility_general}.

\subsubsection{Special case: FIFO News Feed}  \label{sec:fifospecial}
When  all posts are associated to the same TTL $T$, and contents are inserted in the News Feed in the order that they are created, the TTL model behaves as a  FIFO queue, as described in Section~\ref{sec:queues}. If we further allow posts to be filtered, we refer to the resulting model as a \emph{filtered FIFO} model.  Let  $p_{ij}$ be the filtering probability.  In a filtered FIFO model, we let $T=0$ with probability $1-p_{ij}$, and $T=\overline{T}$ otherwise, where $\overline{T}$ is a fixed and given constant, $\overline{T}>0$.

As before, we assume that publisher $j \in \mathcal{J}$  publishes posts according to a Poisson process with rate $\Lambda_j$. Recall that the total publishing rate is given by $\Lambda=\sum_{j=1}^J \Lambda_j$.   In the filtered model, let $\lambda\pubj \leq \Lambda_j$ be the effective arrival rate of posts published by $j$ in the News Feed of user $i$. Then,
 $\lambda_{ij} = p_{ij} \Lambda_j$.

Under the filtered FIFO model,~\eqref{eq:little1} together with~\eqref{eq:K_decomposition} imply that
\begin{equation}
\overline{T} = \frac{K}{ \sum_k \lambda_{ik} }.
\label{eq:ttl}
\end{equation}
The visibility $\pi_{ij}$ is given by \eqref{theo:visibility_general}, where 
\label{eq:visibility_eq_case}
\begin{equation}
N_{ij}=\frac{\lambda_{ij} K}{\sum_l \lambda_{il}}.
\label{eq:occupancy_eq_case}
\end{equation}
If we further assume that $p_{ij}=p$ for all user-publisher pairs, we obtain the uniformly filtered FIFO model, where
\begin{equation}
N_{j}=\frac{\Lambda_j K}{\Lambda}.
\label{eq:occupancy_eq_case_unif}
\end{equation}
As we assume that all users follow the same set of sources, under the uniformly filtered FIFO model the expected occupancy of publisher $j$ at user $i$,  $N_{ij}$, is the same for all users. Therefore, in this case we denote it simply as $N_{j}$.

\subsubsection{Finite size FIFO News Feed}  The analysis presented above assumes an infinite size  News Feed, wherein users are interested, on average, at the topmost $K$ positions.  Alternatively, consider a finite size FIFO News Feed, which can accommodate up to $K$ posts.  We assume that a News Feed has $K$ slots, new posts are inserted at the top of the News Feed and each new arrival shifts older posts one position lower.
A post is evicted from the News Feed when it is shifted from position $K$.

We denote by $\lambda_i$  the aggregate rate of  posts published in the News Feed of user $i$, $\lambda_i=\sum_{j=1}^J \lambda_{ij}$. We further let  $\lambda_{i,-j}$ be the arrival rate of posts in the News Feed of user $i$ from all publishers other than  $j$, $\lambda_{i,-j}= \lambda_i-\lambda_{ij}$.

%
%
%
%
%
The occupancy of contents of publisher $j$ follows from Little's law and  is given by
\begin{equation}\label{eq:occupancy_fifo}
N_{ij}= {\lambda\pubj K}/{\lambda_i}. 
\end{equation}
The visibility of publisher $j$ is given by 
\begin{equation}
\pi_{ij}= 1-\left(\frac{\lambda_{i,-j}}{\lambda_i}\right)^K =  1-\left(1-\frac{N_{ij}}{K}\right)^K, \label{eq:piijfifofinite}
\end{equation} 
and    the rationale goes as follows. After every new arrival,  with probability ${\lambda_{i,-j}}/{\lambda_i}$ the topmost post of publisher $j$ will be shifted down by one unit.  After $K$ consecutive shifts, which occur with probability $({\lambda_{i,-j}}/{\lambda_i})^K$, publisher $j$ will not be visible at the News Feed of user $i$. 
When $K=1$ we have $N_{ij}=\pi_{ij}$. For large values of $K$, \eqref{eq:piijfifofinite} can be approximated by~\eqref{theo:visibility_general}. 

Under the    FIFO instances of the model presented above,  posts are filtered  uniformly at random.  For this reason, such instances yield  simple baselines against which the filtering effects introduced by the News Feed algorithm can be compared. In Section~\ref{sec:mechanism} we revisit  the FIFO model under this perspective.

\begin{table}[h]
\centering
\begin{tabular}{l l}
\hline
\hline
Variable & description \\
\hline
$j$ & $j$-th publisher \\
$i$ & $i$-th News Feed user \\
$\Lambda_j$ & post creation rate by publisher $j$\\
$\lambda_{ij}$ & arrival rate of posts from $j$ at user $i$ \\
$\lambda_i$ & total arrival  rate of posts at user $i$\\
$L(i,j)$ & 1 if user $i$ likes publisher $j$, 0 otherwise \\
\hline
\multicolumn{2}{c}{TTL model variables} \\
\hline
$T_l$ & expected TTL for  posts from class $l$, $l = L(i,j)$\\
$T_{i}^{(l)}$ & expected TTL  for posts from class $l$, \\
& with  user discrimination,  $l = L(i,j)$ \\
$T_{ij}$ & expected TTL for  posts from publisher $j$ at user $i$ \\
$w_{L(i,j)}$ & weight associated to class-$L(i,j)$  \\
\hline
\multicolumn{2}{c}{Metrics of interest as estimated by the model} \\
\hline
$h_{ij}$ & hit probability of publisher $j$ at user $i$\\
$\pi_{ij}$ & visibility of publisher $j$ at user $i$ \\
$N_{ij}$ &  occupancy of publisher $j$ at user $i$ \\
\hline
\multicolumn{2}{c}{Metrics of interest as obtained from measurements} \\
\hline
$\tilde{\pi}_{ij}$ & measured visibility of $j$ at $i$\\
$\tilde{N}_{ij}$ & measured occupancy  of $j$ at $i$\\

\hline
\multicolumn{2}{c}{Measurements} \\
\hline
$I_{ij}$ & number of impressions from publisher $j$ at user $i$ \\
& (counting repeated posts multiple times) \\
$Q_{ij}$ &  number of unique posts from publisher $j$ at user $i$ \\
$S_i$ & number of snapshots collected by user $i$ \\
$C_j$ & number of unique posts created by publisher $j$ \\
$G_l$ & number of user-publisher-post tuples  \\
& corresponding  to user-publisher pairs in class $l$ \\
$I_l$ & number of impressions with user-publisher  in class $l$ \\ 
\hline
\end{tabular}
\caption{Table of notation}
\label{tab:notation}
\vspace{-0.2in}
\end{table}

%% file: validation.tex


\section{A Model-based perspective at the measurement findings}
\label{sec:validation}


In this section, we take a model-based perspective at the measurement findings.  First, we implicitly account for user ``likes'' through their impact on the effective arrival rates in Section~\ref{sec:effrate}. Then, we explicitly account for ``likes'' in Section~\ref{sec:param2class}.

\subsection{Indirectly accounting for   ``likes'': a multi-class perspective on measurements} \label{sec:effrate}

We validate the proposed  model  using data from the 2018 Italian elections, through a multi-class perspective on the measurements.  Each user-publisher pair is associated to a class.  Class $(i,j)$ is associated to the $i$-th user and the $j$-th publisher, and corresponds to an effective arrival rate of $\lambda_{ij}$ posts  per snapshot.

We start by introducing some additional notation regarding our measurements. The notation is summarized in Table~\ref{tab:notation}. Let $Q_{ij}$ be the measured numbered of unique posts from publisher $j$ at user~$i$.  Let $S_i$ be the number of snapshots taken by user $i$. 
Then, the measured effective arrival rate is given by  
\begin{equation}
\tilde{\lambda}\pubj=\frac{Q\pubj }{ S_i}.
\end{equation}
%
Note that $\tilde{\lambda}\pubj$ is modulated through  ``likes''.  
In this section, at each bot $i$  we assume that the same timer is used for all the publishers.  Then,  the TTL model is  equivalent to  FIFO (see Section~\ref{sec:fifospecial}).  Publishers' presence in the News Feed is discriminated upstream by filtering their posts before they arrive to the News Feed, i.e., through rates $\lambda_{ij}$. For this reason, in this section we use  model equations~\eqref{theo:visibility_general}-\eqref{eq:occupancy_eq_case}   
with the  arrival rate at a bot, $\lambda_{ij}$, set to  the measured one, $\tilde{\lambda}_{ij}$. ``Likes'' are indirectly taken into account through rates.

\begin{figure*}[t]
\begin{center}
\begin{tabular}{cc}   \includegraphics[width=0.5\textwidth]{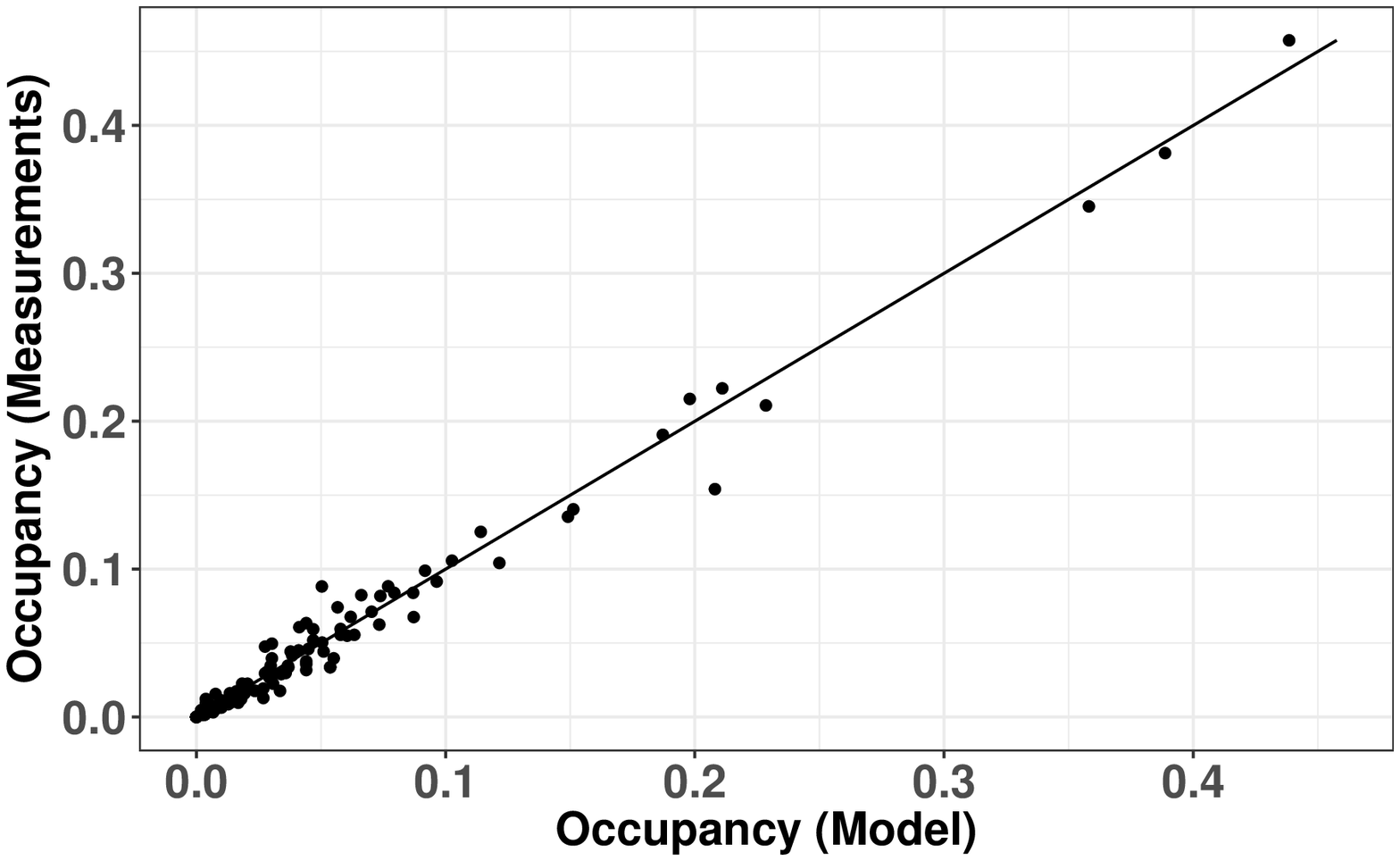} &
\includegraphics[width=0.5\textwidth]{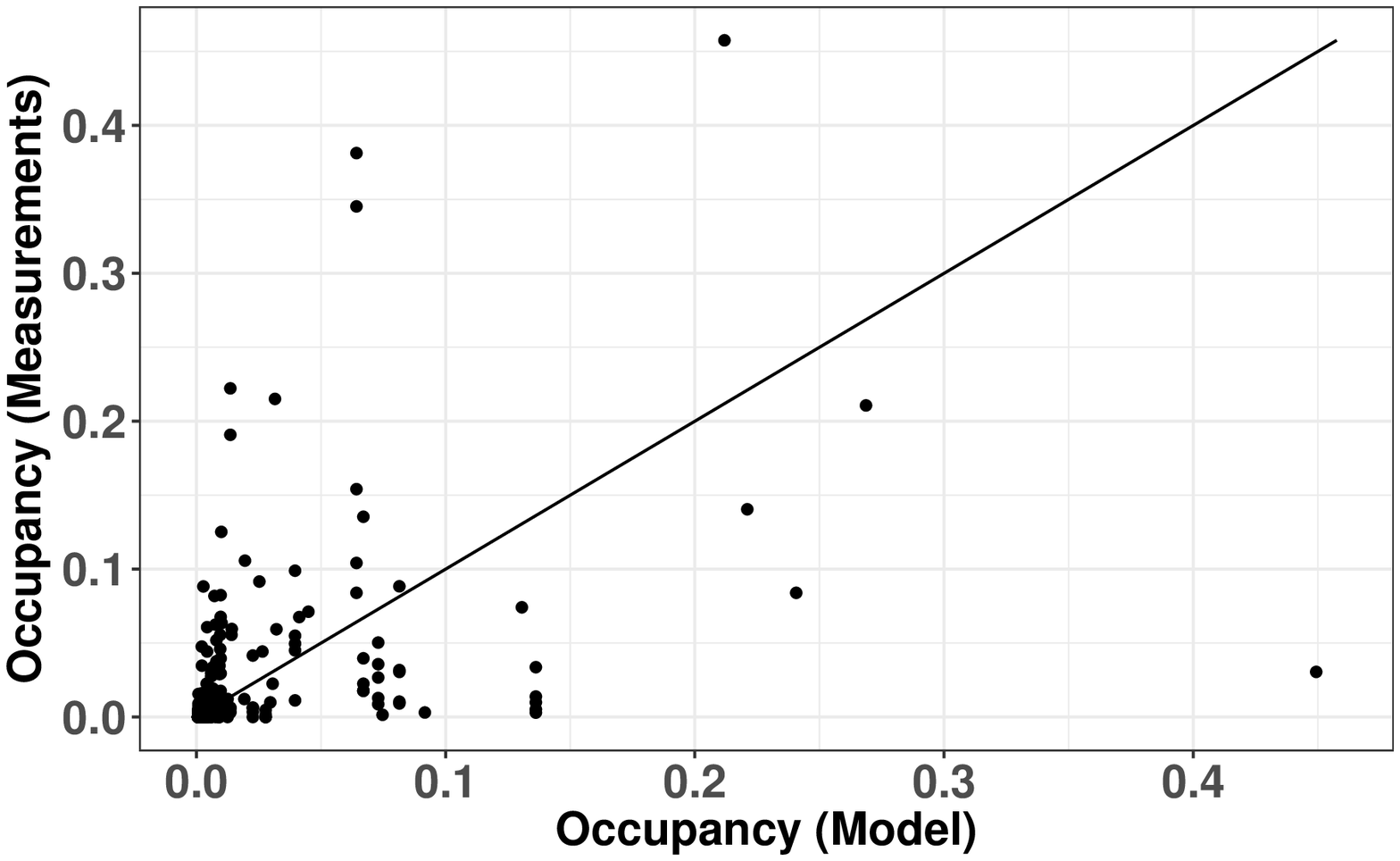}\\ 
        (a) & (b)  
\end{tabular}
\end{center}
\caption{Model validation  of occupancies for $K=1$: $(a)$  multi-class and $(b)$ two-class.}
\label{fig:validation_TTLk1}
\end{figure*}

Figure ~\ref{fig:validation_TTLk1}(a) compares  measured occupancies against
model predictions at the News Feed topmost position ($K=1$).
 Each point  corresponds to a user-publisher pair.  A point $(x={N}_{ij},y=\tilde{N}_{ij})$  indicates that, for the given pair, 
an occupancy   $N_{ij}$ estimated by the proposed model using eq.~\eqref{eq:occupancy_eq_case} corresponds to a  measured occupancy $\tilde{N}_{ij}$.  
Most of the points are close to the $\tilde{N}_{ij}=N_{ij}$ line, indicating the expressive power of the model. Appendix~\ref{sec:addmodelval} contains  results   
for $K>1$, accounting for    visibility in addition to occupancy as the target metric.




Next, our goal is to quantitatively assess the expressive power of  the multi-class model.   To this aim, 
we conduct a linear regression followed by an hypothesis test on the  coefficients produced by the linear regression. 
Let the measured occupancy be given as a function of the model-based occupancy as follows,
\begin{equation}
\tilde{N}_{ij}=\beta_1 N_{ij} +\beta_0.
\end{equation}
The null and alternative hypotheses are given by
\begin{itemize}
\item $H_0$: there is no  relationship between  $\tilde{N}_{ij}$ and $N\pubj$, i.e., $\beta_1=0$;
\item  $H_a$: there is  relationship between  $\tilde{N}_{ij}$ and $N\pubj$, i.e., $\beta_1 \neq 0$
\end{itemize}
The \textit{p-value} for  $\beta_1=0$ is less than $2^{-16}$, allowing us to reject the null hypothesis.  We repeated the test for all values of  $K$ ranging from $1$ to $30$, and obtained similar results as indicated in Table~\ref{tab:summaryhyp}.

\begin{table}
\center 
\begin{tabular}{l|l|lll|lll}
\hline \hline
\multicolumn{2}{c}{} & \multicolumn{3}{|c}{Occupancy} & \multicolumn{3}{|c}{Visibility}  \\
\hline
Model          & $K$ & $p$-value & RMSE & $R^2$   & $p$-value & RMSE & $R^2$ \\
\hline
Multi- &   $1$ &   $<2^{-16}$ & $0.01$  & $0.98$ &   $<2^{-16}$ & $0.02$  & $0.98$\\
  -class          &  $30$ &  $<2^{-16}$ & $0.28$ & $0.93$ &  $<2^{-16}$ & $0.04$ & $0.98$ \\
\hline
Two- &  $1$   & $1.66^{-7}$ & $0.07$   & $0.13$ & $4.56^{-8}$ & $0.06$   & $0.15$\\ 
  -class         &  $30$ &  $<2^{-16}$ &$0.90$  &$0.51$ &  $<2^{-16}$ &$0.25$ & $0.49$\\  
 
\hline
\end{tabular}
\caption{Summary of hypotheses test results} \label{tab:summaryhyp}
\end{table}



\subsection{Directly accounting for  ``likes'': a two-class  measurement analysis} \label{sec:param2class}

In this section, we explicitly account for user ``likes'' in the News Feed occupancies. To this aim, we  divide the publisher-user pairs into two classes, and show the expressive power of the model through a simple parameterization which involves only two parameters, $T_1$ and $T_0$, corresponding to the TTL of posts from publishers that users ``like'' and do not ``like'', respectively.

Let $I_{ij}$ be the number of impressions from publisher $j$ at user $i$.  
Let $I_1$ be the  number of impressions  at News Feeds of users who ``like'' the   publishers of the corresponding impressions  ($I_0$ is similarly defined).   
Then, 
\begin{equation}
I_{l}=\sum_{\forall(i,j) | L(i,j)=l} I_{i,j}.
\end{equation}
Correspondingly, let $G_l$ 
 (resp. $G_0$) be the number of posts  generated, counted as many times as the number of users who ``like'' (do not ``like'') the publishers who generated these posts.
%
Recall that $C_j$ is the number of unique posts created by publisher $j$. 
Then,
\begin{equation}
G_{l}=\sum_{\forall(i,j) | L(i,j)=l} C_j.
\end{equation}
 The  estimate of the TTL associated to class $l$ is given by
 \begin{equation}
 \tilde{T}_{l}={I_l}/{G_l}, \quad l \in \{0,1\}.
 \end{equation}

  In Figure~\ref{fig:validation_TTLk1}(b) each point corresponds to a user-publisher pair. As in Figure~\ref{fig:validation_TTLk1}(a), we let $K=1$ (results for $K=30$ are presented in Appendix~\ref{sec:addmodelval}). 
In Figure~\ref{fig:validation_TTLk1}(b), a point $(x={N}_{ij},y=\tilde{N}_{ij})$ indicates that, for the given pair, 
an occupancy   $N_{ij}$ estimated by the proposed model using eq.~\eqref{eq:little1}  corresponds to a measured occupancy of $\tilde{N}_{ij}$. 
We also resort to the same sort of hypothesis tests described in the previous section, and reject the null hypothesis according to which there is no relationship between the model and the measurements.  A summary of the measurement results, for $K=1$ and $K=30$, is presented in Table~\ref{tab:summaryhyp}. 

The accuracy of the model can also be assessed through the $R^2$ score, which ranges between 1 and 0.  An $R^2$ score of 1 indicates that the variance in the target variable is fully explained by the model. 
As expected, the $R^2$ scores (resp.,  \emph{p-values}) of the two-class model   are smaller (resp., larger) than those of  the multi-class model.  In addition, we observe that the predictive power of the two-class model when applied to the topmost position ($K=1$) is significantly lower when compared against its application to the remainder of the News Feed ($K=30$).  This is partly explained by the fact that the bias is stronger at the topmost position (Section~\ref{sec:topstrong}).
 We leave a more detailed analysis of simple models for the topmost position as subject for future work.

\emph{Summary}
In this section, we evaluated the explanatory power of the model in light of the measurements collected during the Italian 2018 elections. In particular, we have shown that a very simple instance of the model with  two parameters is already able to capture the occupancies experienced by users (Section~\ref{sec:param2class}). In addition, we indicated that a multi-class instance of the model, wherein the number of parameter equals  the number of publishers times the number of bots (180 in the experiment), produces occupancy estimates with higher accuracy, at the expense of additional complexity (Section~\ref{sec:effrate}). In the sections that follow, we leverage the proposed model for mechanism design purposes.



%% file: mechanism.tex
\section{A fairness-based News Feed  mechanism}

\label{sec:mechanism}



Next, we leverage the proposed News Feed model to derive a fairness-based mechanism to design News Feeds.  We present the problem formulation (Section~\ref{sec:utilprob}), followed by its general solution (Section~\ref{sec:utilgeneral}) and by an analysis of $\alpha$-fair utility functions (Section~\ref{sec:utilalpha}). Then, we bridge the utility maximization framework and  measurements to illustrate the applicability of the mechanism (Section~\ref{sec:utilitaly}).



\subsection{Utility maximization formulation} 
\label{sec:utilprob}

We associate to each user-publisher pair a utility function $U_{ij}(h_{ij})$ which is an increasing, strictly concave and continuously differentiable function of $h_{ij}$ over the range $ 0 \leq h_{ij} \leq 1$.  
Recall from Section~\ref{sec:metricsofinterest} that  $h_{ij}$ denotes the hit probability of publisher $j$ at user~$i$, where the hit probability captures the exposure of  publishers to users.  


 We further assume that utilities are additive.  Then, the goal is to maximize the sum of the utilities for each of the individual publishers~\cite{Kelly1997,Dehghan2016a}.
The optimization problem is posed as follows,
%
%
\begin{align}
\max \quad & \sum_{j=1}^J U_{ij}(h_{ij} )  \label{prob:utility_maximization}\\
\mbox{s.t.} \quad & \sum_{j=1}^J N_{ij} = K, \qquad  N_{ij} \geq 0  \label{eq:constraint}
\end{align}
 where $h_{ij}$ are concave and non-decreasing functions of $N_{ij}$. As discussed in Section~\ref{sec:metricsofinterest}, we consider two possible instantiations of the hit probability, 
%
%
corresponding to the normalized occupancy, given by $N_{ij}/K$, and the visibility $\pi_{ij}$,  given by~\eqref{theo:visibility_general} and~\eqref{eq:piijfifofinite} under the TTL and finite capacity FIFO models, respectively.  
%
%
%
In summary,
\begin{subnumcases}{h_{ij}(N_{ij})=} N_{ij}/K, & {     for normalized occupancy,}  \nonumber \\ 
            1-e^{-N_{ij}}, & { for visibility  (TTL model),} \nonumber  \\
            1-\left(1-{N_{ij}}/{K}\right)^K, & {   for visibility (finite FIFO).} \nonumber
\end{subnumcases}
In all cases above, $h_{ij}(N_{ij})$ is an increasing and concave function of $N\pubj$. 
Let $\tilde{U}_{ij}(N_{ij})=U_{ij}(h_{ij}(N_{ij}))$. 
It follows  that 
\begin{itemize}
\item $\tilde{U}_{ij}(N_{ij})$ is 
  non-decreasiing, as $U\pubj$ and $h_{ij}$ are so;
\item $\tilde{U}_{ij}(N_{ij})$ is  concave, as $U\pubj$ is concave and non-decreasing and $h_{ij}$ is concave.
\end{itemize}
Therefore, the objective function~\eqref{prob:utility_maximization} is equivalent to 
\begin{equation}
\max \sum_{j=1}^J \tilde{U}_{ij}(N_{ij}).
\end{equation}
The general solution to the unified problem formulation introduced above is presented in the sequel. 

\subsection{Problem solution} 
\label{sec:utilgeneral}

To solve the convex optimization problem posed above, we introduce  its corresponding Lagrangian, 
%
\begin{equation}
\mathcal{L}({\mathbf{N}},\beta)=\sum_j \tilde U_{ij}(N\pubj)-\beta\sum_j(N\pubj-K)
\end{equation}
where $\mathbf{N}$ is the vector of $N_{ij}$ values, and $\beta$ is the Lagrange multiplier.  
Taking the derivative of the Lagrangian with respect to $N_{ij}$, 
\begin{equation}
\frac{\partial L}{\partial N\pubj}=\tilde{U}^{\prime}\pubj(N \pubj)-\beta. 
\end{equation}

An allocation $N^{\star}\pubj$ is a global optimizer if and only if there exists  $\beta^{\star}$ such that $\mathbf{N^{\star}}$ is feasible  and
\begin{equation}
\label{e:kkt}
\left\{
        \begin{array}{ll}
          \tilde{U}^{\prime}\pubj(N^{\star}\pubj)-\beta^{\star} =0, \quad \mathrm{if} \quad{N_{ij}^{\star}} >0, \\
           \tilde{U}^{\prime}\pubj(N^{\star}\pubj)-\beta^{\star} \leq 0, \quad \mathrm{if} \quad{N_{ij}^{\star}} =0.                        
        \end{array}
    \right.
\end{equation}

At the optimum, all the publishers that appear in the News Feed have assigned a space that equalize their marginal utilities (i.e. $N_{i j}>0$ and $N_{i k}>0$ imply that $\tilde{U}^{\prime}_{i j}(N^{\star}_{i j})=\tilde{U}^{\prime}_{i k}(N^{\star}_{i k})$). The publishers that not appear have smaller marginal utility (i.e., $\tilde{U}^{\prime}_{il}( N^{\star}_{i l})\leq \tilde{U}^{\prime}_{i j}(N^{\star}_{i j}) \quad \mathrm{if} \quad N^{\star}_{i j} >0 $ and $N^{\star}_{il}=0$). Remark that if $\tilde{U}^{\prime} (0)=+\infty$ (e.g. if $U\pubj(h)\propto\log(h)$), then necessarily $N^{\star}\pubj > 0$ for each $j$.

The solution can be found by a water-filling type algorithm: we start from the null vector where no publisher appears in the timeline ($N_{ij}=0$ for all $j$) and we gradually allocate space to the publisher(s) with largest marginal utility(ies).

\subsubsection{Occupancy vs rate-based fairness} \label{sec:occvsrate}  In the problem formulation and solution presented above,  we accounted for occupancy-based  rather than rate-based fairness~\cite{weightedFairCaching}.    To appreciate the distinction between the two sorts of fairness, consider the problem of setting the same average space to different publishers.  Such occupancy-based allocation will penalize   prolific publishers more strongly than a rate-based allocation where publisher rates are uniformly multiplied by a constant factor.  In Section~\ref{sec:utilitaly} we numerically contrast the effect of occupancy-based fairness against a baseline wherein occupancies are proportional to content generating rates.

%% file: fairness.tex
\subsection{News Feed fairness: $\alpha$-fair utilities}
\label{sec:utilalpha}

The optimal allocation of News Feed space to publishers through  problem~\eqref{prob:utility_maximization} depends on the shape of the utility functions. 
The use of a given family of utility functions corresponds to the selection of a fairness criterion. In this section we characterize the optimal allocation under the usual concept of $\alpha$-\textit{fairness} as it is considered in communication networks~\cite{mo2000fair, Kelly1997, destounis2017alpha, Dehghan2016a}.
\begin{equation}
U_{ij}(h_{ij})= \left\{
        \begin{array}{ll}
            w_{ij} \frac{h_{ij}^{1-\alpha}}{1-\alpha},& \quad \alpha \leq 0,\alpha \neq 1. \\
            w_{ij} \log(h_{ij}), & \quad \alpha = 1.\
        \end{array}
    \right.
\end{equation}

In what follows we will consider the case where the hit probability coincides with the normalized occupancy.

\subsubsection{Proportional fairness} Choosing $\alpha=1$ yields  proportional fairness. In this case $\tilde U_{ij}(h_{ij})=w_{ij} \log(N_{ij}/K)$, implying that $N_{ij}^\star>0$ for each publisher. From the first equation in~\eqref{e:kkt}, it follows that 
\begin{equation}
\tilde U^\prime_{ij}\!\left(N_{ij}^\star\right)={w_{ij} }/{N_{ij}^\star}=\beta^\star.
\label{eq:beta_prop_fairnes}
\end{equation}
Imposing the constraint \eqref{eq:constraint} we obtain
\[\beta^\star = {\sum_{j=1}^J w_{ij}}/{K}.\]
Once the value of $\beta^{\star}$ is known, it can be substituted in \eqref{eq:beta_prop_fairnes} to yield
\begin{equation}
N\pubj^\star = \frac{w_{ij} K}{\sum_k w_{ik}}, \qquad T_{ij}^\star=\frac{w_{ij} K}{\Lambda_j \sum_k w_{ik}}. 
\label{eq:beta_prop_fairnes_final}
\end{equation}

\subsubsection{Potential delay fairness}

If $\alpha=2$, $U\pubj(h\pubj)=-w\pubj /h\pubj$, $U^\prime(h\pubj)=w\pubj/h^2\pubj$ and in a similar way we obtain
\begin{equation}
N^{\star}\pubj = K\frac{\sqrt{w\pubj}}{\sum_{j=1}^J \sqrt{w\pubj}}, \qquad T\pubj^\star = \frac{K}{\Lambda_j } \frac{\sqrt{w\pubj}}{ \sum_{j=1}^J \sqrt{w\pubj}}. \end{equation}

\subsubsection{Max-min fairness}

Max-min fairness is the limiting case of $\alpha$-fairness in the limit when $\alpha$ diverges~\cite{Boudec2012}.
In our case a max-min fair allocation corresponds to provide the same occupancy to each publisher. Then, we have:
\begin{equation}
N_{ij}^\star={K}/{ J}, \qquad T_{ij}^\star={K}/({\Lambda_{j} J}).
\end{equation}
Note that the max-min fairness allocation is independent of the weights $w_{ij}$. 

\subsubsection{Summary} In this section, we presented expressions for publishers occupancy and visibility, under three fairness criteria.  In the following section we show how to extend the obtained expressions to account for class-based metrics, and in Section~\ref{sec:utilitaly} we compare the derived class-based metrics against baselines,  illustrating a way  to quantify   biases from real measurements.

\input{classbased}

\input{italian}

%% file: classbased.tex
\subsection{Class-based optimization}

\label{sec:twoclass}

The framework introduced in the previous section for per-publisher optimization can be easily adapted to a per-class optimization.  In this case we divide the user-publisher pairs into classes, and parameters are set in a per-class basis.   To simplify presentation, we specialize the presentation to  classes determined by  user's ``likes'', where for each user $i$, we distinguish the class of publishers $i$ likes (identified by $L(i,j)=1$ in~\eqref{eq:likeij}), and the class of publishers that $i$ follows without expressing likes (identified by $L(i,j)=0$).    
We denote by $\lambda^{(l)}_{i}$ (resp., $N^{(l)}_{ij}$) the aggregate arrival rate (resp., occupancy) of posts of class $l$ in the News Feed of user~$i$,
\begin{equation}
\lambda^{(l)}_{i} = \sum_{j | L(i,j)=l} \Lambda_{j}, \qquad N^{(l)}_{i} = \sum_{j | L(i,j)=l} N_{ij}.
\end{equation}
Let $T_{i}^{(l)}$ denote the  TTL of posts of class $l$ in the News Feed of user~$i$. Then, 
%
%
\begin{equation}
\label{e:occ_proportional}
N\pubj=\Lambda_j T_{i}^{(l)}, \quad \textrm{ where }  l=L(i,j).
\end{equation}
The expression of $T_{i}^{\star (l)}$ for the three  special fairness criteira considered in the previous section can be similarly derived, leading to:
\begin{subnumcases}{T_{i}^{\star (l)}=} \frac{K}{\lambda^{(l)}_i} \frac{w_i^{(l)} }{ \sum_{k=0}^1 w_i^{(k)}}, & \hspace{-0.2in} {proportional fairness,}  \\
\frac{K}{\lambda^{(l)}_i} \frac{\sqrt{w_i^{(l)} } }{ \sum_{k=0}^1 \sqrt{w_i^{(k)}}}, & \hspace{-0.2in} {potential delay fairness,}  \\
\frac{K}{2 \lambda^{(l)}_i}, & \hspace{-0.2in} {max-min fairness.} 
\end{subnumcases}

\subsubsection{Class-based vs publisher-based allocation} \label{sec:classvspub} In a class-based fair allocation, space is allocated fairly across classes:  class  occupancies are determined by the specific fairness criteria, while, inside while inside a given class publisher occupancies are proportional to  publishing rates as indicated by~\eqref{e:occ_proportional}. 
Consider, for instance, a max-min fairness allocation. Then the class of ``liked'' publishers is posed to have the same average occupancy as the remaining publishers, but this does not translate into equal publisher occupancies. For example, in our experimental setting, each of the $6$ publishers ``liked'' by bot $i$ will on average occupy $1/12$-th of bot-$i$ timeline, while any of the other $24$ publishers will on average get $1/48$-th of it.  Hence, the ``liked publishers'' are overall advantaged under max-min fairness allocation. 




%% file: italian.tex
\subsection{Italian election case study}
\label{sec:utilitaly}

\begin{figure*}
 \includegraphics[width=1.0\textwidth]{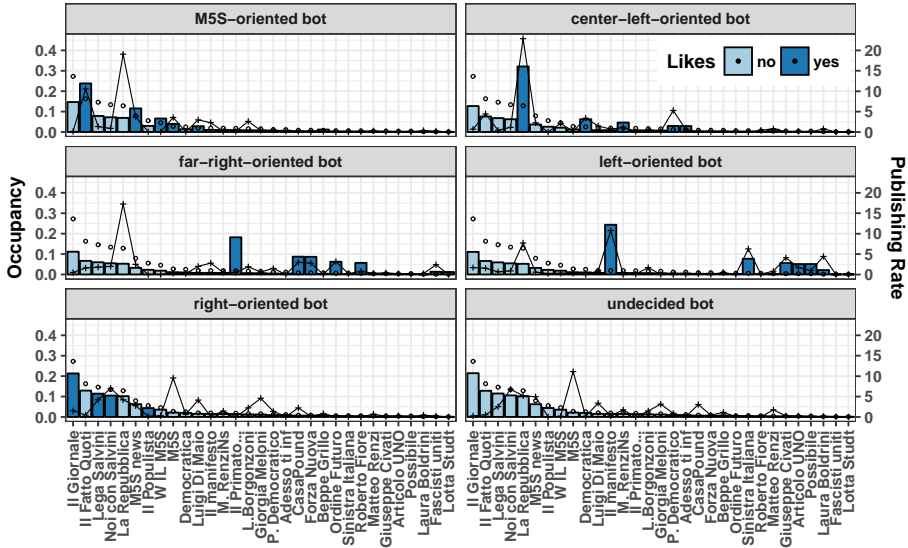} 
\caption{Publishers' occupancies under proportional fairness (bars colored by preferences), Facebook measured occupancies (crosses and lines) and publishing rates (circles),  for the topmost position, at the six bots. 
At the undecided bot,  proportional fairness occupancies decrease  with respect to publishing rates, but measured occupancies deviate from baseline.}
\label{fig:propfairness}
\end{figure*}

Next, we illustrate how the proposed utility-based framework can shed further insights into the Italian election dataset. Throughout this section, we consider a proportional fairness allocation for the two-class model (Section~\ref{sec:twoclass}). In particular here we show results for the topmost position and
$w_i^{(l)}=1$ for each $i$ and $l$, while  in Appendix~\ref{sec:sensitiveweights} we  report results  for a broader set of weights and values of $K$.As discussed in Section~\ref{sec:classvspub},  in this case the ``liked'' publishers get collectively as much timeline space as all the others.

Figure~\ref{fig:propfairness} shows different publishers' occupancies at the $6$ bots when $K=1$: occupancies measured at the bots (crosses and lines), occupancies computed by our model to maximize 2-class proportional fairness (bars) and  publishers' posting rates (circles).

At the undecided bot, all  publishers belong to the same class.
Then, an allocation under proportional fairness yields occupancies proportional to publishing rates. In our measurements, in contrast, we observed that occupancies aren't proportional to rates (Sec~\ref{sec:effec}), which indicates that the filtering effects are non-trivial even for neutral users. 
For all the other bots (except the right-oriented one), we observe that the ``liked publishers'' are favored under the 2-class fairness model. This is due to the fact that the 6 ``liked'' publishers get collectively as much space as all other 24 bots. Under each class, occupancies are split proportionally to publishing rates. Therefore, the ``liked'' publishers end up having less competition for space than other ``non-liked'' publishers with similar publishing rates. The situation is different at the right bot.  Among  the 6 right-oriented  publishers ``liked'' by the right-oriented bot,  we find some of the most prolific publishers in our dataset:   their aggregate rate is almost equal to the aggregate rate of all the other 24 publishers. As a consequence, all the publishers get occupancies almost proportional to their publishing rate.

We observe that our simple 2-class model already qualitatively predicts some of the results observed in our traces: most of the ``liked'' publishers indeed also have larger measured occupancies (crosses and lines in Fig.~\ref{fig:propfairness}). Nonetheless, there are still a number  of exceptions. 
For example, ``La Repubblica'' exhibits a quite large occupancy independently of the orientation of the corresponding bot. This may be justified by the fact that ``La Repubblica'' is  a newspaper, rather than  a party or a candidate, and Facebook News Feed algorithm may be filtering less news. The same, however, does not hold for other newspapers appearing in the list. The occupancy of ``Il Giornale'' at the different News Feeds, for instance,  is quite small (except at the right bot), even though it is the source with the largest publishing rate.

In order to quantify the discrimination among different providers introduced by our utility maximization allocation or by Facebook News Feed filtering algorithm we introduce a new metric.
We define the (occupancy) bias as the difference between a given normalized occupancy estimate  and the normalized occupancy that would have been obtained if timelines were operated according to the FIFO model without content filtering. We denote by $b_{ij}^{(m)} $ the occupancy bias  of publisher $j$ at user $i$. The symbol $m$ refers to the scenario against which the FIFO baseline model is compared. It equals Face, PropF, MaxMinF and PotentF respectively when referring  to the occupancies derived from raw Facebook measurements, proportional fairness, max-min fairness and potential fairness models.  

\begin{definition}[bias] The bias incurred by posts of publisher $j$ at the News Feed of user $i$ is given by 
\begin{equation}
b^{(m)}_{ij}=\frac{N^{(m)}_{ij} - {N}_{j}}{K}. \label{eq:bijm30}
\end{equation}
\end{definition}
In the definition above, $N_j$ is the baseline occupancy under FIFO as given by~\eqref{eq:occupancy_eq_case_unif} and does not depend on the specific bot. We observe that 
$\sum_{j} b^{(m)}_{ij}=0$, as the sum of  occupancies at a given bot equals $K$.
Note that the definition of bias is general, and can be coupled with different baseline models for occupancy (see Appendix~\ref{sec:potential}).

\emph{Discussion} Statistical bias is a systemic deviation, e.g., of an estimator, with respect to the true value of a parameter. 
In this paper  we use the term bias with a different, albeit intuitively related, meaning.  
%
%
  We assume that  users explicitly express their preferences by selecting which publishers they follow and like. Any deviation from the unfiltered occupancy, as estimated by the proposed model, is attributed to  bias. This kind of bias is called \emph{social bias}~\cite{Bias_on_the_web}.   Note that our definition of bias does not necessarily entail a negative connotation, as biasing users  towards their interests might be a desired feature.

\begin{figure*}[h!]
        \includegraphics[width=1.0\textwidth]     {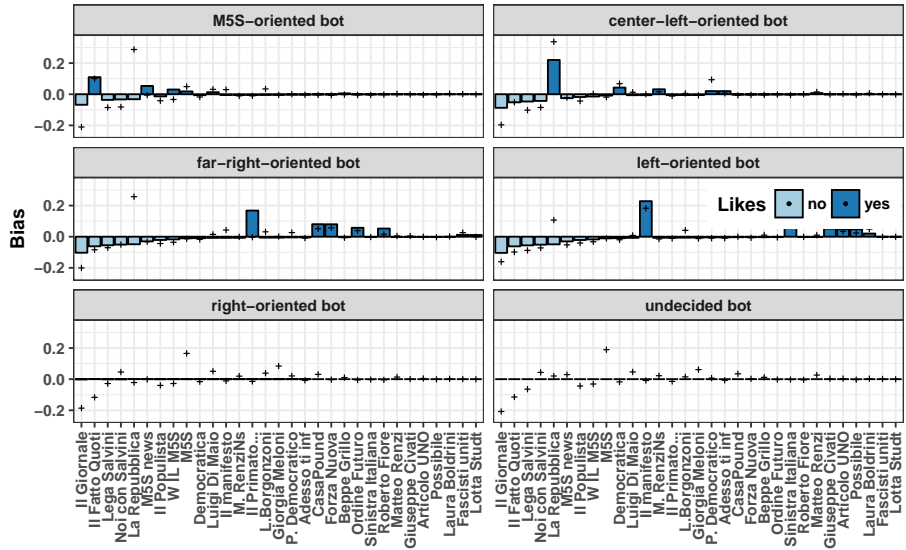} 

\caption{Proportional fairness  bias (bars) and  Facebook News Feed bias (crosses),  for  the six bots }
\label{fig:bias_top}
\end{figure*}

\emph{Example}
Figure \ref{fig:bias_top} numerically illustrate the behavior of  bias, 
for the top publishers in the 2018 Italian election dataset, letting $K=1$, respectively (recall that $b_{ij}^{\textnormal{(PropF)}}=b_{ij}^{\textnormal{(MaxMinF)}}=b_{ij}^{\textnormal{(PotentF)}}$ as in this section we assume $w_i^{(l)}=1$, and see Appendix~\ref{sec:potential} for additional results). 
Figure \ref{fig:bias_top} supports the observations reported from Fig. \ref{fig:propfairness}. In particular we observe that $b_{ij}^{\textnormal{(PropF)}}$ is null or almost null for the undecided bot and for the right bot, while for the others there are significant positive (resp., negative) biases for the ``liked'' (resp., ``non-liked'') contents. Facebook biases $b_{ij}^{\textnormal{(Face)}}$ appear to be qualitatively aligned with those estimated through the 2-class proportional fairness model, , particularly at the center-left and left bots.   Nonetheless, there are a number of exceptions, as discussed above.

We observe that the Facebook News Feed algorithm produces high biases even at the undecided bot. In addition, the bias at the topmost position  does not reflect  user preferences, specially at the far-right, right and undecided bots. Our analysis sheds light on some peculiar choices of Facebook that are difficult to explain even taking into account users' preferences as expressed through ``likes.''

\subsection{Practical implications}

The proposed mechanisms evidence the challenges involved in building a fair News Feed.  Any fair solution must trade between conflicting goals such as 1) giving more exposure to favorite publishers, 2) reserving space for non-preferred publishers, 3) penalizing publishers that produce irrelevant content at high rates (e.g., spam) and 4)  avoiding   bias towards  publishers that produce at low rates.  We believe that the proposed mechanisms constitute a principled way to cope with such conflicting goals, through the use of utility functions.   

Ultimately, users should be aware of the filtering that they are exposed to, and tune their utilities based on their needs.  
By promoting user awareness, 
the risk of amplifying  filter bubbles should be mitigated.
A brief discussion of the compliance of the proposed methodology to Facebook policies is presented in a short preliminary version of this work~\cite{eduardofosint}.

One way to implement the proposed mechanisms  is through tools such as MIT Gobo~\cite{gobo} or FeedVis~\cite{Eslami2015a}. 
Such social media aggregators can be fed by posts from different sources, e.g., Facebook, Twitter and Youtube, providing filters that users can  control. Therefore, users can explicitly set mechanisms to decide what is  edited out of their News Feed. 
We envision that  
these tools can be coupled with  variants of the News Feed control  mechanisms proposed in this paper. Then,   A/B tests  may be used to select which mechanisms are best suited to different users based on their explicit and implicit feedback, e.g., obtained  through questionnaires and click-rates.

%% file: related.tex
\section{Related work}

\label{sec:related}
The literature on Facebook News Feed includes topological aspects related to cascading structures and growth ~\cite{Gjoka2010,anatomy_fB,Cheng2018} and its effects on the creation of echo chambers and  polarization~\cite{BESSI2016319,bessi2016users}. 

\subsection{Social networks, TTL-counters and utility-based allocation}

The proposed News Feed model relies on TTL counters.  TTL-based caching mechanisms   are versatile and flexible, and can be used to reproduce the behavior of traditional caching policies such as LRU and FIFO   \cite{fofack1,tortellimodel,Dehghan2016a,Ferragut,Neglia2017}. In this paper, we leverage the analytical tractability of  TTL-based caches, showing how to adapt them 
for the purposes of modeling and  optimization of the News Feed.

The implications  of the limited budget of attention of users in OSNs have been previously studied by Reiffers-Masson \emph{et al.}~\cite{masson2017posting} and  Jiang \emph{et al.}~\cite{jiang2013optimally}.  In these two papers, the authors consider the problem of optimally   deciding what  to post and to  read, 
 respectively.  Such works are complementary to ours. To the best of our knowledge, none of the previous works considered the problem of inferring the visibility of publishers from News Feed measurements, and using such measurements to parameterize models and propose utility-based mechanisms  for News Feed design.





\subsection{Fairness, accountability and transparency}

The literature on fairness, accountability and transparency (FAT) is rapidly growing~\cite{fatconference, fate, fates}, accounting for its implications on  social networks~\cite{valenzuela2009there}, risk score  estimation~\cite{kleinberg2016inherent, barabas2017interventions}, recommender system~\cite{Krishnasamy2016,Zehlike2017,Celis2017,Singh2018}, resource allocation~\cite{Lan2010}, individuals classification in order to prevent discrimination~\cite{dwork2012fairness} and computational policy~\cite{stanfordcomp, gilbert2018computational}, using tools such as causal analysis~\cite{zhangfairness}, quantitative input influence~\cite{datta2016algorithmic} and  machine learning~\cite{zafar2015learning, pleiss2017fairness}.

Surveys and books on notions of fairness include  those by Moulin~\cite{moulin_fair_division}, 
Zliobaite~\cite{vzliobaite2017measuring}, 
Rmoei and Ruggieri~\cite{romei2014multidisciplinary},   Narayanan~\cite{talk21defs} and~Drosou \emph{et al.}~\cite{drosou2017diversity}.  The later is a  survey about  diversity concepts from the  information retrieval literature~\cite{an2013individuals}. The mechanism design proposed in this  paper is  both  a diversity-aware and fairness-aware allocation scheme, taking users' and publishers' perspectives, respectively.

ACM~\cite{Garfinkel2017} introduces a set of principles intended to ensure fairness in the evolving policy and technology ecosystem: awareness, access and redness, accountability, explanation, data provenance, auditability, and validation and testing. We particularly focus on  awareness, explanation and auditability, as we do not rely on the Facebook API to collect impressions. Algorithmic transparency is one of the cornerstones of the 
 General Data Protection Regulation (GDPR), which stresses the importance of  providing explanations for automatic recommendations~\cite{goodman2016eu}.  The   measurements, models and mechanisms proposed in this paper  contribute to the development of   GDPR-compliant policies, as the allocations derived from the proposed  model-based mechanism are built on top of first principles.

Most of the previous literature on social fairness   assumes  that utility functions are non-parametric~\cite{zhangfairness,abebe2017fair, varian1983non} or, in classification problems, that cost functions  are linear combinations of false positive and false negative rates
 \cite{pleiss2017fairness,kleinberg2016inherent,corbett2017algorithmic}.  Convex  utility functions, such as the $\alpha$-fair family of utilities,   are prevalent in the literature of networking and computer systems~\cite{mo2000fair,ghodsi2011dominant, Kelly1997, Dehghan2016a, shakkottai2008network, palomar2006tutorial, bonald2001impact}. In this paper, we identify how parameterized utilities can be applicable to the analysis of social networks.  We believe that such connection is a step towards  promoting more dialogue
between the networking and the online social network communities on the issue of fairness.

Algorithmic bias and forms to audit it were investigated in~\cite{Bias_on_the_web,Diakopoulos2013,Sandvig2014}. In~\cite{Epstein2015,Kulshrestha2017} it was shown that search engine rank  manipulation can influence the vote of undecided citizens. 
The models proposed in this paper  further foster  accountability, by quantifying  bias in the Facebook News Feed.  

\subsubsection{Public datasets and reproducible methodologies}

The behavior  of users searching for visibility was studied in ~\cite{eslami2016first,Bucher2012,Sleeper2013}. Such studies are primarily based on small datasets.  A notable exception is~\cite{Bakshy2012,Bakshy2015,Guillory2014,Sun2009,bond201261}, who   considered a massive dataset provided by Facebook through restrictive non-disclosure agreements.  Datasets   to assess Facebook publishers'  visibilities are  usually not made publicly  available. Our work aims to contribute by filling that gap.

 It is out of the scope of this paper to present a nuts-and-bolts perspective on how Facebook News Feed works. Instead, our goal is to provide a simple model that can  explain the occupancy and visibility of different publishers, given a reproducible measurement framework. We  profile Facebook, which is taken as a black box to be   scientifically analyzed. 
This approach dates back to Skinner tests~\cite{skinner_box}, and   
 has been gaining significant attention in the literature of social networks~\cite{Tan2017}. 





 


%% file: conclusion.tex
\section{Conclusion}

We presented a framework encompassing  reproducible measurements, analytical models and utility-based mechanisms for the  analysis of Facebook News Feed algorithm. 
The   analytical model   enables quantitative what-if analysis to assess the bias introduced by the News Feed algorithm.  The utility-based mechanisms shed light into novel directions towards the control of the News~Feed.

Our measurements indicate that the News Feed  algorithm currently tends to reinforce the orientation indicated by users about the pages they ``like'', by filtering  posts and creating biases among the set of followed publishers. The effects of filtering are stronger at the topmost position where only a fraction of the set of publishers followed  by the users was represented. We observed that a neutral user that did not ``like'' any page was also exposed to a noticeable bias.

Facebook mission  is to ``give people the power to build community.'' We believe that the measurements, model and tools presented in this work are one step further towards that goal, as they help evaluating algorithms' transparency and promote user awareness about the filtering process they are submitted to.   Ultimately, such awareness is key to protect and empower Facebook users,  communities, society and democracy as a whole~\cite{MIT_TR}.  

\section*{Acknowledgements} 
This work was partially funded by the Brazilian-French project THANES (co-sponsored by FAPERJ and INRIA). 
D.~Menasché was sponsored by a JCNE fellowship (FAPERJ), CNPq and in part by the Coordenação de
Aperfeiçoamento de Pessoal de Nível Superior - Brasil (CAPES) -
Finance Code 001. This work has also been supported by the French government, through the UCA 
JEDI "Investments in the Future" project managed by the National 
Research Agency (ANR) with the reference number ANR-15-IDEX-01.

%% file: appendix.tex
\begin{appendix}
\pagebreak

\section{List of publishers used in the experiment} 
\label{appendix:list_publishers}

\begin{table*}[]
\centering
\caption{Lists of the publishers followed in the Italian experiment with their respective labels.}
\label{table:publishers}
\begin{tabular}{|l|l|l|l|}
\hline
Orientation	&	Page URL	&	Publisher Name	\\
\hline
\hline
Right	&	NoiconSalviniUfficiale	&	Noi con Salvini	\\
Right	&	ilpopulista.it	&	Il Populista	\\
Right	&	ilGiornale	&	Il Giornale	\\
Right	&	legasalvinipremier	&	Lega - Salvini Premier	\\
Right	&	rivogliobologna	&	Lucia Borgonzoni	\\
Right	&	giorgiameloni.paginaufficiale	&	Giorgia Meloni	\\
Far-right	&	Fascisti-uniti-per-Litalia-411675765615435	&	Fascisti uniti per L'italia	\\
Far-right	&	Lotta-Studentesca-257153365332	&	Lotta Studentesca	\\
Far-right	&	OrdineFuturo	&	Ordine Futuro	\\
Far-right	&	ilprimatonatsionale	&	Il Primato Nazionale	\\
Far-right	&	ForzaNuovaPaginaUfficiale	&	Forza Nuova	\\
Far-right	&	casapounditalia	&	CasaPound Italia	\\
Far-right	&	RobertoFiorePaginaUfficiale	&	Roberto Fiore	\\
Left	&	Articolo1Modempro	&	Articolo UNO 	\\
Left	&	sinistraitalianaSI	&	Sinistra Italiana	\\
Left	&	ilmanifesto	&	il manifesto	\\
Left	&	Possibile.it	&	Possibile	\\
Left	&	giuseppecivati	&	Giuseppe Civati	\\
Left	&	Laura-Boldrini-325228170920721	&	Laura Boldrini	\\
Center-left	&	Adessotiinformo	&	Adesso ti informo	\\
Center-left	&	matteorenzinews	&	Matteo Renzi News	\\
Center-left	&	Repubblica	&	la Repubblica	\\
Center-left	&	democratica	&	Democratica	\\
Center-left	&	matteorenziufficiale	&	Matteo Renzi	\\
Center-left	&	partitodemocratico.it	&	Partito Democratico	\\
M5S	&	news.m5s	&	M5S news	\\
M5S	&	WIlM5s	&	W IL M5S	\\
M5S	&	ilFattoQuotidiano	&	Il Fatto Quotidiano	\\
M5S	&	movimentocinquestelle	&	MoVimento 5 Stelle	\\
M5S	&	LuigiDiMaio	&	Luigi Di Maio	\\
M5S	&	beppegrillo.it	&	Beppe Grillo	\\
\hline
\end{tabular}
\end{table*}

Table \ref{table:publishers} contains a list of  publishers followed in the Italian experiment with their respective orientations.
During the experiments, the page entitled \emph{Fasciti uniti per L'italia} was shutdown by Facebook and it was replaced by \emph{Lotta Studentesca}.

\section{Publishers' visibilities}

\label{sec:visibilities}
\begin{figure*}[h!]
        \includegraphics[width=1.0\textwidth]{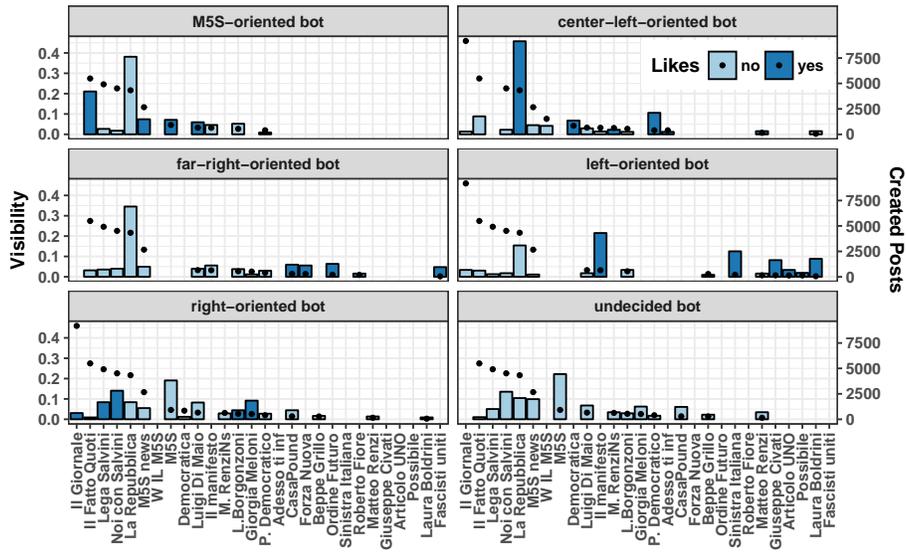} 
\caption{Publishers' visibilities at the six bots  (bars colored by preferences) and number of created posts (black dots), for $K=1$.}
\label{fig:pubfontes1visA}
\end{figure*}

\begin{figure*}[h!]
        \includegraphics[width=1.0\textwidth] {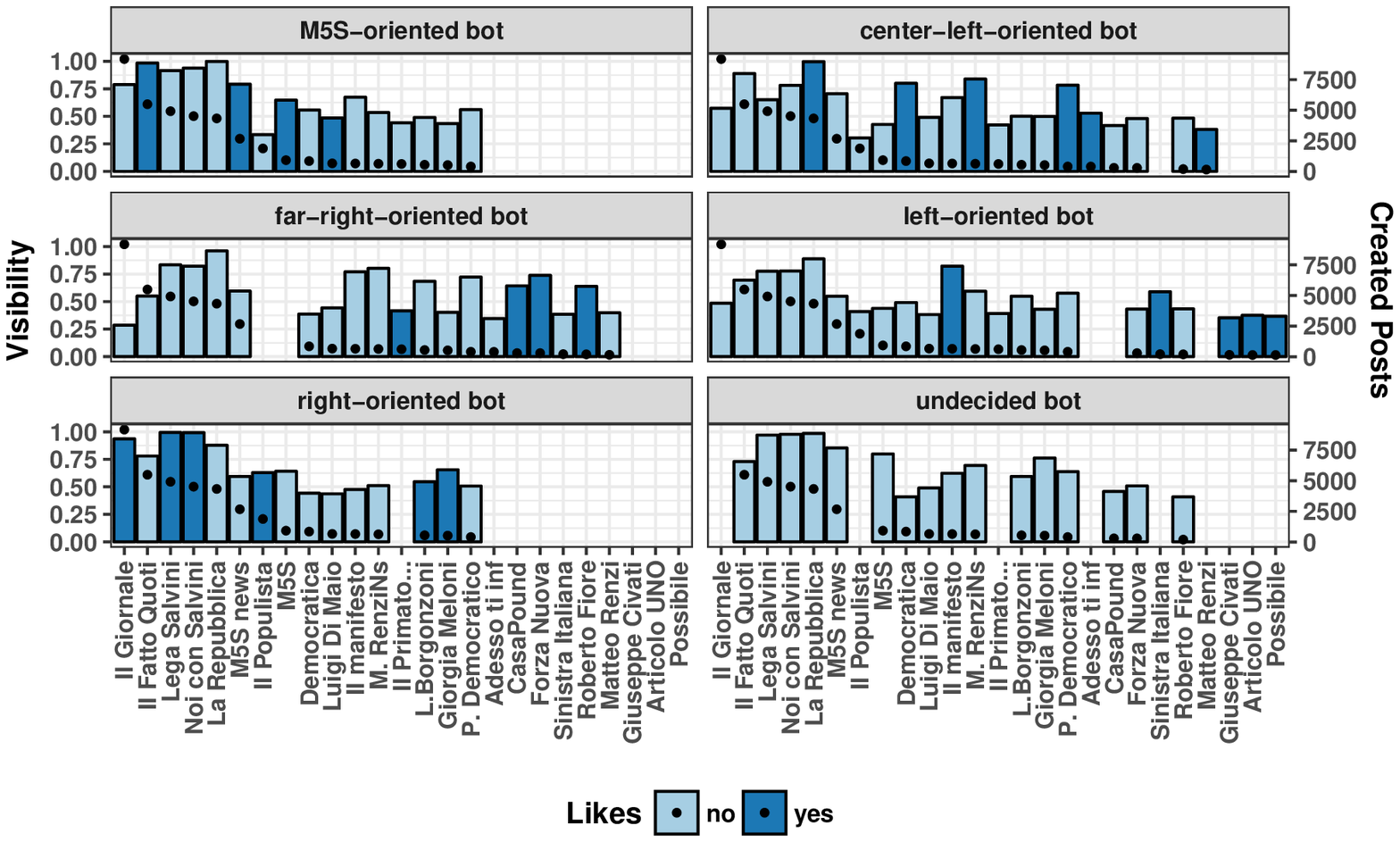} 
\caption{Publishers visibilities at the six bots  (bars colored by preferences) and number of created posts (black dots), for $K=30$.}
\label{fig:pubfontes1visB}
\end{figure*}

In what follows we complement results presented in Section~\ref{sec:effec}.  Whereas in Section~\ref{sec:effec} we showed how  \emph{occupancies} varied as a function of the rate at which publishers create posts, in this appendix we focus on \emph{visibilities}.  Figure~\ref{fig:pubfontes1visA} shows the visibility for the top publishers, at the topmost News Feed position ($K=1$), under the six considered bots.  It indicates, for instance, that even at the undecided bot, visibilities do  not vary monotonically with respect to publishers' post creation rates.  Figure~\ref{fig:pubfontes1visB} considers the case $K=30$.  As expected, the visibilities increase as $K$ grows from 1 to 30. Nonetheless, when $K=30$ we still find some top publishers that have almost negligible visibility at a number of bots.  In particular, eight out of the thirty top publishers have negligible visibility at the undecided bot.

\section{Model validation}

\subsection{Validation for $K=30$}
\label{sec:addmodelval}

\begin{figure*}[t]
\begin{center}
\begin{tabular}{cc}   \includegraphics[width=0.5\textwidth]{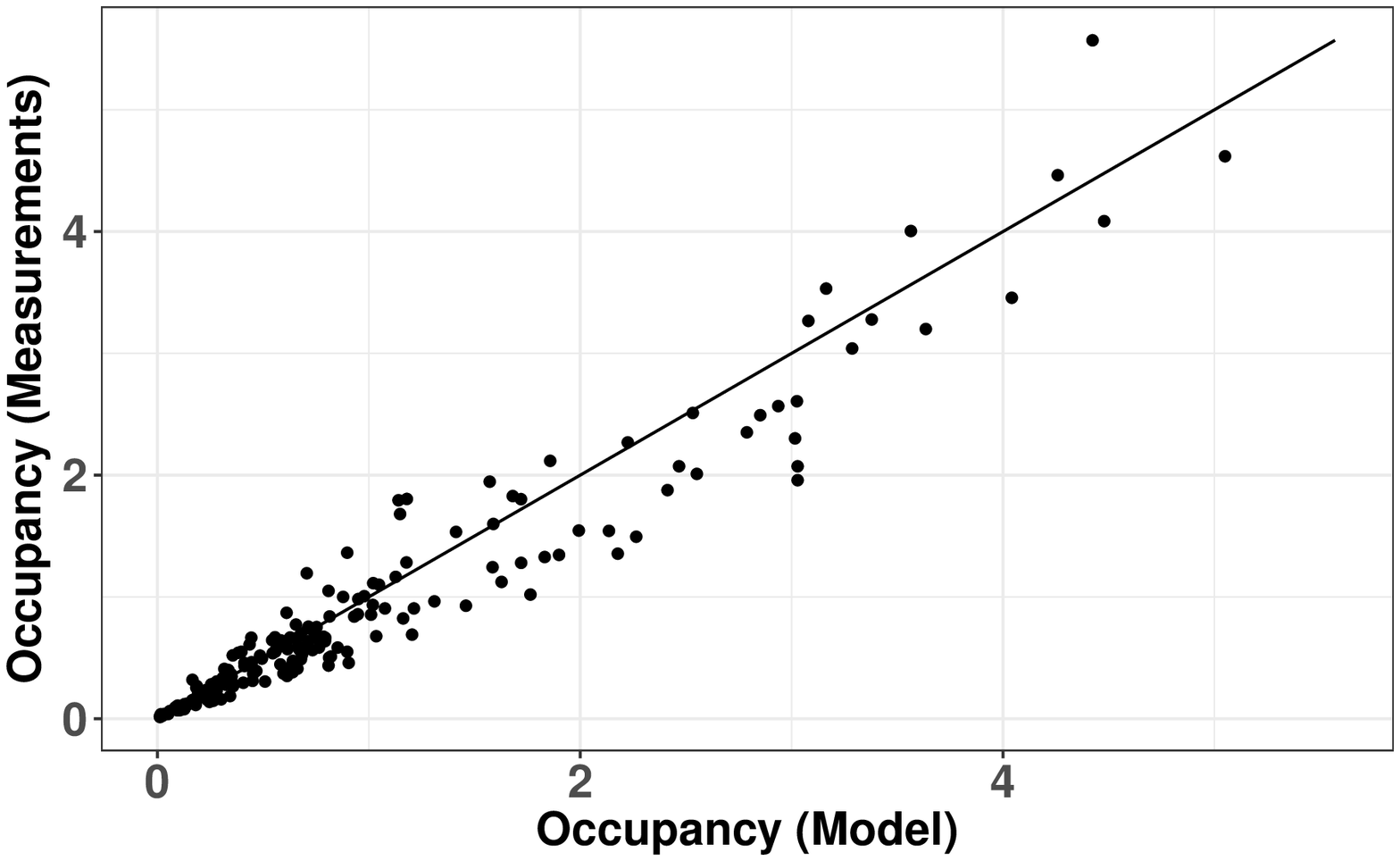} &
\includegraphics[width=0.5\textwidth]{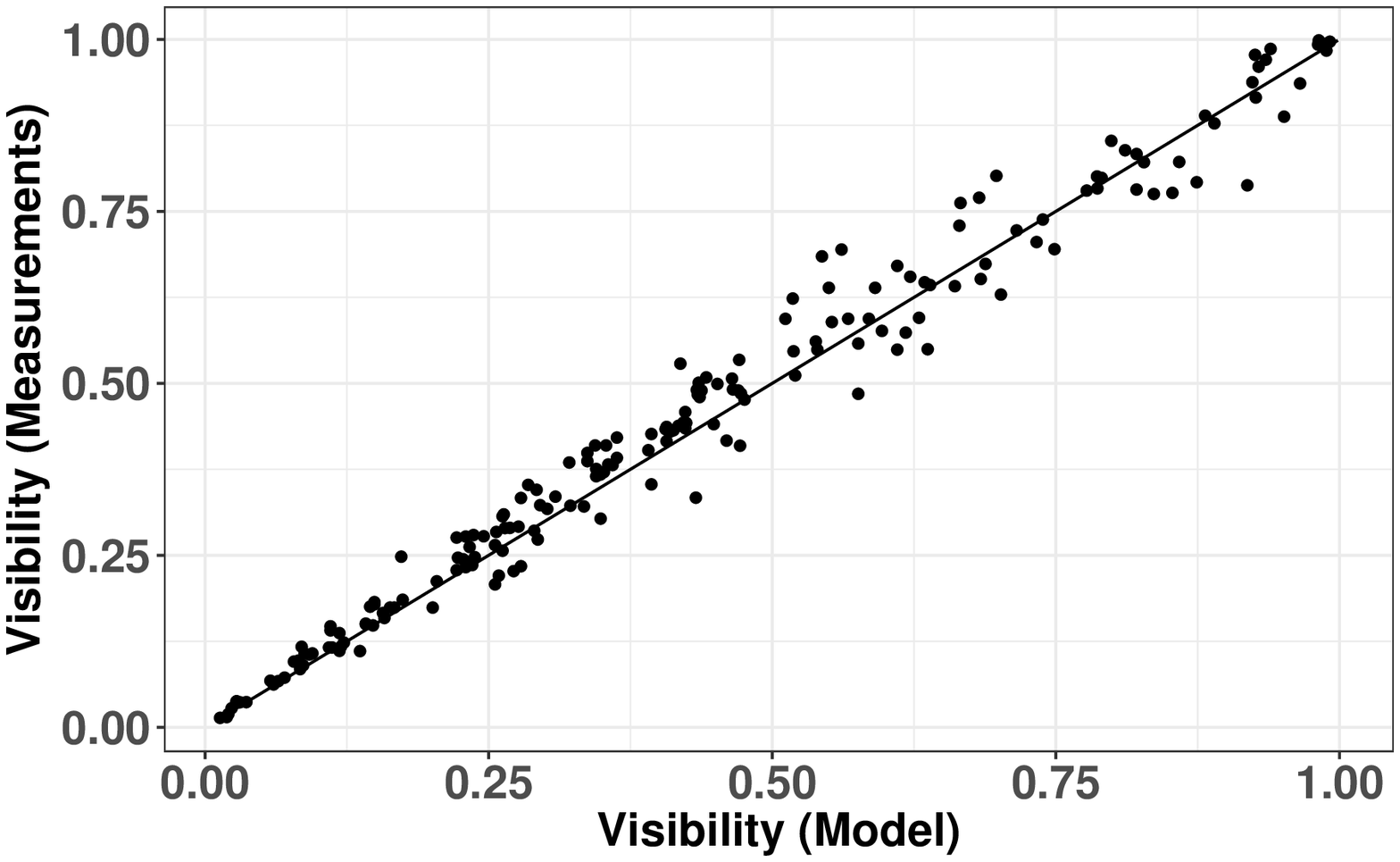} \\
(a) & (b)
\end{tabular}
\end{center}
\caption{Multi-class model validation for the $(a)$ occupancy and $(b)$ visibility metrics, for $K=30$. } 
\label{fig:Validation_multiclass}
\end{figure*}

\begin{figure*}[t]
\begin{center}
\begin{tabular}{cc}   \includegraphics[width=0.5\textwidth]{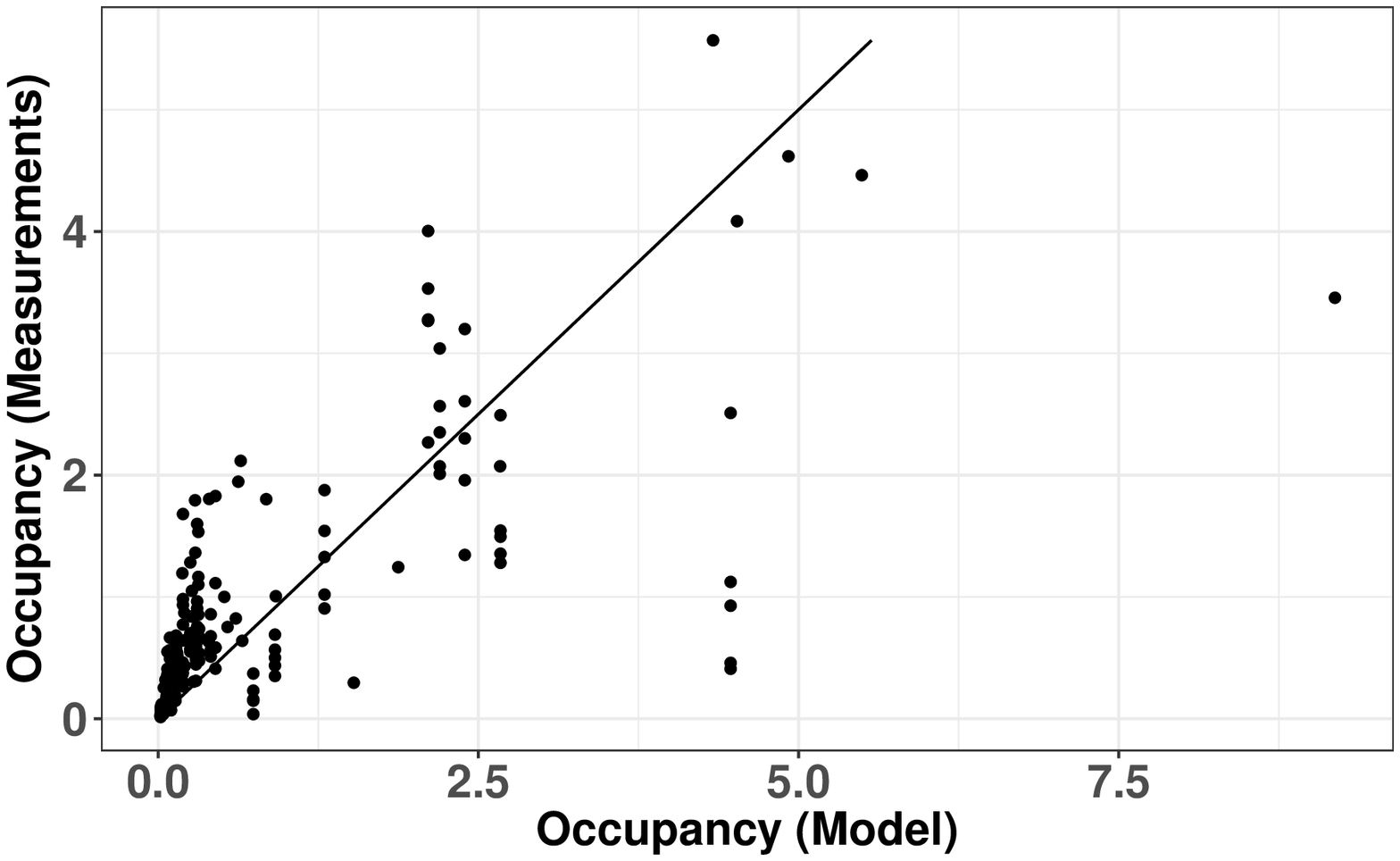} &
\includegraphics[width=0.5\textwidth]{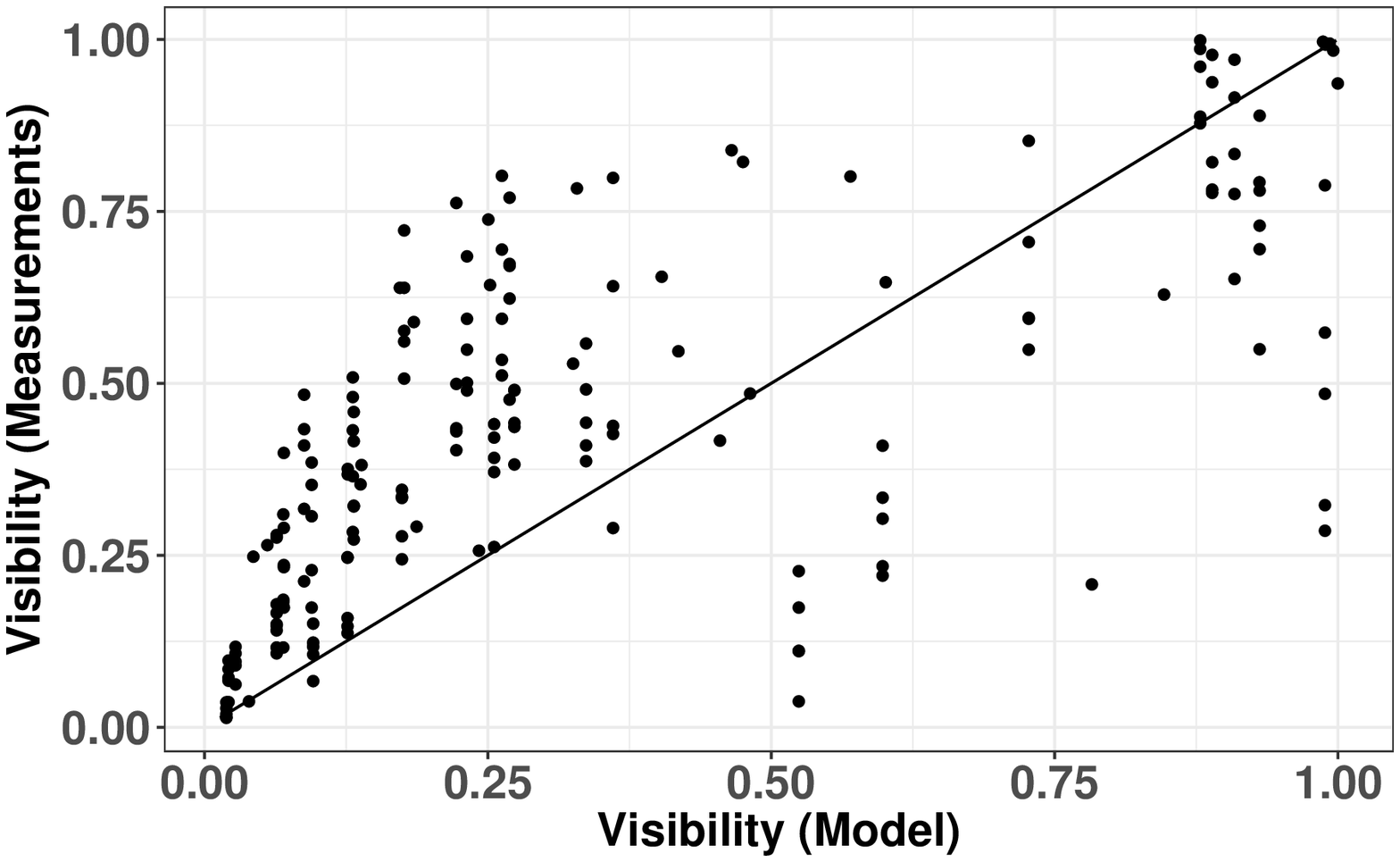} \\
(a) & (b) 
\end{tabular}
\end{center}
\caption{Two-class model validation for the $(a)$ occupancy and $(b)$ visibility metrics, for $K=30$.}
\label{fig:Validation_twoclass}
\end{figure*}

From Figures~\ref{fig:Validation_multiclass}(a) and \ref{fig:Validation_multiclass}(b) to Figures~\ref{fig:Validation_twoclass}(a) and (b), each point corresponds to a user-publisher pair. We let $K=30$ (results for $K=1$ are presented in Section~\ref{sec:validation}). 
In Figure~\ref{fig:Validation_multiclass}(a) (resp., Fig.~\ref{fig:Validation_multiclass}(b)), a point $(x={N}_{ij},y=\tilde{N}_{ij})$ (resp., $x={\pi}_{ij},y=\tilde{\pi}_{ij}$) indicates that, for the given pair, 
an occupancy   $N_{ij}$ (resp., visibility $\pi_{ij}$) estimated by the multi-class model using eq.~\eqref{eq:little1} (resp.,~\eqref{theo:visibility_general}) corresponds to a measured occupancy of $\tilde{N}_{ij}$ (resp., measured visibility of~$\tilde{\pi}_{ij}$). 
Most of the points are close to the $\tilde{N}_{ij}=N_{ij}$ line, indicating the expressive power of the model.
In Figure~\ref{fig:Validation_twoclass}(a) (resp., Fig.~\ref{fig:Validation_twoclass}(b)), a point $(x={N}_{ij},y=\tilde{N}_{ij})$ (resp., $x={\pi}_{ij},y=\tilde{\pi}_{ij}$) indicates that, for the given pair, 
an occupancy   $N_{ij}$ (resp., visibility $\pi_{ij}$) estimated by the two-class model using eq.~\eqref{eq:little1} (resp.,~\eqref{theo:visibility_general}) corresponds to a measured occupancy of $\tilde{N}_{ij}$ (resp., measured visibility of~$\tilde{\pi}_{ij}$). 
The two-class model has two parameters, while the number of parameters in the multi-class model is equal to the number of publishers times the number of bots (180 in the experiment). For this reason, the accuracy of the former is significantly lower than the later.

\subsection{2017 French presidential elections }

\begin{figure*}[t]
\begin{center}
\begin{tabular}{cc}   \includegraphics[width=0.5\textwidth]{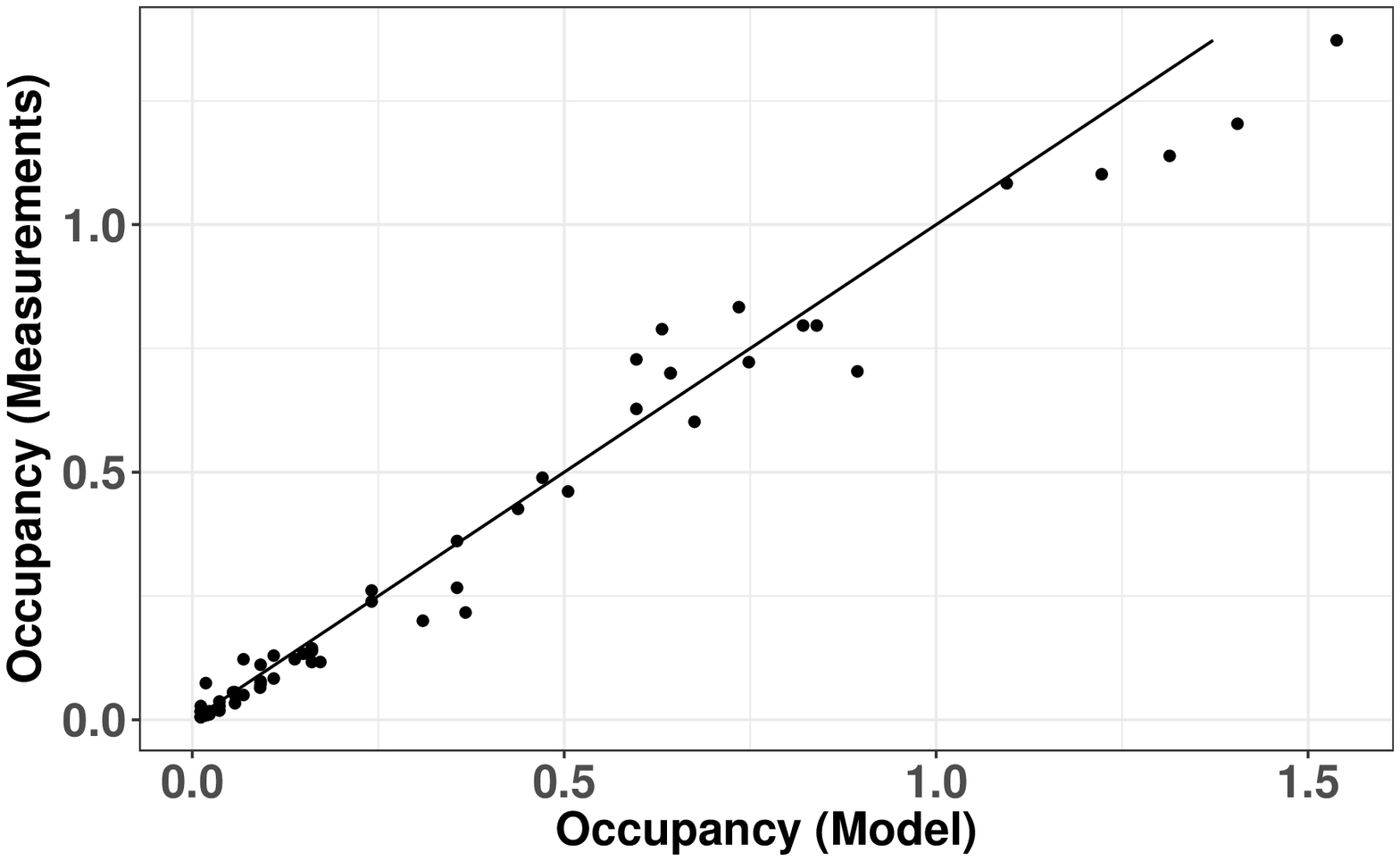} &
\includegraphics[width=0.5\textwidth]{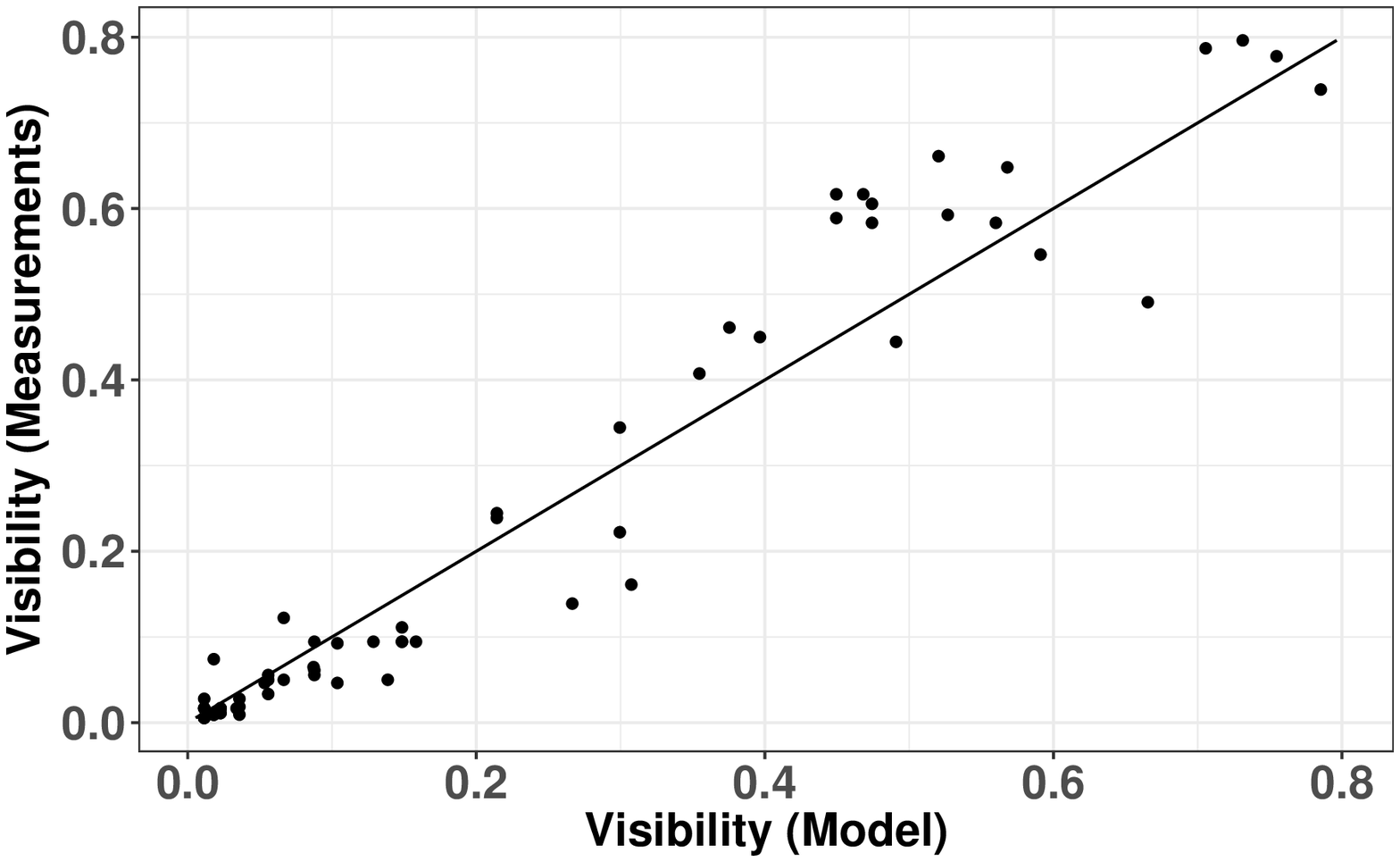} \\
(a) & (b) 
\end{tabular}
\end{center}
\caption{Model Validation for the $(a)$ occupancy and $(b)$ visibility metrics, for $K=10$ using the 2017 French Elections Dataset.}
\label{fig:Validation_French_elections_K=10}
\end{figure*}

\begin{figure*}[t]
\begin{center}
\begin{tabular}{cc}   \includegraphics[width=0.5\textwidth]{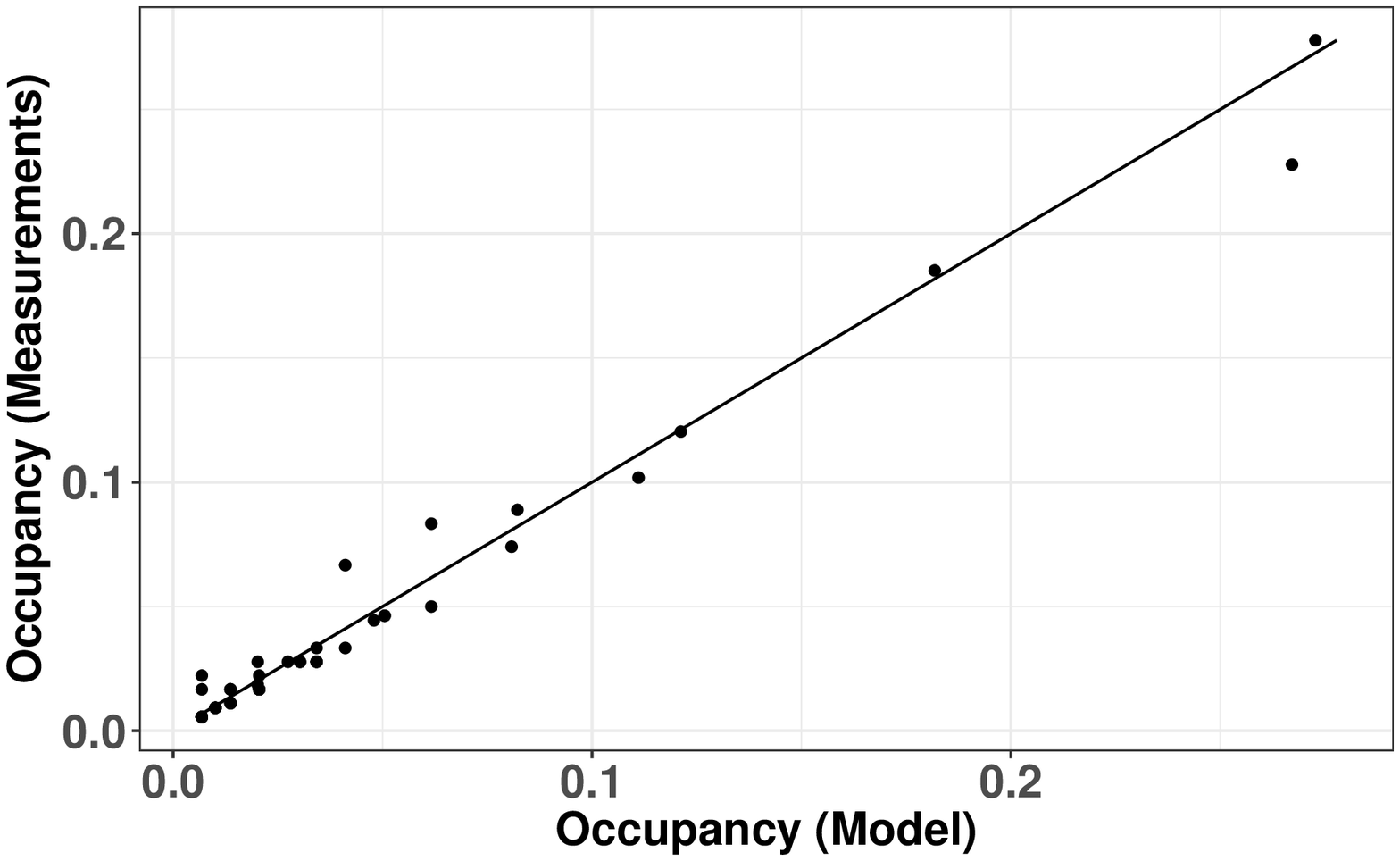} &
\includegraphics[width=0.5\textwidth]{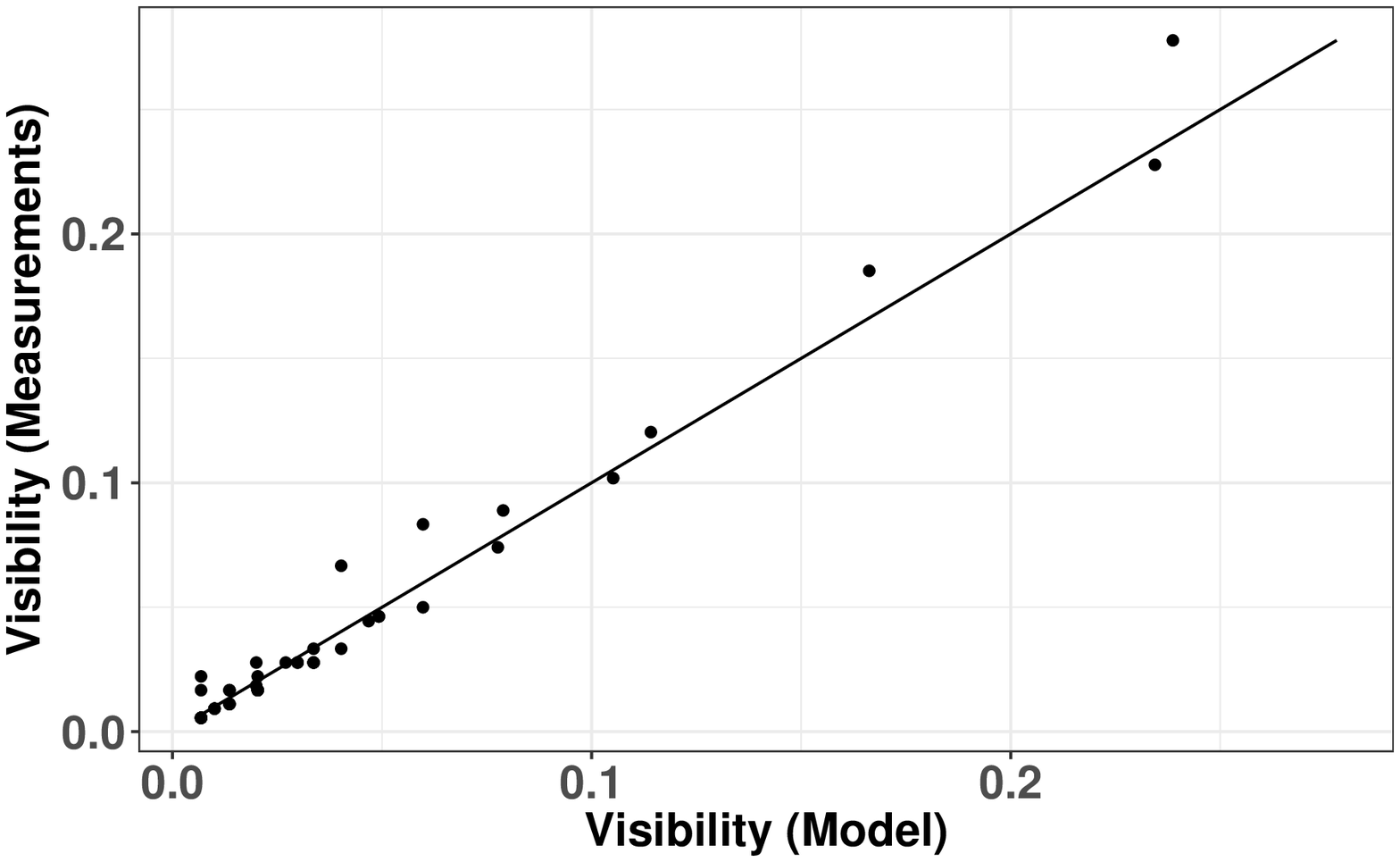} \\
(a) & (b) 
\end{tabular}
\end{center}
\caption{Model Validation for the $(a)$ occupancy and $(b)$ visibility metrics, for the topmost position using the 2017 French Elections Dataset.}
\label{fig:Validation_French_elections_K=1}
\end{figure*}

Another experiment was conducted during the 2017 French presidential elections where four Facebook bots were created and monitored. The experiments started in April 28, 2017, and ended in May 08, 2017. Our profiles were kept with no friends, and they all followed the same group of 13 pages in a addition to a number of random pages. We adopted the multi-class approach (Section~\ref{sec:effrate}) to parametrize and validate the model with such dataset. Figures~\ref{fig:Validation_French_elections_K=10} and \ref{fig:Validation_French_elections_K=1} show our model validation for $K=10$ and $K=1$ (topmost position), respectively. The values predicted by the model are very close to the measured points indicating once again the expressive power of the model.  The dataset corresponding to the French elections is publicly available.\footnote{https://github.com/EduardoHargreaves/Effect-of-the-OSN-on-the-elections}

\pagebreak

\section{Potential delay and max-min fairness}

\label{sec:potential}

\begin{figure*}[h!]
        \includegraphics[width=1.0\textwidth]{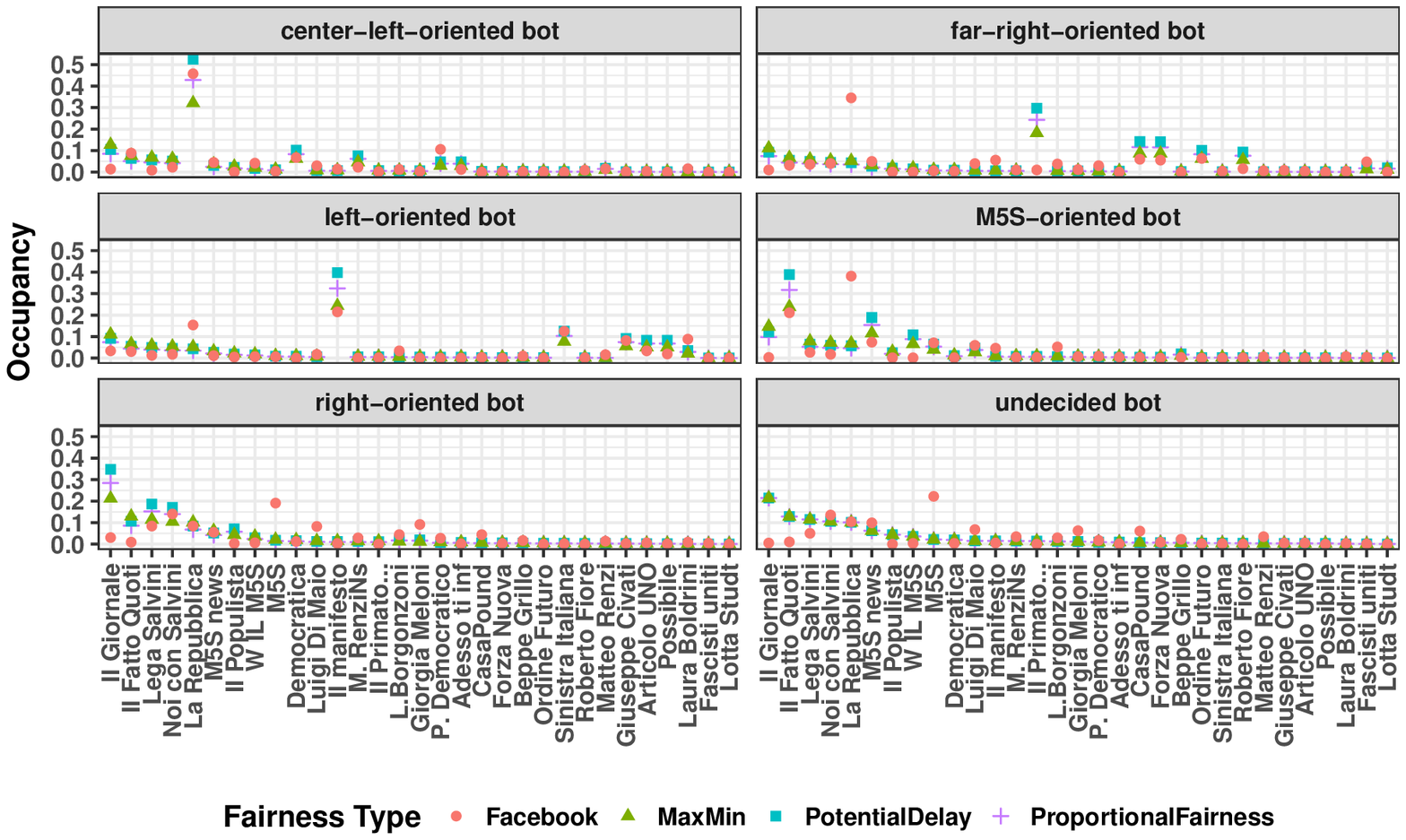} 
\caption{Publishers' occupancies under the three types of fairness, in addition to measured Facebook occupancies, for $K=1$, $w_i^{(1)}=2$ and $w_i^{(0)}=1$ for all users, at the six bots.}
\label{fig:multifairnessk1}
\end{figure*}

\begin{figure*}[h!]
        \includegraphics[width=1.0\textwidth]{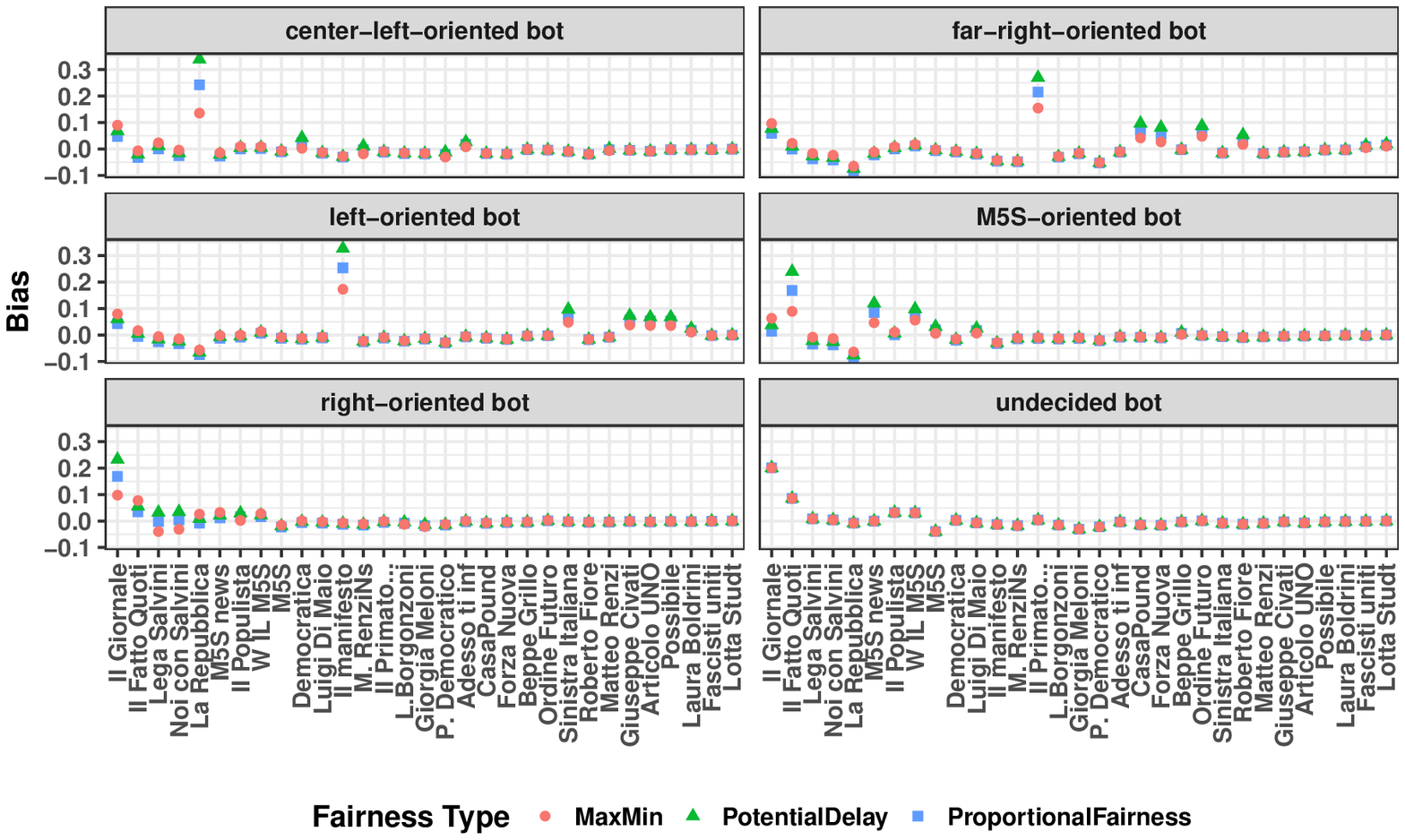} 
\caption{Publishers' bias under the three types of fairness:  Facebook occupancy set as baseline, for $K=30$, $w_i^{(1)}=2$ and $w_i^{(0)}=1$ for all users, at the six bots.  Potential delay fairness (green tringles)  tends to favor ``liked'' publishers more than  proportional fairness (blue squares).}
\label{fig:multifairnessbiask1}
\end{figure*}

\begin{figure*}[h!]
        \includegraphics[width=1.0\textwidth]{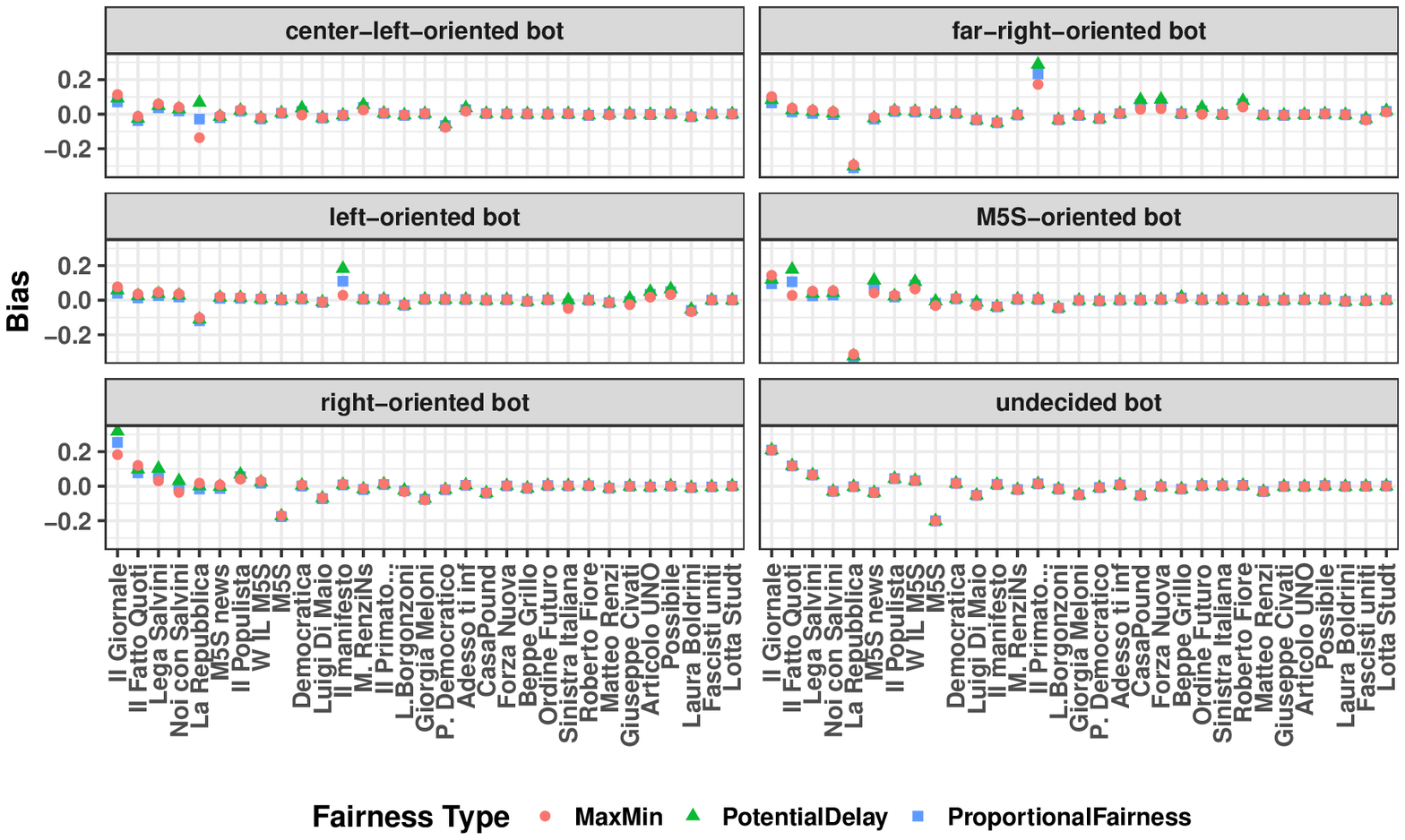} 
\caption{Publishers' bias under the three types of fairness: Facebook occupancy set as  baseline, for $K=1$, $w_i^{(1)}=2$ and $w_i^{(0)}=1$ for all users, at the six bots.  Potential delay fairness (green tringles)  tends to favor ``liked'' publishers more than  proportional fairness (blue squares).}
\label{fig:multifairnessbiask30}
\end{figure*}

Next, we report results on the potential delay, max-min and proportional fairness.  In Figure~\ref{fig:multifairnessk1} we set  uniform filtering as our baseline, whereas in 
Figures~\ref{fig:multifairnessbiask1} and~\ref{fig:multifairnessbiask30}
  we set the Facebook measurements as the baseline in equation~\eqref{eq:bijm30}.  We let $w_i^{(1)}=2$ and $w_i^{(0)}=1$.
Figure~\ref{fig:multifairnessk1} shows the occupancy for the top publishers, for $K=1$, under the Facebook measurements,  potential fairness, max-min fairness and proportional fairness. Note that the general trends of proportional fairness, potential delay fairness and max-min fairness are similar.  Potential delay fairness tends to favor publishers that bots ``like'' more than the other fairness functions, while max-min  tends to favor less.
 Facebook occupancies, in contrast, do not reflect any of the two considered fairness criteria. At the undecided bot, all fairness criteria yields the same occupancies and biases, reflecting the lack of preferences of this bot.

Figures~\ref{fig:multifairnessbiask1} and~\ref{fig:multifairnessbiask30} show  the biases of the considered fairness criteria   using the Facebook as baseline, for $K=30$ and $K=1$. 
In both cases, we note that 
  potential delay fairness (green triangles)  tends to favor ``liked'' publishers more than  proportional fairness (blue squares). 
 When $K=1$, Facebook tends to penalize the publishers that produced more posts at the undecided bot. Note also that there is a positive negative bias towards M5S posts at the undecided bot, meaning that Facebook allocated far more occupancy to M5S posts than the proposed methods would allocate.


\pagebreak

\section{Sensitivity analysis with respect of weights and News Feed size}

\label{sec:sensitiveweights}


In Section~\ref{sec:utilitaly}
we assumed  
$w_i^{(l)}=1$ when considering the two-class model.  
Recall that $w_i^{(0)}$ (resp., $w_i^{(1)}$) correspond to publishers that a bot does not ``like'' (resp., ``likes'').

\begin{figure*}[h!]
        \includegraphics[width=1.0\textwidth]{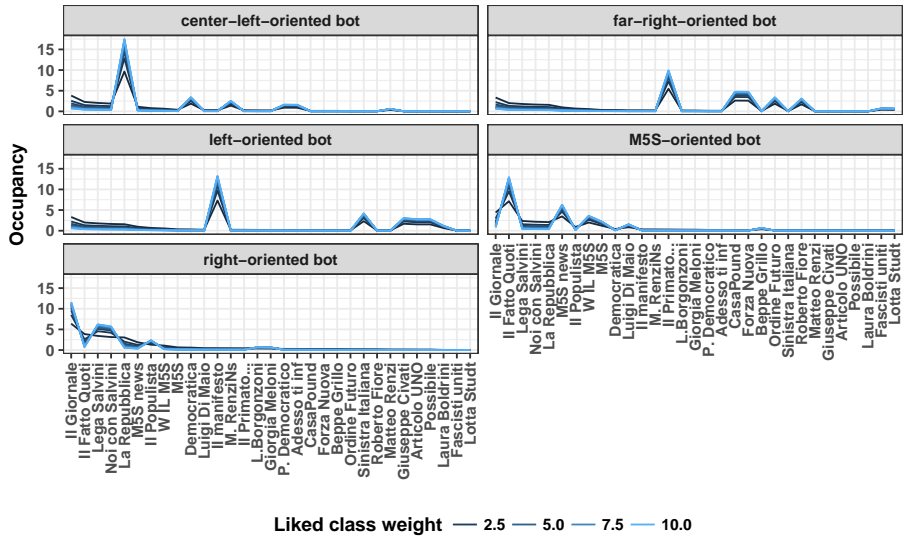} 
\caption{Publishers' occupancies under proportional fairness for $K=30$ at the six bots  (bars colored by preferences) , $w_i^{(1)}$ ranging from 1 to 10 and $w_i^{(0)}=1$ . }
\label{fig:propfairnesssensi}
\end{figure*}

In Figure~\ref{fig:propfairnesssensi}, we  keep $w_i^{(0)}=1$ and vary $w_i^{(1)}$ from $1$ to $10$ to show the impact of the weights on the occupancies. We consider  proportional fairness allocations, with $K=30$.  Figure~\ref{fig:propfairnesssensi} shows that a ten-fold increase in $w_i^{(1)}$, from 1 to 10, may lead to an up to two-fold increase in the occupancies of publishers that bots ``like''.  This is the case, for instance, with ``Il Fatto Quotidiano'', which was classified as a M5S source, and which significantly benefited from the increase of $w_i^{(1)}$ at the M5S-oriented bot.

\begin{figure*}[h!]
\center
  \includegraphics[width=1\textwidth]{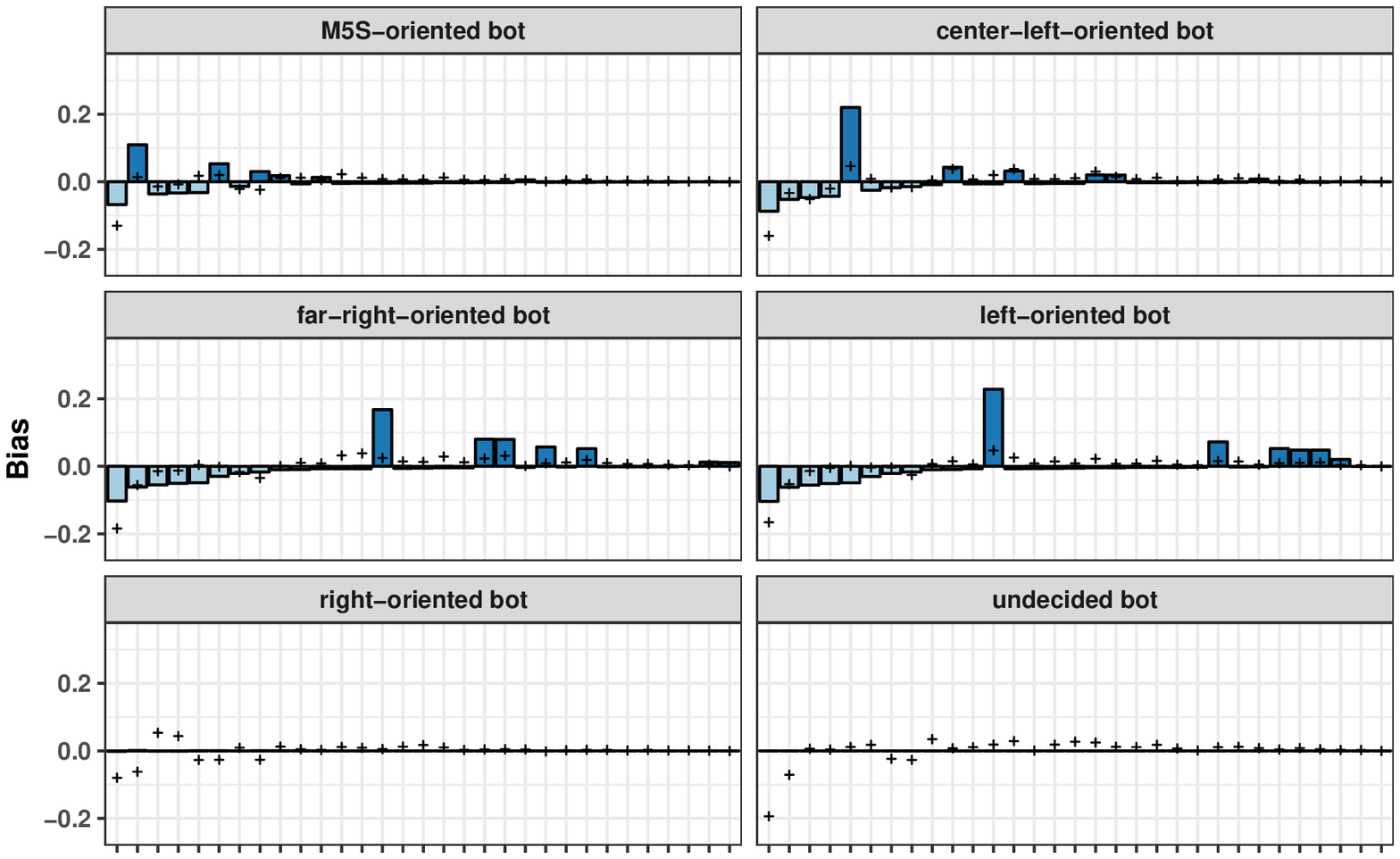} \\ 
       (a) $K=30$\\         \includegraphics[width=1\textwidth]{Bias_top.eps} \\ 
(b) $K=1$ \\
\caption{ Proportional fairness  bias (bars) and  Facebook News Feed bias (crosses),  for  the six bots. 
Note that the bias experienced by the Facebook News Feed  significantly differs between $K=30$ and $K=1$, and tends to be stronger at the topmost position.}
\label{fig:bias_30}
\end{figure*}

Next, we consider the impact of $K$ on our results. We observe that our utility optimization framework produces occupancies that are directly proportional to $K$ (see, for example,  \eqref{eq:beta_prop_fairnes_final}), and the corresponding biases are independent from $K$. For this reason $b_{ij}^{\textnormal{(PropF)}}$ does not change between Figures \ref{fig:bias_30}(a) and \ref{fig:bias_30}(b). 
In contrast, the shape of the biases accounting for the Facebook measurements, $b_{ij}^{\textnormal{(Face)}}$, are substantially different for $K=1$ and $K=30$, with stronger biases for $K=1$
(in agreement with the empirical findings reported in Section~\ref{sec:effec}).

\end{appendix}

%% file: main.bbl

\begin{thebibliography}{91}


\ifx \showCODEN    \undefined \def \showCODEN     #1{\unskip}     \fi
\ifx \showDOI      \undefined \def \showDOI       #1{#1}\fi
\ifx \showISBNx    \undefined \def \showISBNx     #1{\unskip}     \fi
\ifx \showISBNxiii \undefined \def \showISBNxiii  #1{\unskip}     \fi
\ifx \showISSN     \undefined \def \showISSN      #1{\unskip}     \fi
\ifx \showLCCN     \undefined \def \showLCCN      #1{\unskip}     \fi
\ifx \shownote     \undefined \def \shownote      #1{#1}          \fi
\ifx \showarticletitle \undefined \def \showarticletitle #1{#1}   \fi
\ifx \showURL      \undefined \def \showURL       {\relax}        \fi
\providecommand\bibfield[2]{#2}
\providecommand\bibinfo[2]{#2}
\providecommand\natexlab[1]{#1}
\providecommand\showeprint[2][]{arXiv:#2}

\bibitem[\protect\citeauthoryear{Abebe, Kleinberg, and Parkes}{Abebe
  et~al\mbox{.}}{2017}]%
        {abebe2017fair}
\bibfield{author}{\bibinfo{person}{Rediet Abebe}, \bibinfo{person}{Jon
  Kleinberg}, {and} \bibinfo{person}{David~C Parkes}.}
  \bibinfo{year}{2017}\natexlab{}.
\newblock \showarticletitle{Fair division via social comparison}. In
  \bibinfo{booktitle}{\emph{Proceedings of the 16th Conference on Autonomous
  Agents and MultiAgent Systems}}. International Foundation for Autonomous
  Agents and Multiagent Systems, \bibinfo{pages}{281--289}.
\newblock


\bibitem[\protect\citeauthoryear{Adam D. I.~Kramer and Hancock}{Adam D.
  I.~Kramer and Hancock}{2014}]%
        {Guillory2014}
\bibfield{author}{\bibinfo{person}{Jamie E.~Guillory Adam D. I.~Kramer} {and}
  \bibinfo{person}{Jeffrey~T. Hancock}.} \bibinfo{year}{2014}\natexlab{}.
\newblock \showarticletitle{{Experimental evidence of massive-scale emotional
  contagion through social networks}}.
\newblock \bibinfo{journal}{\emph{Proceedings of the National Academy of
  Sciences}} (\bibinfo{year}{2014}).
\newblock
\showISBNx{1091-6490 (Electronic)$\backslash$r0027-8424 (Linking)}
\showISSN{0027-8424}
\urldef\tempurl%
\url{https://doi.org/10.1073/pnas.1412469111}
\showDOI{\tempurl}


\bibitem[\protect\citeauthoryear{Agosti}{Agosti}{2018}]%
        {fbtrex}
\bibfield{author}{\bibinfo{person}{Claudio Agosti}.}
  \bibinfo{year}{2018}\natexlab{}.
\newblock \bibinfo{title}{Facebook tracking exposed}.
\newblock
\newblock
\newblock
\shownote{\url{ https://facebook.tracking.exposed}.}


\bibitem[\protect\citeauthoryear{Altman, Kumar, Venkatramanan, and
  Kumar}{Altman et~al\mbox{.}}{2013}]%
        {altman2013competition}
\bibfield{author}{\bibinfo{person}{Eitan Altman}, \bibinfo{person}{Parmod
  Kumar}, \bibinfo{person}{Srinivasan Venkatramanan}, {and}
  \bibinfo{person}{Anurag Kumar}.} \bibinfo{year}{2013}\natexlab{}.
\newblock \showarticletitle{{Competition over timeline in social networks}}.
\newblock \bibinfo{journal}{\emph{Proceedings of the 2013 IEEE/ACM
  International Conference on Advances in Social Networks Analysis and Mining -
  ASONAM '13}} (\bibinfo{year}{2013}).
\newblock


\bibitem[\protect\citeauthoryear{Baeza-Yates}{Baeza-Yates}{2018}]%
        {Bias_on_the_web}
\bibfield{author}{\bibinfo{person}{Ricardo Baeza-Yates}.}
  \bibinfo{year}{2018}\natexlab{}.
\newblock \showarticletitle{{Bias on the web}}.
\newblock \bibinfo{journal}{\emph{Commun. ACM}} \bibinfo{volume}{61},
  \bibinfo{number}{6} (\bibinfo{date}{may} \bibinfo{year}{2018}),
  \bibinfo{pages}{54--61}.
\newblock
\showISSN{00010782}
\urldef\tempurl%
\url{https://doi.org/10.1145/3209581}
\showDOI{\tempurl}


\bibitem[\protect\citeauthoryear{Bakshy, Messing, and Adamic}{Bakshy
  et~al\mbox{.}}{2015}]%
        {Bakshy2015}
\bibfield{author}{\bibinfo{person}{Eytan Bakshy}, \bibinfo{person}{Solomon
  Messing}, {and} \bibinfo{person}{Lada~A Adamic}.}
  \bibinfo{year}{2015}\natexlab{}.
\newblock \showarticletitle{{Exposure to ideologically diverse news and opinion
  on Facebook}}.
\newblock \bibinfo{journal}{\emph{Science}} \bibinfo{volume}{348},
  \bibinfo{number}{6239} (\bibinfo{year}{2015}), \bibinfo{pages}{1130--1132}.
\newblock
\showISBNx{1460-2466}
\showISSN{0036-8075}


\bibitem[\protect\citeauthoryear{Bakshy, Rosenn, Marlow, and Adamic}{Bakshy
  et~al\mbox{.}}{2012}]%
        {Bakshy2012}
\bibfield{author}{\bibinfo{person}{Eytan Bakshy}, \bibinfo{person}{Itamar
  Rosenn}, \bibinfo{person}{Cameron Marlow}, {and} \bibinfo{person}{Lada
  Adamic}.} \bibinfo{year}{2012}\natexlab{}.
\newblock \showarticletitle{{The role of social networks in information
  diffusion}}. In \bibinfo{booktitle}{\emph{Proceedings of the 21st
  international conference on World Wide Web - WWW '12}}.
  \bibinfo{publisher}{ACM Press}, \bibinfo{address}{New York, NY, USA},
  \bibinfo{pages}{519}.
\newblock
\showISBNx{9781450312295}
\showISSN{1450312292}
\showeprint[arxiv]{1201.4145}


\bibitem[\protect\citeauthoryear{Barabas, Dinakar, Virza, Zittrain,
  et~al\mbox{.}}{Barabas et~al\mbox{.}}{2017}]%
        {barabas2017interventions}
\bibfield{author}{\bibinfo{person}{Chelsea Barabas}, \bibinfo{person}{Karthik
  Dinakar}, \bibinfo{person}{Joichi~Ito Virza}, \bibinfo{person}{Jonathan
  Zittrain}, {et~al\mbox{.}}} \bibinfo{year}{2017}\natexlab{}.
\newblock \showarticletitle{Interventions over Predictions: Reframing the
  Ethical Debate for Actuarial Risk Assessment}.
\newblock \bibinfo{journal}{\emph{arXiv preprint arXiv:1712.08238}}
  (\bibinfo{year}{2017}).
\newblock


\bibitem[\protect\citeauthoryear{Bastug, Bennis, and Debbah}{Bastug
  et~al\mbox{.}}{2014}]%
        {bastug2014living}
\bibfield{author}{\bibinfo{person}{Ejder Bastug}, \bibinfo{person}{Mehdi
  Bennis}, {and} \bibinfo{person}{M{\'e}rouane Debbah}.}
  \bibinfo{year}{2014}\natexlab{}.
\newblock \showarticletitle{Living on the edge: The role of proactive caching
  in 5G wireless networks}.
\newblock \bibinfo{journal}{\emph{IEEE Communications Magazine}}
  \bibinfo{volume}{52}, \bibinfo{number}{8} (\bibinfo{year}{2014}),
  \bibinfo{pages}{82--89}.
\newblock


\bibitem[\protect\citeauthoryear{Bessi}{Bessi}{2016}]%
        {BESSI2016319}
\bibfield{author}{\bibinfo{person}{Alessandro Bessi}.}
  \bibinfo{year}{2016}\natexlab{}.
\newblock \showarticletitle{Personality traits and echo chambers on
  {F}acebook}.
\newblock \bibinfo{journal}{\emph{Computers in Human Behavior}}
  \bibinfo{volume}{65} (\bibinfo{year}{2016}), \bibinfo{pages}{319 -- 324}.
\newblock
\showISSN{0747-5632}
\urldef\tempurl%
\url{https://doi.org/10.1016/j.chb.2016.08.016}
\showDOI{\tempurl}


\bibitem[\protect\citeauthoryear{Bessi, Zollo, Del~Vicario, Puliga, Scala,
  Caldarelli, Uzzi, and Quattrociocchi}{Bessi et~al\mbox{.}}{2016}]%
        {bessi2016users}
\bibfield{author}{\bibinfo{person}{Alessandro Bessi}, \bibinfo{person}{Fabiana
  Zollo}, \bibinfo{person}{Michela Del~Vicario}, \bibinfo{person}{Michelangelo
  Puliga}, \bibinfo{person}{Antonio Scala}, \bibinfo{person}{Guido Caldarelli},
  \bibinfo{person}{Brian Uzzi}, {and} \bibinfo{person}{Walter Quattrociocchi}.}
  \bibinfo{year}{2016}\natexlab{}.
\newblock \showarticletitle{Users polarization on {F}acebook and {Y}outube}.
\newblock \bibinfo{journal}{\emph{PloS one}} \bibinfo{volume}{11},
  \bibinfo{number}{8} (\bibinfo{year}{2016}), \bibinfo{pages}{e0159641}.
\newblock


\bibitem[\protect\citeauthoryear{Bonald and Massouli{\'e}}{Bonald and
  Massouli{\'e}}{2001}]%
        {bonald2001impact}
\bibfield{author}{\bibinfo{person}{Thomas Bonald} {and}
  \bibinfo{person}{Laurent Massouli{\'e}}.} \bibinfo{year}{2001}\natexlab{}.
\newblock \showarticletitle{Impact of fairness on Internet performance}. In
  \bibinfo{booktitle}{\emph{ACM SIGMETRICS Performance Evaluation Review}},
  Vol.~\bibinfo{volume}{29}. ACM, \bibinfo{pages}{82--91}.
\newblock


\bibitem[\protect\citeauthoryear{Bond, Fariss, Jones, Kramer, Marlow, Settle,
  and Fowler}{Bond et~al\mbox{.}}{2012}]%
        {bond201261}
\bibfield{author}{\bibinfo{person}{Robert~M Bond},
  \bibinfo{person}{Christopher~J Fariss}, \bibinfo{person}{Jason~J Jones},
  \bibinfo{person}{Adam~DI Kramer}, \bibinfo{person}{Cameron Marlow},
  \bibinfo{person}{Jaime~E Settle}, {and} \bibinfo{person}{James~H Fowler}.}
  \bibinfo{year}{2012}\natexlab{}.
\newblock \showarticletitle{A 61-million-person experiment in social influence
  and political mobilization}.
\newblock \bibinfo{journal}{\emph{Nature}} \bibinfo{volume}{489},
  \bibinfo{number}{7415} (\bibinfo{year}{2012}), \bibinfo{pages}{295}.
\newblock


\bibitem[\protect\citeauthoryear{Bucher}{Bucher}{2012}]%
        {Bucher2012}
\bibfield{author}{\bibinfo{person}{Taina Bucher}.}
  \bibinfo{year}{2012}\natexlab{}.
\newblock \showarticletitle{{Want to be on the top? Algorithmic power and the
  threat of invisibility on Facebook}}.
\newblock \bibinfo{journal}{\emph{New Media {\&} Society}}
  (\bibinfo{year}{2012}).
\newblock
\showISBNx{1461-4448$\backslash$r1461-7315}
\showISSN{1461-4448}
\urldef\tempurl%
\url{https://doi.org/10.1177/1461444812440159}
\showDOI{\tempurl}


\bibitem[\protect\citeauthoryear{Celis, Straszak, and Vishnoi}{Celis
  et~al\mbox{.}}{2017}]%
        {Celis2017}
\bibfield{author}{\bibinfo{person}{L~Elisa Celis}, \bibinfo{person}{Damian
  Straszak}, {and} \bibinfo{person}{Nisheeth~K Vishnoi}.}
  \bibinfo{year}{2017}\natexlab{}.
\newblock \showarticletitle{{Ranking with Fairness Constraints}}.
\newblock  (\bibinfo{year}{2017}), \bibinfo{pages}{1--32}.
\newblock
\showISBNx{9783959770767}
\showISSN{0306-4530}
\urldef\tempurl%
\url{https://doi.org/10.4230/LIPIcs.ICALP.2018.28}
\showDOI{\tempurl}
\showeprint[arxiv]{1704.06840}


\bibitem[\protect\citeauthoryear{Cheng, Kleinberg, Leskovec, Liben-Nowell,
  State, Subbian, and Adamic}{Cheng et~al\mbox{.}}{2018}]%
        {Cheng2018}
\bibfield{author}{\bibinfo{person}{Justin Cheng}, \bibinfo{person}{Jon
  Kleinberg}, \bibinfo{person}{Jure Leskovec}, \bibinfo{person}{David
  Liben-Nowell}, \bibinfo{person}{Bogdan State}, \bibinfo{person}{Karthik
  Subbian}, {and} \bibinfo{person}{Lada Adamic}.}
  \bibinfo{year}{2018}\natexlab{}.
\newblock \showarticletitle{{Do Diffusion Protocols Govern Cascade Growth?}}
\newblock \bibinfo{journal}{\emph{Icwsm}} (\bibinfo{year}{2018}).
\newblock
\urldef\tempurl%
\url{https://www-cs.stanford.edu/people/jure/pubs/diffusion-icwsm18.pdf}
\showURL{%
\tempurl}


\bibitem[\protect\citeauthoryear{Corbett-Davies, Pierson, Feller, Goel, and
  Huq}{Corbett-Davies et~al\mbox{.}}{2017}]%
        {corbett2017algorithmic}
\bibfield{author}{\bibinfo{person}{Sam Corbett-Davies}, \bibinfo{person}{Emma
  Pierson}, \bibinfo{person}{Avi Feller}, \bibinfo{person}{Sharad Goel}, {and}
  \bibinfo{person}{Aziz Huq}.} \bibinfo{year}{2017}\natexlab{}.
\newblock \showarticletitle{Algorithmic decision making and the cost of
  fairness}. In \bibinfo{booktitle}{\emph{Proceedings of the 23rd ACM SIGKDD
  International Conference on Knowledge Discovery and Data Mining}}. ACM,
  \bibinfo{pages}{797--806}.
\newblock


\bibitem[\protect\citeauthoryear{Cruch}{Cruch}{2016}]%
        {How_Facebook_NewsFeed_Works}
\bibfield{author}{\bibinfo{person}{Tech Cruch}.}
  \bibinfo{year}{2016}\natexlab{}.
\newblock \bibinfo{title}{How Facebook News Feed Works}.
\newblock
  \bibinfo{howpublished}{\url{https://techcrunch.com/2016/09/06/ultimate-guide-to-the-news-feed/}}.
\newblock


\bibitem[\protect\citeauthoryear{Cruch}{Cruch}{2017}]%
        {Facebook_2bi}
\bibfield{author}{\bibinfo{person}{Tech Cruch}.}
  \bibinfo{year}{2017}\natexlab{}.
\newblock \bibinfo{title}{Facebook now has 2 billion monthly users}.
\newblock
\newblock
\newblock
\shownote{\url{https://techcrunch.com}.}


\bibitem[\protect\citeauthoryear{Datta, Sen, and Zick}{Datta
  et~al\mbox{.}}{2016}]%
        {datta2016algorithmic}
\bibfield{author}{\bibinfo{person}{Anupam Datta}, \bibinfo{person}{Shayak Sen},
  {and} \bibinfo{person}{Yair Zick}.} \bibinfo{year}{2016}\natexlab{}.
\newblock \showarticletitle{Algorithmic transparency via quantitative input
  influence: Theory and experiments with learning systems}. In
  \bibinfo{booktitle}{\emph{Security and Privacy (SP), 2016 IEEE Symposium
  on}}. IEEE, \bibinfo{pages}{598--617}.
\newblock


\bibitem[\protect\citeauthoryear{Dehghan, Massoulie, Towsley, Menasche, and
  Tay}{Dehghan et~al\mbox{.}}{2016}]%
        {Dehghan2016a}
\bibfield{author}{\bibinfo{person}{Mostafa Dehghan}, \bibinfo{person}{Laurent
  Massoulie}, \bibinfo{person}{Don Towsley}, \bibinfo{person}{Daniel Menasche},
  {and} \bibinfo{person}{Y.~C. Tay}.} \bibinfo{year}{2016}\natexlab{}.
\newblock \showarticletitle{{A utility optimization approach to network cache
  design}}. In \bibinfo{booktitle}{\emph{Proceedings of IEEE INFOCOM 2016}}.
\newblock
\showISBNx{9781467399531}
\showISSN{0743166X}


\bibitem[\protect\citeauthoryear{Destounis, Kobayashi, Paschos, and
  Ghorbel}{Destounis et~al\mbox{.}}{2017}]%
        {destounis2017alpha}
\bibfield{author}{\bibinfo{person}{Apostolos Destounis}, \bibinfo{person}{Mari
  Kobayashi}, \bibinfo{person}{Georgios Paschos}, {and} \bibinfo{person}{Asma
  Ghorbel}.} \bibinfo{year}{2017}\natexlab{}.
\newblock \showarticletitle{Alpha fair coded caching}. In
  \bibinfo{booktitle}{\emph{Modeling and Optimization in Mobile, Ad Hoc, and
  Wireless Networks (WiOpt), 2017 15th International Symposium on}}. IEEE,
  \bibinfo{pages}{1--8}.
\newblock


\bibitem[\protect\citeauthoryear{Dhounchak, Kavitha, and Altman}{Dhounchak
  et~al\mbox{.}}{2017}]%
        {dhounchak2017viral}
\bibfield{author}{\bibinfo{person}{Ranbir Dhounchak},
  \bibinfo{person}{Veeraruna Kavitha}, {and} \bibinfo{person}{Eitan Altman}.}
  \bibinfo{year}{2017}\natexlab{}.
\newblock \showarticletitle{{A Viral Timeline Branching Process to study a
  Social Network}}. In \bibinfo{booktitle}{\emph{29th International Teletraffic
  Congress PhD Workshop, {\{}ITC{\}} PhD Workshop 2017}}.
\newblock


\bibitem[\protect\citeauthoryear{Diakopoulos}{Diakopoulos}{2013}]%
        {Diakopoulos2013}
\bibfield{author}{\bibinfo{person}{Nicholas Diakopoulos}.}
  \bibinfo{year}{2013}\natexlab{}.
\newblock \showarticletitle{{Algorithmic accountability reporting: On the
  investigation of black boxes}}.
\newblock \bibinfo{journal}{\emph{Tow Center for Digital Journalism A
  Tow/Knight Brief}} (\bibinfo{year}{2013}), \bibinfo{pages}{1--33}.
\newblock
\showISBNx{9781107671812}
\showISSN{0196-6553}
\urldef\tempurl%
\url{https://doi.org/10.1002/ejoc.201200111}
\showDOI{\tempurl}
\showeprint[arxiv]{arXiv:1011.1669v3}


\bibitem[\protect\citeauthoryear{Drosou, Jagadish, Pitoura, and
  Stoyanovich}{Drosou et~al\mbox{.}}{2017}]%
        {drosou2017diversity}
\bibfield{author}{\bibinfo{person}{Marina Drosou}, \bibinfo{person}{HV
  Jagadish}, \bibinfo{person}{Evaggelia Pitoura}, {and} \bibinfo{person}{Julia
  Stoyanovich}.} \bibinfo{year}{2017}\natexlab{}.
\newblock \showarticletitle{Diversity in big data: A review}.
\newblock \bibinfo{journal}{\emph{Big data}} \bibinfo{volume}{5},
  \bibinfo{number}{2} (\bibinfo{year}{2017}), \bibinfo{pages}{73--84}.
\newblock


\bibitem[\protect\citeauthoryear{Dwork, Hardt, Pitassi, Reingold, and
  Zemel}{Dwork et~al\mbox{.}}{2012}]%
        {dwork2012fairness}
\bibfield{author}{\bibinfo{person}{Cynthia Dwork}, \bibinfo{person}{Moritz
  Hardt}, \bibinfo{person}{Toniann Pitassi}, \bibinfo{person}{Omer Reingold},
  {and} \bibinfo{person}{Richard Zemel}.} \bibinfo{year}{2012}\natexlab{}.
\newblock \showarticletitle{Fairness through awareness}. In
  \bibinfo{booktitle}{\emph{Proceedings of the 3rd innovations in theoretical
  computer science conference}}. ACM, \bibinfo{pages}{214--226}.
\newblock


\bibitem[\protect\citeauthoryear{Epstein and Robertson}{Epstein and
  Robertson}{2015}]%
        {Epstein2015}
\bibfield{author}{\bibinfo{person}{Robert Epstein} {and}
  \bibinfo{person}{Ronald~E Robertson}.} \bibinfo{year}{2015}\natexlab{}.
\newblock \showarticletitle{{The search engine manipulation effect (SEME) and
  its possible impact on the outcomes of elections.}}
\newblock \bibinfo{journal}{\emph{Proceedings of the National Academy of
  Sciences of the United States of America}} \bibinfo{volume}{112},
  \bibinfo{number}{33} (\bibinfo{year}{2015}), \bibinfo{pages}{E4512--21}.
\newblock
\showISBNx{0027-8424}
\showISSN{1091-6490}
\urldef\tempurl%
\url{https://doi.org/10.1073/pnas.1419828112}
\showDOI{\tempurl}


\bibitem[\protect\citeauthoryear{Epstein, Payne, Shen, Dubey, Felbo, Groh,
  Obradovich, Cebri{\'{a}}n, and Rahwan}{Epstein et~al\mbox{.}}{2018}]%
        {skinner_box}
\bibfield{author}{\bibinfo{person}{Ziv Epstein}, \bibinfo{person}{Blakeley~H.
  Payne}, \bibinfo{person}{Judy~Hanwen Shen}, \bibinfo{person}{Abhimanyu
  Dubey}, \bibinfo{person}{Bjarke Felbo}, \bibinfo{person}{Matthew Groh},
  \bibinfo{person}{Nick Obradovich}, \bibinfo{person}{Manuel Cebri{\'{a}}n},
  {and} \bibinfo{person}{Iyad Rahwan}.} \bibinfo{year}{2018}\natexlab{}.
\newblock \showarticletitle{Closing the {AI} Knowledge Gap}.
\newblock \bibinfo{journal}{\emph{CoRR}}  \bibinfo{volume}{abs/1803.07233}
  (\bibinfo{year}{2018}).
\newblock
\showeprint[arxiv]{1803.07233}
\urldef\tempurl%
\url{http://arxiv.org/abs/1803.07233}
\showURL{%
\tempurl}


\bibitem[\protect\citeauthoryear{Eslami, Aleyasen, Karahalios, Hamilton, and
  Sandvig}{Eslami et~al\mbox{.}}{2015a}]%
        {Eslami2015}
\bibfield{author}{\bibinfo{person}{Motahhare Eslami},
  \bibinfo{person}{Amirhossein Aleyasen}, \bibinfo{person}{Karrie Karahalios},
  \bibinfo{person}{Kevin Hamilton}, {and} \bibinfo{person}{Christian Sandvig}.}
  \bibinfo{year}{2015}\natexlab{a}.
\newblock \showarticletitle{Feedvis: A path for exploring news feed curation
  algorithms}. In \bibinfo{booktitle}{\emph{Proceedings of the 18th ACM
  Conference Companion on Computer Supported Cooperative Work \& Social
  Computing}}. ACM, \bibinfo{pages}{65--68}.
\newblock


\bibitem[\protect\citeauthoryear{Eslami, Karahalios, Sandvig, Vaccaro, Rickman,
  Hamilton, and Kirlik}{Eslami et~al\mbox{.}}{2016}]%
        {eslami2016first}
\bibfield{author}{\bibinfo{person}{Motahhare Eslami}, \bibinfo{person}{Karrie
  Karahalios}, \bibinfo{person}{Christian Sandvig}, \bibinfo{person}{Kristen
  Vaccaro}, \bibinfo{person}{Aimee Rickman}, \bibinfo{person}{Kevin Hamilton},
  {and} \bibinfo{person}{Alex Kirlik}.} \bibinfo{year}{2016}\natexlab{}.
\newblock \showarticletitle{{First I like it, then I hide it: Folk theories of
  social feeds}}. In \bibinfo{booktitle}{\emph{Proceedings of the 2016 cHI
  conference on human factors in computing systems}}. ACM,
  \bibinfo{pages}{2371--2382}.
\newblock


\bibitem[\protect\citeauthoryear{Eslami, Rickman, Vaccaro, Aleyasen, Vuong,
  Karahalios, Hamilton, and Sandvig}{Eslami et~al\mbox{.}}{2015b}]%
        {Eslami2015a}
\bibfield{author}{\bibinfo{person}{Motahhare Eslami}, \bibinfo{person}{Aimee
  Rickman}, \bibinfo{person}{Kristen Vaccaro}, \bibinfo{person}{Amirhossein
  Aleyasen}, \bibinfo{person}{Andy Vuong}, \bibinfo{person}{Karrie Karahalios},
  \bibinfo{person}{Kevin Hamilton}, {and} \bibinfo{person}{Christian Sandvig}.}
  \bibinfo{year}{2015}\natexlab{b}.
\newblock \showarticletitle{{"I always assumed that I wasn't really that close
  to [her]"}}.
\newblock \bibinfo{journal}{\emph{Proceedings of the 33rd Annual ACM Conference
  on Human Factors in Computing Systems - CHI '15}} (\bibinfo{year}{2015}),
  \bibinfo{pages}{153--162}.
\newblock
\showISBNx{9781450331456}
\urldef\tempurl%
\url{https://doi.org/10.1145/2702123.2702556}
\showDOI{\tempurl}


\bibitem[\protect\citeauthoryear{Facebook}{Facebook}{2017}]%
        {Facebook_top_stories}
\bibfield{author}{\bibinfo{person}{Facebook}.} \bibinfo{year}{2017}\natexlab{}.
\newblock \bibinfo{title}{Controlling What You See in News Feed}.
\newblock
  \bibinfo{howpublished}{\url{https://www.facebook.com/help/335291769884272}}.
\newblock
\newblock
\shownote{(\url{https://developers.facebook.com/docs/graph-api/}).}


\bibitem[\protect\citeauthoryear{Facebook}{Facebook}{2018a}]%
        {insidenewsfeed}
\bibfield{author}{\bibinfo{person}{Facebook}.}
  \bibinfo{year}{2018}\natexlab{a}.
\newblock
\newblock
\newblock
\shownote{\url{https://newsroom.fb.com/news/2018/05/inside-feed-news-feed-ranking}
  (2:30).}


\bibitem[\protect\citeauthoryear{Facebook}{Facebook}{2018b}]%
        {Facebook_API_home}
\bibfield{author}{\bibinfo{person}{Facebook}.}
  \bibinfo{year}{2018}\natexlab{b}.
\newblock \bibinfo{title}{Graph API Reference}.
\newblock
\newblock
\newblock
\shownote{\url{http://bit.ly/2HB1vzq}.}


\bibitem[\protect\citeauthoryear{{FAT conference}}{{FAT conference}}{2018}]%
        {fatconference}
\bibfield{author}{\bibinfo{person}{{FAT conference}}.}
  \bibinfo{year}{2018}\natexlab{}.
\newblock \bibinfo{title}{Conference on Fairness, Accountability, and
  Transparency (FAT*)}.
\newblock
\newblock
\newblock
\shownote{\url{https://fatconference.org/}.}


\bibitem[\protect\citeauthoryear{Ferragut, Rodr{\'\i}guez, and
  Paganini}{Ferragut et~al\mbox{.}}{2016}]%
        {Ferragut}
\bibfield{author}{\bibinfo{person}{Andr{\'e}s Ferragut},
  \bibinfo{person}{Ismael Rodr{\'\i}guez}, {and} \bibinfo{person}{Fernando
  Paganini}.} \bibinfo{year}{2016}\natexlab{}.
\newblock \showarticletitle{Optimizing TTL caches under heavy-tailed demands}.
  In \bibinfo{booktitle}{\emph{ACM SIGMETRICS Performance Evaluation Review}},
  Vol.~\bibinfo{volume}{44}. ACM, \bibinfo{pages}{101--112}.
\newblock


\bibitem[\protect\citeauthoryear{Fofack, Nain, Neglia, and Towsley}{Fofack
  et~al\mbox{.}}{2012}]%
        {fofack1}
\bibfield{author}{\bibinfo{person}{Nicaise~Choungmo Fofack},
  \bibinfo{person}{Philippe Nain}, \bibinfo{person}{Giovanni Neglia}, {and}
  \bibinfo{person}{Don Towsley}.} \bibinfo{year}{2012}\natexlab{}.
\newblock \showarticletitle{{Analysis of TTL-based cache networks}}. In
  \bibinfo{booktitle}{\emph{Performance Evaluation Methodologies and Tools
  (VALUETOOLS), 2012 6th International Conference on}}. \bibinfo{pages}{1--10}.
\newblock
\showISBNx{978-1-936968-63-3}
\showISSN{1389-1286}
\urldef\tempurl%
\url{https://doi.org/10.1016/j.comnet.2014.03.006}
\showDOI{\tempurl}


\bibitem[\protect\citeauthoryear{for Civic~Media}{for Civic~Media}{2018}]%
        {gobo}
\bibfield{author}{\bibinfo{person}{MIT Media Lab's~Center for Civic~Media}.}
  \bibinfo{year}{2018}\natexlab{}.
\newblock \bibinfo{title}{GOBO}.
\newblock \bibinfo{howpublished}{\url{https://gobo.social/}}.
\newblock


\bibitem[\protect\citeauthoryear{Garfinkel, Matthews, Shapiro, and
  Smith}{Garfinkel et~al\mbox{.}}{2017}]%
        {Garfinkel2017}
\bibfield{author}{\bibinfo{person}{Simson Garfinkel}, \bibinfo{person}{Jeanna
  Matthews}, \bibinfo{person}{Stuart~S. Shapiro}, {and}
  \bibinfo{person}{Jonathan~M. Smith}.} \bibinfo{year}{2017}\natexlab{}.
\newblock \showarticletitle{{Toward algorithmic transparency and
  accountability}}.
\newblock \bibinfo{journal}{\emph{Commun. ACM}} \bibinfo{volume}{60},
  \bibinfo{number}{9} (\bibinfo{year}{2017}), \bibinfo{pages}{5--5}.
\newblock
\showISSN{00010782}
\urldef\tempurl%
\url{https://doi.org/10.1145/3125780}
\showDOI{\tempurl}


\bibitem[\protect\citeauthoryear{Ghodsi, Zaharia, Hindman, Konwinski, Shenker,
  and Stoica}{Ghodsi et~al\mbox{.}}{2011}]%
        {ghodsi2011dominant}
\bibfield{author}{\bibinfo{person}{Ali Ghodsi}, \bibinfo{person}{Matei
  Zaharia}, \bibinfo{person}{Benjamin Hindman}, \bibinfo{person}{Andy
  Konwinski}, \bibinfo{person}{Scott Shenker}, {and} \bibinfo{person}{Ion
  Stoica}.} \bibinfo{year}{2011}\natexlab{}.
\newblock \showarticletitle{Dominant Resource Fairness: Fair Allocation of
  Multiple Resource Types.}. In \bibinfo{booktitle}{\emph{Nsdi}},
  Vol.~\bibinfo{volume}{11}. \bibinfo{pages}{24--24}.
\newblock


\bibitem[\protect\citeauthoryear{Gilbert, Ahrweiler, Barbrook-Johnson,
  Narasimhan, and Wilkinson}{Gilbert et~al\mbox{.}}{2018}]%
        {gilbert2018computational}
\bibfield{author}{\bibinfo{person}{Geoffrey Gilbert}, \bibinfo{person}{Petra
  Ahrweiler}, \bibinfo{person}{Peter Barbrook-Johnson}, \bibinfo{person}{Kavin
  Narasimhan}, {and} \bibinfo{person}{Helen Wilkinson}.}
  \bibinfo{year}{2018}\natexlab{}.
\newblock \showarticletitle{Computational Modelling of Public Policy:
  Reflections on Practice}.
\newblock \bibinfo{journal}{\emph{Journal of Artificial Societies and Social
  Simulation}} \bibinfo{volume}{21}, \bibinfo{number}{1}
  (\bibinfo{year}{2018}), \bibinfo{pages}{1--14}.
\newblock


\bibitem[\protect\citeauthoryear{Gjoka, Kurant, Butts, and Markopoulou}{Gjoka
  et~al\mbox{.}}{2010}]%
        {Gjoka2010}
\bibfield{author}{\bibinfo{person}{Minas Gjoka}, \bibinfo{person}{Maciej
  Kurant}, \bibinfo{person}{Carter~T. Butts}, {and} \bibinfo{person}{Athina
  Markopoulou}.} \bibinfo{year}{2010}\natexlab{}.
\newblock \showarticletitle{{Walking in facebook: A case study of unbiased
  sampling of OSNs}}.
\newblock \bibinfo{journal}{\emph{Proceedings - IEEE INFOCOM}}
  (\bibinfo{year}{2010}).
\newblock
\showISBNx{9781424458363}
\showISSN{0743166X}
\urldef\tempurl%
\url{https://doi.org/10.1109/INFCOM.2010.5462078}
\showDOI{\tempurl}
\showeprint[arxiv]{0906.0060}


\bibitem[\protect\citeauthoryear{Golrezaei, Molisch, Dimakis, and
  Caire}{Golrezaei et~al\mbox{.}}{2013}]%
        {golrezaei2013femtocaching}
\bibfield{author}{\bibinfo{person}{Negin Golrezaei}, \bibinfo{person}{Andreas~F
  Molisch}, \bibinfo{person}{Alexandros~G Dimakis}, {and}
  \bibinfo{person}{Giuseppe Caire}.} \bibinfo{year}{2013}\natexlab{}.
\newblock \showarticletitle{Femtocaching and device-to-device collaboration: A
  new architecture for wireless video distribution}.
\newblock \bibinfo{journal}{\emph{IEEE Communications Magazine}}
  \bibinfo{volume}{51}, \bibinfo{number}{4} (\bibinfo{year}{2013}),
  \bibinfo{pages}{142--149}.
\newblock


\bibitem[\protect\citeauthoryear{Goodman and Flaxman}{Goodman and
  Flaxman}{2016}]%
        {goodman2016eu}
\bibfield{author}{\bibinfo{person}{Bryce Goodman} {and} \bibinfo{person}{Seth
  Flaxman}.} \bibinfo{year}{2016}\natexlab{}.
\newblock \showarticletitle{EU regulations on algorithmic decision-making and a
  “right to explanation”}. In \bibinfo{booktitle}{\emph{ICML workshop on
  human interpretability in machine learning (WHI 2016), New York, NY.
  http://arxiv. org/abs/1606.08813 v1}}.
\newblock


\bibitem[\protect\citeauthoryear{Hargreaves, Agosti, Menasché, Neglia,
  Reiffers-Masson, and Altman}{Hargreaves et~al\mbox{.}}{2018}]%
        {eduardofosint}
\bibfield{author}{\bibinfo{person}{Eduardo Hargreaves},
  \bibinfo{person}{Claudio Agosti}, \bibinfo{person}{Daniel Menasché},
  \bibinfo{person}{Giovanni Neglia}, \bibinfo{person}{Alexandre
  Reiffers-Masson}, {and} \bibinfo{person}{Eitan Altman}.}
  \bibinfo{year}{2018}\natexlab{}.
\newblock \showarticletitle{Biases in the Facebook News Feed: a Case Study on
  the Italian Elections}.
\newblock \bibinfo{journal}{\emph{Proceedings of the IEEE/ACM International
  Conference on Social Networks Analysis and Mining (ASONAM 2018)}}.
\newblock
\urldef\tempurl%
\url{https://doi.org/10.1109/ASONAM.2018.8508659}
\showDOI{\tempurl}
\showeprint[arxiv]{1807.08346}


\bibitem[\protect\citeauthoryear{Jiang, Hegde, Massouli{\'e}, and
  Towsley}{Jiang et~al\mbox{.}}{2013}]%
        {jiang2013optimally}
\bibfield{author}{\bibinfo{person}{Bo Jiang}, \bibinfo{person}{Nidhi Hegde},
  \bibinfo{person}{Laurent Massouli{\'e}}, {and} \bibinfo{person}{Don
  Towsley}.} \bibinfo{year}{2013}\natexlab{}.
\newblock \showarticletitle{How to optimally allocate your budget of attention
  in social networks}. In \bibinfo{booktitle}{\emph{INFOCOM, 2013 Proceedings
  IEEE}}. IEEE, \bibinfo{pages}{2373--2381}.
\newblock


\bibitem[\protect\citeauthoryear{Jones, Bond, Bakshy, Eckles, and Fowler}{Jones
  et~al\mbox{.}}{2017}]%
        {jones2017social}
\bibfield{author}{\bibinfo{person}{Jason~J Jones}, \bibinfo{person}{Robert~M
  Bond}, \bibinfo{person}{Eytan Bakshy}, \bibinfo{person}{Dean Eckles}, {and}
  \bibinfo{person}{James~H Fowler}.} \bibinfo{year}{2017}\natexlab{}.
\newblock \showarticletitle{Social influence and political mobilization:
  Further evidence from a randomized experiment in the 2012 US presidential
  election}.
\newblock \bibinfo{journal}{\emph{PloS one}} \bibinfo{volume}{12},
  \bibinfo{number}{4} (\bibinfo{year}{2017}), \bibinfo{pages}{e0173851}.
\newblock


\bibitem[\protect\citeauthoryear{Jung, Berger, and {Hari Balakrishnan}}{Jung
  et~al\mbox{.}}{2003}]%
        {Jung2003a}
\bibfield{author}{\bibinfo{person}{Jaeyeon Jung}, \bibinfo{person}{A.W.
  Berger}, {and} \bibinfo{person}{{Hari Balakrishnan}}.}
  \bibinfo{year}{2003}\natexlab{}.
\newblock \showarticletitle{{Modeling TTL-based Internet caches}}. In
  \bibinfo{booktitle}{\emph{Proceedings of IEEE INFOCOM 2003}},
  Vol.~\bibinfo{volume}{1}. \bibinfo{publisher}{IEEE},
  \bibinfo{pages}{417--426}.
\newblock
\showISBNx{0-7803-7752-4}
\showISSN{0743-166X}
\urldef\tempurl%
\url{http://ieeexplore.ieee.org/document/1208693/}
\showURL{%
\tempurl}


\bibitem[\protect\citeauthoryear{Kelly}{Kelly}{1997}]%
        {Kelly1997}
\bibfield{author}{\bibinfo{person}{Frank Kelly}.}
  \bibinfo{year}{1997}\natexlab{}.
\newblock \showarticletitle{{Charging and rate control for elastic traffic}}.
\newblock \bibinfo{journal}{\emph{European Transactions on Telecommunications}}
  \bibinfo{volume}{8}, \bibinfo{number}{1} (\bibinfo{year}{1997}),
  \bibinfo{pages}{33--37}.
\newblock
\showISBNx{1541-8251}
\showISSN{1124318X}
\urldef\tempurl%
\url{https://doi.org/10.1002/ett.4460080106}
\showDOI{\tempurl}


\bibitem[\protect\citeauthoryear{Kleinberg, Mullainathan, and
  Raghavan}{Kleinberg et~al\mbox{.}}{2017}]%
        {kleinberg2016inherent}
\bibfield{author}{\bibinfo{person}{Jon Kleinberg}, \bibinfo{person}{Sendhil
  Mullainathan}, {and} \bibinfo{person}{Manish Raghavan}.}
  \bibinfo{year}{2017}\natexlab{}.
\newblock \showarticletitle{{Inherent Trade-Offs in the Fair Determination of
  Risk Scores}}. In \bibinfo{booktitle}{\emph{Proceeding of the 8th Innovations
  in Theoretical Computer Science Conference, ITCS 2017}}.
\newblock


\bibitem[\protect\citeauthoryear{Krishnasamy, Sen, Shakkottai, and
  Oh}{Krishnasamy et~al\mbox{.}}{2016}]%
        {Krishnasamy2016}
\bibfield{author}{\bibinfo{person}{Subhashini Krishnasamy},
  \bibinfo{person}{Rajat Sen}, \bibinfo{person}{Sanjay Shakkottai}, {and}
  \bibinfo{person}{Sewoong Oh}.} \bibinfo{year}{2016}\natexlab{}.
\newblock \showarticletitle{{Detecting Sponsored Recommendations}}.
\newblock \bibinfo{journal}{\emph{ACM Transactions on Modeling and Performance
  Evaluation of Computing Systems}} \bibinfo{volume}{2}, \bibinfo{number}{1}
  (\bibinfo{year}{2016}), \bibinfo{pages}{1--29}.
\newblock
\showISBNx{978-1-4503-3486-0}
\showISSN{23763639}
\urldef\tempurl%
\url{https://doi.org/10.1145/2988543}
\showDOI{\tempurl}
\showeprint[arxiv]{arXiv:1504.03713v1}


\bibitem[\protect\citeauthoryear{Kulshrestha, Eslami, Messias, Zafar, Ghosh,
  Gummadi, and Karahalios}{Kulshrestha et~al\mbox{.}}{2017}]%
        {Kulshrestha2017}
\bibfield{author}{\bibinfo{person}{Juhi Kulshrestha},
  \bibinfo{person}{Motahhare Eslami}, \bibinfo{person}{Johnnatan Messias},
  \bibinfo{person}{Muhammad~Bilal Zafar}, \bibinfo{person}{Saptarshi Ghosh},
  \bibinfo{person}{Krishna~P Gummadi}, {and} \bibinfo{person}{Karrie
  Karahalios}.} \bibinfo{year}{2017}\natexlab{}.
\newblock \showarticletitle{{Quantifying Search Bias: Investigating Sources of
  Bias for Political Searches in Social Media}}.
\newblock \bibinfo{journal}{\emph{20th ACM Conference on Computer-Supported
  Cooperative Work and Social Computing (CSCW 2017)}} (\bibinfo{year}{2017}),
  \bibinfo{pages}{417--432}.
\newblock
\showISBNx{9781450343350}


\bibitem[\protect\citeauthoryear{Kulshrestha, Zafar, Noboa, Gummadi, and
  Ghosh}{Kulshrestha et~al\mbox{.}}{2015}]%
        {juhiInfoDiets}
\bibfield{author}{\bibinfo{person}{Juhi Kulshrestha},
  \bibinfo{person}{Muhammad~Bilal Zafar}, \bibinfo{person}{Lisette~Espin
  Noboa}, \bibinfo{person}{Krishna~P Gummadi}, {and} \bibinfo{person}{Saptarshi
  Ghosh}.} \bibinfo{year}{2015}\natexlab{}.
\newblock \showarticletitle{Characterizing Information Diets of Social Media
  Users}. In \bibinfo{booktitle}{\emph{Proceedings of the International AAAI
  Conference on Web and Social Media}} \emph{(\bibinfo{series}{ICWSM'15})}.
  \bibinfo{pages}{218--227}.
\newblock


\bibitem[\protect\citeauthoryear{Lan, Kao, Chiang, and Sabharwal}{Lan
  et~al\mbox{.}}{2010}]%
        {Lan2010}
\bibfield{author}{\bibinfo{person}{Tian Lan}, \bibinfo{person}{David Kao},
  \bibinfo{person}{Mung Chiang}, {and} \bibinfo{person}{Ashutosh Sabharwal}.}
  \bibinfo{year}{2010}\natexlab{}.
\newblock \showarticletitle{{An Axiomatic Theory of Fairness in Network
  Resource Allocation}}. In \bibinfo{booktitle}{\emph{Proceedings of IEEE
  INFOCOM}}. \bibinfo{pages}{1--9}.
\newblock
\showISBNx{978-1-4244-5836-3}
\urldef\tempurl%
\url{https://doi.org/10.1109/INFCOM.2010.5461911}
\showDOI{\tempurl}
\showeprint[arxiv]{arXiv:0906.0557v4}


\bibitem[\protect\citeauthoryear{Le~Boudec}{Le~Boudec}{2016}]%
        {Boudec2012}
\bibfield{author}{\bibinfo{person}{Jean-Yves Le~Boudec}.}
  \bibinfo{year}{2016}\natexlab{}.
\newblock \bibinfo{booktitle}{\emph{{Rate adaptation, congestion control and
  fairness: A tutorial}}}.
\newblock \bibinfo{type}{{T}echnical {R}eport}.
\newblock
\urldef\tempurl%
\url{http://moodle.epfl.ch/file.php/523/CC{\_}Tutorial/cc.pdf}
\showURL{%
\tempurl}


\bibitem[\protect\citeauthoryear{Lianjie~Shi}{Lianjie~Shi}{2018}]%
        {weightedFairCaching}
\bibfield{author}{\bibinfo{person}{Richard T. B. Ma Y. C.~Tay Lianjie~Shi,
  Xin~Wang}.} \bibinfo{year}{2018}\natexlab{}.
\newblock \showarticletitle{Weighted Fair Caching: Occupancy-Centric Allocation
  for Space-Shared Resources}. In \bibinfo{booktitle}{\emph{IFIP WG Performance
  2018}}.
\newblock


\bibitem[\protect\citeauthoryear{Lichfield}{Lichfield}{2018}]%
        {MIT_TR}
\bibfield{author}{\bibinfo{person}{Gideon Lichfield}.}
  \bibinfo{year}{2018}\natexlab{}.
\newblock \showarticletitle{Technology is threatening our democracy. How do we
  save it?}
\newblock \bibinfo{journal}{\emph{MIT Technology Review}}
  (\bibinfo{date}{September/October} \bibinfo{year}{2018}).
\newblock


\bibitem[\protect\citeauthoryear{Lua}{Lua}{2018}]%
        {facebook1}
\bibfield{author}{\bibinfo{person}{Alfred Lua}.}
  \bibinfo{year}{2018}\natexlab{}.
\newblock \bibinfo{title}{Decoding the Facebook Algorithm: A Fully Up-to-Date
  List of the Algorithm Factors and Changes}.
\newblock
\newblock
\newblock
\shownote{\url{https://blog.bufferapp.com/facebook-news-feed-algorithm}.}


\bibitem[\protect\citeauthoryear{{Microsoft Research}}{{Microsoft
  Research}}{2018}]%
        {fate}
\bibfield{author}{\bibinfo{person}{{Microsoft Research}}.}
  \bibinfo{year}{2018}\natexlab{}.
\newblock \bibinfo{title}{FATE: Fairness, Accountability, Transparency, and
  Ethics in AI}.
\newblock
\newblock
\newblock
\shownote{\url{https://www.microsoft.com/en-us/research/group/fate/}.}


\bibitem[\protect\citeauthoryear{Mo and Walrand}{Mo and Walrand}{2000}]%
        {mo2000fair}
\bibfield{author}{\bibinfo{person}{Jeonghoon Mo} {and} \bibinfo{person}{Jean
  Walrand}.} \bibinfo{year}{2000}\natexlab{}.
\newblock \showarticletitle{Fair end-to-end window-based congestion control}.
\newblock \bibinfo{journal}{\emph{IEEE/ACM Transactions on networking}}
  \bibinfo{volume}{8}, \bibinfo{number}{5} (\bibinfo{year}{2000}),
  \bibinfo{pages}{556--567}.
\newblock


\bibitem[\protect\citeauthoryear{Mosseri}{Mosseri}{2018}]%
        {facebook2}
\bibfield{author}{\bibinfo{person}{Adam Mosseri}.}
  \bibinfo{year}{2018}\natexlab{}.
\newblock \bibinfo{title}{News Feed Ranking in Three Minutes Flat}.
\newblock
\newblock
\newblock
\shownote{\url{https://newsroom.fb.com/news/2018/04/inside-feed-news-feed-ranking/}.}


\bibitem[\protect\citeauthoryear{Moulin}{Moulin}{2004}]%
        {moulin_fair_division}
\bibfield{author}{\bibinfo{person}{Herv{\'{e}} Moulin}.}
  \bibinfo{year}{2004}\natexlab{}.
\newblock \bibinfo{booktitle}{\emph{{Fair Division and Collective Welfare}}}.
\newblock \bibinfo{publisher}{MIT Press}.
\newblock
\showISBNx{9780262134231}


\bibitem[\protect\citeauthoryear{Narayanan}{Narayanan}{2018}]%
        {talk21defs}
\bibfield{author}{\bibinfo{person}{Arvind Narayanan}.}
  \bibinfo{year}{2018}\natexlab{}.
\newblock \bibinfo{title}{Tutorial: 21 Fairness Definitions and Their
  Politics}.
\newblock
\newblock
\newblock
\shownote{\url{https://fatconference.org/2018/livestream_vh220.html}.}


\bibitem[\protect\citeauthoryear{Neglia, Carra, and Michiardi}{Neglia
  et~al\mbox{.}}{2017}]%
        {Neglia2017}
\bibfield{author}{\bibinfo{person}{Giovanni Neglia}, \bibinfo{person}{Damiano
  Carra}, {and} \bibinfo{person}{Pietro Michiardi}.}
  \bibinfo{year}{2017}\natexlab{}.
\newblock \showarticletitle{{Cache policies for linear utility maximization}}.
  In \bibinfo{booktitle}{\emph{Proceedings of IEEE INFOCOM 2017}}.
\newblock
\showISBNx{9781509053360}
\showISSN{10636692}


\bibitem[\protect\citeauthoryear{Palomar and Chiang}{Palomar and
  Chiang}{2006}]%
        {palomar2006tutorial}
\bibfield{author}{\bibinfo{person}{Daniel~P{\'e}rez Palomar} {and}
  \bibinfo{person}{Mung Chiang}.} \bibinfo{year}{2006}\natexlab{}.
\newblock \showarticletitle{A tutorial on decomposition methods for network
  utility maximization}.
\newblock \bibinfo{journal}{\emph{IEEE Journal on Selected Areas in
  Communications}} \bibinfo{volume}{24}, \bibinfo{number}{8}
  (\bibinfo{year}{2006}), \bibinfo{pages}{1439--1451}.
\newblock


\bibitem[\protect\citeauthoryear{Pariser}{Pariser}{2011}]%
        {pariser2011filter}
\bibfield{author}{\bibinfo{person}{Eli Pariser}.}
  \bibinfo{year}{2011}\natexlab{}.
\newblock \bibinfo{booktitle}{\emph{The Filter Bubble: What the Internet Is
  Hiding from You}}.
\newblock \bibinfo{publisher}{The Penguin Group}.
\newblock
\showISBNx{1594203008, 9781594203008}


\bibitem[\protect\citeauthoryear{Pleiss, Raghavan, Wu, Kleinberg, and
  Weinberger}{Pleiss et~al\mbox{.}}{2017}]%
        {pleiss2017fairness}
\bibfield{author}{\bibinfo{person}{Geoff Pleiss}, \bibinfo{person}{Manish
  Raghavan}, \bibinfo{person}{Felix Wu}, \bibinfo{person}{Jon Kleinberg}, {and}
  \bibinfo{person}{Kilian~Q Weinberger}.} \bibinfo{year}{2017}\natexlab{}.
\newblock \showarticletitle{On fairness and calibration}. In
  \bibinfo{booktitle}{\emph{Advances in Neural Information Processing
  Systems}}. \bibinfo{pages}{5684--5693}.
\newblock


\bibitem[\protect\citeauthoryear{Reiffers-Masson, Hargreaves, Altman, Caarls,
  and Menasch{\'e}}{Reiffers-Masson et~al\mbox{.}}{2017a}]%
        {netecon}
\bibfield{author}{\bibinfo{person}{Alexandre Reiffers-Masson},
  \bibinfo{person}{Eduardo Hargreaves}, \bibinfo{person}{Eitan Altman},
  \bibinfo{person}{Wouter Caarls}, {and} \bibinfo{person}{Daniel~S.
  Menasch{\'e}}.} \bibinfo{year}{2017}\natexlab{a}.
\newblock \showarticletitle{Timelines Are Publisher-Driven Caches: Analyzing
  and Shaping Timeline Networks}.
\newblock \bibinfo{journal}{\emph{SIGMETRICS Perform. Eval. Rev.}}
  \bibinfo{volume}{44}, \bibinfo{number}{3} (\bibinfo{date}{Jan.}
  \bibinfo{year}{2017}), \bibinfo{pages}{26--29}.
\newblock
\showISSN{0163-5999}


\bibitem[\protect\citeauthoryear{Reiffers-Masson, Hayel, and
  Altman}{Reiffers-Masson et~al\mbox{.}}{2017b}]%
        {masson2017posting}
\bibfield{author}{\bibinfo{person}{Alexandre Reiffers-Masson},
  \bibinfo{person}{Yezekael Hayel}, {and} \bibinfo{person}{Eitan Altman}.}
  \bibinfo{year}{2017}\natexlab{b}.
\newblock \showarticletitle{Posting Behavior Dynamics and Active Filtering for
  Content Diversity in Social Networks}.
\newblock \bibinfo{journal}{\emph{IEEE transactions on Signal and Information
  Processing over Networks}} \bibinfo{volume}{3}, \bibinfo{number}{2}
  (\bibinfo{year}{2017}), \bibinfo{pages}{376--387}.
\newblock


\bibitem[\protect\citeauthoryear{Ribeiro, Henrique, Benevenuto, Chakraborty,
  Kulshrestha, Babaei, and Gummadi}{Ribeiro et~al\mbox{.}}{2018}]%
        {mediabiasmonitor}
\bibfield{author}{\bibinfo{person}{Filipe Ribeiro}, \bibinfo{person}{Lucas
  Henrique}, \bibinfo{person}{Fabrício Benevenuto}, \bibinfo{person}{Abhijnan
  Chakraborty}, \bibinfo{person}{Juhi Kulshrestha},
  \bibinfo{person}{Mahmoudreza Babaei}, {and} \bibinfo{person}{Krishna~P.
  Gummadi}.} \bibinfo{year}{2018}\natexlab{}.
\newblock \showarticletitle{Media Bias Monitor: Quantifying Biases of Social
  Media News Outlets at Large-Scale}. In \bibinfo{booktitle}{\emph{Proceedings
  of the International AAAI Conference on Web and Social Media}}
  \emph{(\bibinfo{series}{ICWSM'18})}.
\newblock


\bibitem[\protect\citeauthoryear{Romei and Ruggieri}{Romei and
  Ruggieri}{2014}]%
        {romei2014multidisciplinary}
\bibfield{author}{\bibinfo{person}{Andrea Romei} {and}
  \bibinfo{person}{Salvatore Ruggieri}.} \bibinfo{year}{2014}\natexlab{}.
\newblock \showarticletitle{A multidisciplinary survey on discrimination
  analysis}.
\newblock \bibinfo{journal}{\emph{The Knowledge Engineering Review}}
  \bibinfo{volume}{29}, \bibinfo{number}{5} (\bibinfo{year}{2014}),
  \bibinfo{pages}{582--638}.
\newblock


\bibitem[\protect\citeauthoryear{Rossi, Polderman, and Frasca}{Rossi
  et~al\mbox{.}}{2018}]%
        {Rossi2018}
\bibfield{author}{\bibinfo{person}{Wilbert~Samuel Rossi},
  \bibinfo{person}{Jan~Willem Polderman}, {and} \bibinfo{person}{Paolo
  Frasca}.} \bibinfo{year}{2018}\natexlab{}.
\newblock \showarticletitle{{The closed loop between opinion formation and
  personalised recommendations}}.
\newblock  (\bibinfo{year}{2018}), \bibinfo{pages}{1--24}.
\newblock
\showeprint[arxiv]{arXiv:1809.04644v1}


\bibitem[\protect\citeauthoryear{Sandvig, Hamilton, Karahalios, and
  Langbort}{Sandvig et~al\mbox{.}}{2014}]%
        {Sandvig2014}
\bibfield{author}{\bibinfo{person}{Christian Sandvig}, \bibinfo{person}{Kevin
  Hamilton}, \bibinfo{person}{Karrie Karahalios}, {and} \bibinfo{person}{Cedric
  Langbort}.} \bibinfo{year}{2014}\natexlab{}.
\newblock \showarticletitle{{Auditing Algorithms : Research Methods for
  Detecting Discrimination on Internet Platforms}}.
\newblock \bibinfo{journal}{\emph{International Communication Association}}
  (\bibinfo{year}{2014}), \bibinfo{pages}{1--20}.
\newblock
\showISBNx{9781450329682}
\showISSN{00404411}
\urldef\tempurl%
\url{https://doi.org/10.1109/DEXA.2009.55}
\showDOI{\tempurl}


\bibitem[\protect\citeauthoryear{Shakkottai and Srikant}{Shakkottai and
  Srikant}{2008}]%
        {shakkottai2008network}
\bibfield{author}{\bibinfo{person}{Srinivas Shakkottai} {and}
  \bibinfo{person}{Rayadurgam Srikant}.} \bibinfo{year}{2008}\natexlab{}.
\newblock \showarticletitle{Network optimization and control}.
\newblock \bibinfo{journal}{\emph{Foundations and Trends in Networking}}
  \bibinfo{volume}{2}, \bibinfo{number}{3} (\bibinfo{year}{2008}),
  \bibinfo{pages}{271--379}.
\newblock


\bibitem[\protect\citeauthoryear{Singh and Joachims}{Singh and
  Joachims}{2018}]%
        {Singh2018}
\bibfield{author}{\bibinfo{person}{Ashudeep Singh} {and}
  \bibinfo{person}{Thorsten Joachims}.} \bibinfo{year}{2018}\natexlab{}.
\newblock \showarticletitle{{Fairness of Exposure in Rankings}}. In
  \bibinfo{booktitle}{\emph{Proceedings of the 24th ACM SIGKDD International
  Conference on Knowledge Discovery {\&} Data Mining - KDD '18}}.
  \bibinfo{publisher}{ACM Press}, \bibinfo{address}{New York, New York, USA},
  \bibinfo{pages}{2219--2228}.
\newblock
\showISBNx{9781450355520}
\urldef\tempurl%
\url{https://doi.org/10.1145/3219819.3220088}
\showDOI{\tempurl}
\showeprint[arxiv]{arXiv:1802.07281v2}


\bibitem[\protect\citeauthoryear{Sleeper, Balebako, Das, McConahy, Wiese, and
  Cranor}{Sleeper et~al\mbox{.}}{2013}]%
        {Sleeper2013}
\bibfield{author}{\bibinfo{person}{Manya Sleeper}, \bibinfo{person}{Rebecca
  Balebako}, \bibinfo{person}{Sauvik Das}, \bibinfo{person}{Amber~Lynn
  McConahy}, \bibinfo{person}{Jason Wiese}, {and} \bibinfo{person}{Lorrie~Faith
  Cranor}.} \bibinfo{year}{2013}\natexlab{}.
\newblock \showarticletitle{{The post that wasn't: exploring self-censorship on
  facebook}}.
\newblock \bibinfo{journal}{\emph{Proceedings of the 2013 conference on
  Computer supported cooperative work}} (\bibinfo{year}{2013}),
  \bibinfo{pages}{793}.
\newblock
\showISBNx{9781450313315}
\urldef\tempurl%
\url{https://doi.org/10.1145/2441776.2441865}
\showDOI{\tempurl}


\bibitem[\protect\citeauthoryear{{Stanford computational policy lab}}{{Stanford
  computational policy lab}}{2018}]%
        {stanfordcomp}
\bibfield{author}{\bibinfo{person}{{Stanford computational policy lab}}.}
  \bibinfo{year}{2018}\natexlab{}.
\newblock \bibinfo{title}{Stanford computational policy lab}.
\newblock
\newblock
\newblock
\shownote{\url{https://policylab.stanford.edu/}.}


\bibitem[\protect\citeauthoryear{Sun, Rosenn, Marlow, and Lento}{Sun
  et~al\mbox{.}}{2009}]%
        {Sun2009}
\bibfield{author}{\bibinfo{person}{Eric Sun}, \bibinfo{person}{Itamar Rosenn},
  \bibinfo{person}{Cameron~A. Marlow}, {and} \bibinfo{person}{Thomas~M.
  Lento}.} \bibinfo{year}{2009}\natexlab{}.
\newblock \showarticletitle{{Gesundheit ! Modeling Contagion through Facebook
  News Feed Mechanics of Facebook Page Diffusion}}.
\newblock \bibinfo{journal}{\emph{Proceedings of the 3rd International ICWSM
  Conference}} \bibinfo{number}{2000} (\bibinfo{year}{2009}),
  \bibinfo{pages}{146--153}.
\newblock
\showISBNx{978-1-57735-421-5}
\urldef\tempurl%
\url{https://doi.org/no DOI. URL correct}
\showDOI{\tempurl}


\bibitem[\protect\citeauthoryear{Tadrous and Eryilmaz}{Tadrous and
  Eryilmaz}{2016}]%
        {tadrous2016optimal}
\bibfield{author}{\bibinfo{person}{John Tadrous} {and} \bibinfo{person}{Atilla
  Eryilmaz}.} \bibinfo{year}{2016}\natexlab{}.
\newblock \showarticletitle{On optimal proactive caching for mobile networks
  with demand uncertainties}.
\newblock \bibinfo{journal}{\emph{IEEE/ACM Transactions on Networking}}
  \bibinfo{volume}{24}, \bibinfo{number}{5} (\bibinfo{year}{2016}),
  \bibinfo{pages}{2715--2727}.
\newblock


\bibitem[\protect\citeauthoryear{Tan, Caruana, Hooker, and Lou}{Tan
  et~al\mbox{.}}{2017}]%
        {Tan2017}
\bibfield{author}{\bibinfo{person}{Sarah Tan}, \bibinfo{person}{Rich Caruana},
  \bibinfo{person}{Giles Hooker}, {and} \bibinfo{person}{Yin Lou}.}
  \bibinfo{year}{2017}\natexlab{}.
\newblock \showarticletitle{{Detecting Bias in Black-Box Models Using
  Transparent Model Distillation}}.
\newblock  \bibinfo{number}{1} (\bibinfo{year}{2017}).
\newblock
\showeprint[arxiv]{1710.06169}
\urldef\tempurl%
\url{http://arxiv.org/abs/1710.06169}
\showURL{%
\tempurl}


\bibitem[\protect\citeauthoryear{Tortelli, Rossi, and Leonardi}{Tortelli
  et~al\mbox{.}}{2017}]%
        {tortellimodel}
\bibfield{author}{\bibinfo{person}{Michele Tortelli}, \bibinfo{person}{Dario
  Rossi}, {and} \bibinfo{person}{Emilio Leonardi}.}
  \bibinfo{year}{2017}\natexlab{}.
\newblock \showarticletitle{{ModelGraft: Accurate, scalable, and flexible
  performance evaluation of general cache networks}}. In
  \bibinfo{booktitle}{\emph{Proceedings of the 28th International Teletraffic
  Congress, ITC 2016}}, Vol.~\bibinfo{volume}{1}. \bibinfo{pages}{304--312}.
\newblock
\showISBNx{9780988304512}


\bibitem[\protect\citeauthoryear{Tufekci}{Tufekci}{2015}]%
        {message}
\bibfield{author}{\bibinfo{person}{Zeynep Tufekci}.}
  \bibinfo{year}{2015}\natexlab{}.
\newblock \bibinfo{title}{How Facebook’s Algorithm Suppresses Content
  Diversity (Modestly) and How the Newsfeed Rules Your Clicks}.
\newblock \bibinfo{howpublished}{\url{https://medium.com/message}}.
\newblock


\bibitem[\protect\citeauthoryear{Ugander, Karrer, Backstrom, and
  Marlow}{Ugander et~al\mbox{.}}{2011}]%
        {anatomy_fB}
\bibfield{author}{\bibinfo{person}{Johan Ugander}, \bibinfo{person}{Brian
  Karrer}, \bibinfo{person}{Lars Backstrom}, {and} \bibinfo{person}{Cameron
  Marlow}.} \bibinfo{year}{2011}\natexlab{}.
\newblock \showarticletitle{{The Anatomy of the Facebook Social Graph}}.
\newblock \bibinfo{journal}{\emph{CoRR}} (\bibinfo{year}{2011}).
\newblock
\showISBNx{9781450324656}
\showISSN{15525996}
\urldef\tempurl%
\url{https://doi.org/10.1145/2532508.2532512}
\showDOI{\tempurl}
\showeprint[arxiv]{1111.4503v1}


\bibitem[\protect\citeauthoryear{Valenzuela, Park, and Kee}{Valenzuela
  et~al\mbox{.}}{2009}]%
        {valenzuela2009there}
\bibfield{author}{\bibinfo{person}{Sebasti{\'a}n Valenzuela},
  \bibinfo{person}{Namsu Park}, {and} \bibinfo{person}{Kerk~F Kee}.}
  \bibinfo{year}{2009}\natexlab{}.
\newblock \showarticletitle{Is there social capital in a social network site?:
  Facebook use and college students' life satisfaction, trust, and
  participation}.
\newblock \bibinfo{journal}{\emph{Journal of computer-mediated communication}}
  \bibinfo{volume}{14}, \bibinfo{number}{4} (\bibinfo{year}{2009}),
  \bibinfo{pages}{875--901}.
\newblock


\bibitem[\protect\citeauthoryear{Varian}{Varian}{1983}]%
        {varian1983non}
\bibfield{author}{\bibinfo{person}{Hal~R Varian}.}
  \bibinfo{year}{1983}\natexlab{}.
\newblock \showarticletitle{Non-parametric tests of consumer behaviour}.
\newblock \bibinfo{journal}{\emph{The review of economic studies}}
  \bibinfo{volume}{50}, \bibinfo{number}{1} (\bibinfo{year}{1983}),
  \bibinfo{pages}{99--110}.
\newblock


\bibitem[\protect\citeauthoryear{Wang, Tyson, Kangasharju, and Crowcroft}{Wang
  et~al\mbox{.}}{2017}]%
        {an2013individuals}
\bibfield{author}{\bibinfo{person}{Liang Wang}, \bibinfo{person}{Gareth Tyson},
  \bibinfo{person}{Jussi Kangasharju}, {and} \bibinfo{person}{Jon Crowcroft}.}
  \bibinfo{year}{2017}\natexlab{}.
\newblock \showarticletitle{Milking the Cache Cow With Fairness in Mind}.
\newblock \bibinfo{journal}{\emph{IEEE/ACM Transactions on Networking}}
  \bibinfo{volume}{25}, \bibinfo{number}{5} (\bibinfo{year}{2017}).
\newblock
\urldef\tempurl%
\url{https://doi.org/10.1109/TNET.2017.2707131}
\showDOI{\tempurl}


\bibitem[\protect\citeauthoryear{Wing}{Wing}{2018}]%
        {fates}
\bibfield{author}{\bibinfo{person}{Jeannette Wing}.}
  \bibinfo{year}{2018}\natexlab{}.
\newblock \bibinfo{title}{Data for Good: Scary AI and the Dangers of Big Data}.
\newblock
\newblock
\newblock
\shownote{\url{https://www.youtube.com/watch?v=YL6zXHa5-RY }.}


\bibitem[\protect\citeauthoryear{Zafar, Valera, Rodriguez, and Gummadi}{Zafar
  et~al\mbox{.}}{2015}]%
        {zafar2015learning}
\bibfield{author}{\bibinfo{person}{Muhammad~Bilal Zafar},
  \bibinfo{person}{Isabel Valera}, \bibinfo{person}{Manuel~Gomez Rodriguez},
  {and} \bibinfo{person}{Krishna~P Gummadi}.} \bibinfo{year}{2015}\natexlab{}.
\newblock \showarticletitle{Learning fair classifiers}.
\newblock \bibinfo{journal}{\emph{arXiv preprint arXiv:1507.05259}}
  (\bibinfo{year}{2015}).
\newblock


\bibitem[\protect\citeauthoryear{Zehlike, Bonchi, Castillo, Hajian, Megahed,
  and Baeza-Yates}{Zehlike et~al\mbox{.}}{2017}]%
        {Zehlike2017}
\bibfield{author}{\bibinfo{person}{Meike Zehlike}, \bibinfo{person}{Francesco
  Bonchi}, \bibinfo{person}{Carlos Castillo}, \bibinfo{person}{Sara Hajian},
  \bibinfo{person}{Mohamed Megahed}, {and} \bibinfo{person}{Ricardo
  Baeza-Yates}.} \bibinfo{year}{2017}\natexlab{}.
\newblock \showarticletitle{{FA*IR: A Fair Top-k Ranking Algorithm}}. In
  \bibinfo{booktitle}{\emph{Proceedings of the 2017 ACM on Conference on
  Information and Knowledge Management - CIKM '17}}. \bibinfo{publisher}{ACM
  Press}, \bibinfo{address}{New York, NY, USA}, \bibinfo{pages}{1569--1578}.
\newblock
\showISBNx{9781450349185}
\urldef\tempurl%
\url{https://doi.org/10.1145/3132847.3132938}
\showDOI{\tempurl}
\showeprint[arxiv]{1706.06368}


\bibitem[\protect\citeauthoryear{Zhang and Bareinboim}{Zhang and
  Bareinboim}{2017}]%
        {zhangfairness}
\bibfield{author}{\bibinfo{person}{Junzhe Zhang} {and} \bibinfo{person}{Elias
  Bareinboim}.} \bibinfo{year}{2017}\natexlab{}.
\newblock \showarticletitle{Fairness in Decision-Making: The Causal Explanation
  Formula}.
\newblock \bibinfo{journal}{\emph{Purdue technical report R-30-L}}
  (\bibinfo{year}{2017}).
\newblock


\bibitem[\protect\citeauthoryear{{\v{Z}}liobait{\.e}}{{\v{Z}}liobait{\.e}}{2017}]%
        {vzliobaite2017measuring}
\bibfield{author}{\bibinfo{person}{Indr{\.e} {\v{Z}}liobait{\.e}}.}
  \bibinfo{year}{2017}\natexlab{}.
\newblock \showarticletitle{Measuring discrimination in algorithmic decision
  making}.
\newblock \bibinfo{journal}{\emph{Data Mining and Knowledge Discovery}}
  \bibinfo{volume}{31}, \bibinfo{number}{4} (\bibinfo{year}{2017}),
  \bibinfo{pages}{1060--1089}.
\newblock


\end{thebibliography}
